\documentclass{article}
\usepackage{amssymb}
\usepackage{amsmath}
\usepackage{amsfonts}
\usepackage{hyperref}
\usepackage{graphicx}
\usepackage{subfigure}

\input{tcilatex}

\begin{document}

\title{Pricing of vanilla and first generation exotic options in the local
stochastic volatility framework: survey and new results}
\author{Alexander Lipton \\
Bank of America Merrill Lynch and Imperial College \and Andrey Gal \\
Bank of America Merrill Lynch \and Andris Lasis \\
Bank of America Merrill Lynch}
\maketitle

\begin{abstract}
Stochastic volatility (SV) and local stochastic volatility (LSV) processes
can be used to model the evolution of various financial variables such as FX
rates, stock prices, and so on. Considerable efforts have been devoted to
pricing derivatives written on underliers governed by such processes. Many
issues remain, though, including the efficacy of the standard alternating
direction implicit (ADI) numerical methods for solving SV and LSV pricing
problems. In general, the amount of required computations for these methods
is very substantial. In this paper we address some of these issues and
propose a viable alternative to the standard ADI methods based on
Galerkin-Ritz ideas. We also discuss various approaches to solving the
corresponding pricing problems in a semi-analytical fashion. We use the fact
that in the zero correlation case some of the pricing problems can be solved
analytically, and develop a closed-form series expansion in powers of
correlation. We perform a thorough benchmarking of various numerical
solutions by using analytical and semi-analytical solutions derived in the
paper.
\end{abstract}

\tableofcontents

\section{Introduction\label{Intro}}

In the standard European option pricing model of Black-Scholes and Merton
(BSM) (see \cite{bs} and \cite{merton-1}), forward price processes are
assumed to be log-normal and characterized by a single volatility $\sigma $.
The corresponding SDE has the form%
\begin{equation}
dF_{t}=\sigma F_{t}dW_{t},\ \ \ \ \ F_{0}=F,  \label{ln_sde}
\end{equation}%
where $F_{t}$ is the observable forward price for a particular maturity $%
\hat{T}$, $\sigma $ is a constant volatility, and $W_{t}$ is a Brownian
motion. Note that Eq. (\ref{ln_sde}) assumes that the asset price $F_{t}$ is
a risk-neutral martingale. Such dynamics immediately leads to a closed-form
formula for the price of a call option on an asset $F_{t}$ paying $%
(F_{T}-K)^{+}$ at expiration time $T\geq t$, $T\leq \hat{T}$. At time $t$,
the undiscounted price $C^{BS}(t,F_{t};T,K)$ is given by 
\begin{equation}
\frac{C^{BS}(t,F_{t};T,K;\sigma )}{F_{t}}=\Phi (d_{+})-e^{X_{K}}\Phi (d_{-}),
\label{BS_formula}
\end{equation}%
where $\Phi (\cdot )$ is the cumulative Gaussian distribution function, and 
\begin{equation}
d_{\pm }=\frac{-X_{K}\pm \QTOVERD. . {1}{2}\sigma ^{2}\tau }{\sigma \sqrt{%
\tau }},  \label{d_pm}
\end{equation}%
with $X_{K}\triangleq \ln (K/F_{t})$, $\tau \triangleq T-t$. Here and below,
as usual,%
\begin{equation}
\left( x\right) ^{\pm }=\pm \max \left( \pm x,0\right) .  \label{x_pm}
\end{equation}

\begin{figure}[h]
\center%
\includegraphics[width=1.00\textwidth, angle=0]
{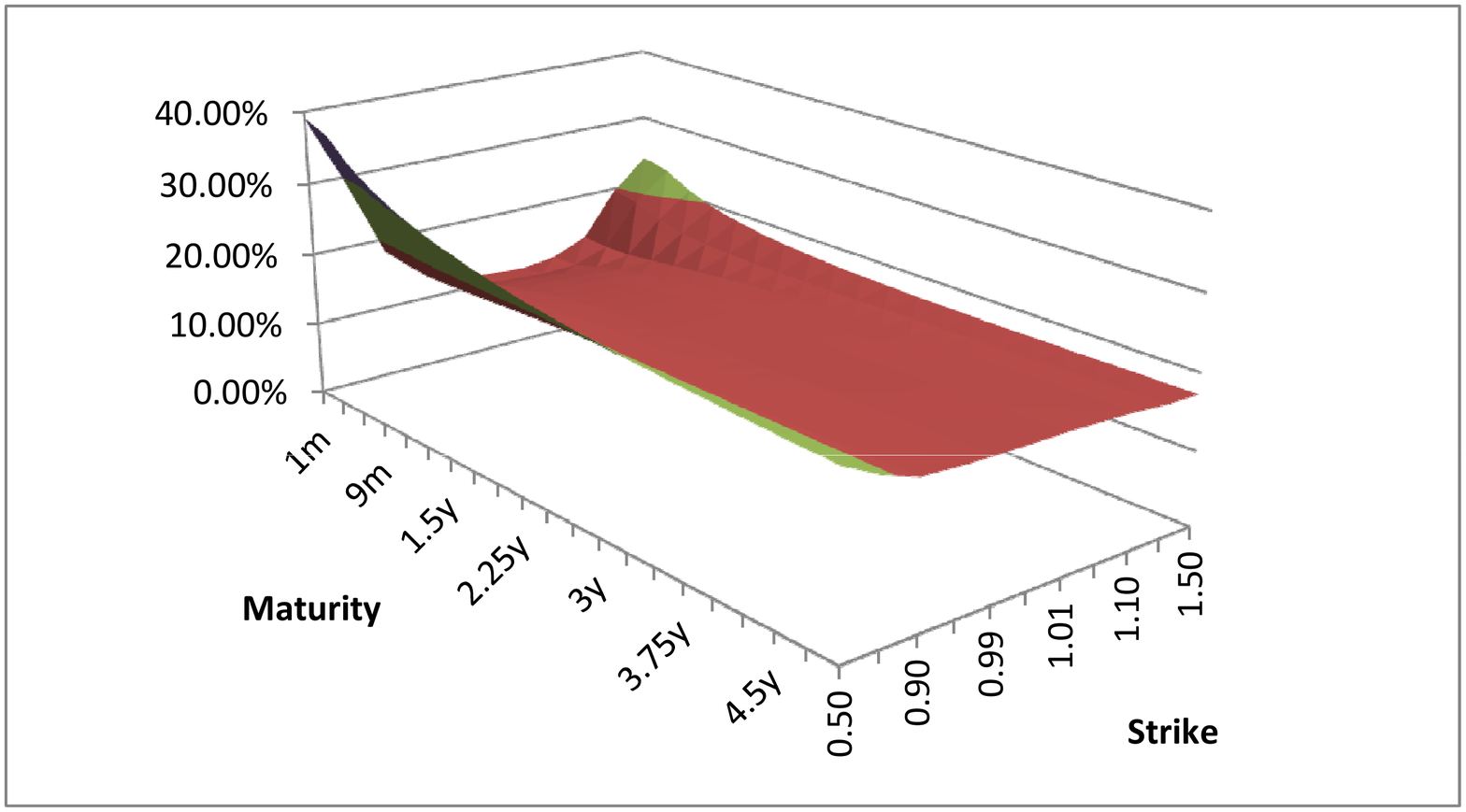}
\caption{AUDJPY implied volatility surface $\protect\sigma _{imp}\left( 
\protect\tau ,K\right) $. The quotes are for the 22nd of August, 2012.}
\label{fig:AUDJPY}
\end{figure}

In reality, the market prices of call options rarely agree with their
theoretical values, so, in order to make the BSM formula (\ref{BS_formula})
work, practitioners are forced to introduce the so-called implied volatility 
$\sigma _{imp}(t,F_{t};\tau ,K)$, which depends on option maturity $(\tau )$
and strike $(K)$. In virtually all option markets a strike- and
maturity-dependent \textit{implied volatility surface}, $\sigma
_{imp}(t,F_{t};\tau ,K)$ is of paramount importance. By using this surface,
we can write the price of a call option with strike $K$ and expiration time $%
T\geq t$ in the form (\ref{BS_formula}) with $d_{\pm }$ of the form%
\begin{equation}
d_{\pm }=\frac{-X_{K}\pm \QTOVERD. . {1}{2}\sigma _{imp}^{2}(t,F_{t};\tau
,K)\tau }{\sigma _{imp}(t,F_{t};\tau ,K)\sqrt{\tau }}.  \label{d_pm1}
\end{equation}%
A typical volatility surface for the AUDJPY currency pair is shown in Figure %
\ref{fig:AUDJPY}.%
\begin{equation*}
\text{Fig \ref{fig:AUDJPY} near here.}
\end{equation*}

In order to explain the existence and behavior of the implied volatility,
various alternatives to the dynamics (\ref{ln_sde}) have been proposed in
the literature, see, e.g., \cite{merton-2}, \cite{dupire}, \cite{heston}, 
\cite{bates}, \cite{jex}, \cite{and-and}, \cite{britten}, \cite{blacher}, 
\cite{lipton-2}, \cite{hagan et al}, \cite{cont-tankov}, among others.
Broadly speaking, the following approaches have been discussed in the
literature:(A) Local volatility (LV) models, assuming that $\sigma $ is a
deterministic function of $t$ and $F_{t}$; (B) Stochastic volatility (SV)
models, assuming that $\sigma $ is a random variable, possibly correlated
with $F_{t}$, but not depending on $F_{t}$ directly; (C) Local stochastic
volatility (LSV) models, combining local and stochastic volatility dynamics;
(D) Jump diffusion (JD) models, assuming that the process for $F_{t}$
incorporates jumps; (E) Universal volatility (UV) models, combining LV, SV,
and JD models, and adding volatility jumps.

Whilst theoretically appealing, full-blown UV models are seldom used in
practice because of their complexity; instead, different asset classes tend
to use simpler models reflecting the most relevant features of their
respective underliers. For instance, equity-linked products are
predominantly priced via LV models, while LSV are \textit{de facto} standard
for pricing FX options; credit products are often priced with JD models. In
all cases, values of options are given by partial differential equations
supplemented with initial and boundary conditions. These equations are
derived directly from stochastic volatility dynamics using standard It\^{o}
calculus techniques. They are typically solved by combining numerical,
analytical, and asymptotic methods.

In this paper we review some familiar and widely used numerical methods for
solving PDEs for the classical Heston stochastic volatility model and its
generalizations; we also propose some new numerical and analytical
techniques. Specifically, we study a variety of finite difference (FD)
methods applied to the Heston PDE: an explicit finite difference (EFD)
scheme based on Fast Exponentiation, which can be viewed as a simplified
version of the scheme due to \cite{osullivan}, and four
alternating-directions implicit (ADI) schemes, due to \cite{douglas}, \cite%
{craig}, \cite{hunsdorfer}, and \cite{inthout2}. Building on this, we
introduce the Galerkin method (or, perhaps more accurately, the
Galerkin-Ritz method), which allows us to obtain a good representation of
the correlation term without the time--averaging step, as in the FD
approach. To the best of our knowledge, this method has not been used for
solving LSV before. This method has significant advantages compared to ADI
methods because, as we shall demonstrate later, it treats the pricing
problem in a more natural fashion. Following this, we present a method of
analytical expansion in powers of $\rho $, which allows us to obtain a
close--to--analytical solution of a pricing problem. We also briefly discuss
the Monte Carlo (MC) method in the Heston model context.

The paper is organized as follows. In Section \ref{LSVFormulation} we
introduce the LSV model and apply the Liouville transform to write it in a
simple and uniform way. We place particular emphasis on the so-called
quadratic LSV (QLSV) model. We show that the standard Heston and the
displaced Heston models can be viewed as special cases of the QLSV\ model.
In Section \ref{NumSol} we discuss various numerical methods for solving the
pricing problem for vanilla and first generation exotic options for LSV
models in general, and the QLSV model in particular. In Section \ref%
{PricingCall} we formulate the Liouville transformed pricing problem for the
call option and show how it can be solved both analytically and numerically.
Section \ref{DNT}, which is dedicated to the analysis of double no-touch
(DNT) options, constitutes the heart of the paper. We compute and compare
the prices of such options obtained by the various methods described in
Section \ref{NumSol} and conclude that these prices are in agreement. In
order to get additional confirmation of the validity of the Galerkin method,
we dedicate Section \ref{Two_corr_BM} to studying a related (but not
identical) pricing problem for two-dimensional Brownian motion in a positive
quadrant and in a rectangle with absorbing boundaries. We find that, as
before, we have very good agreement among solutions computed by the
different methods. We draw our conclusions in Section \ref{Conclusion}.
Finally, in the Appendices we derive some of the formulas used in the main
body of the paper, and make some additional comments.

\section{Local stochastic volatility pricing problem\label{LSVFormulation}}

Assuming for simplicity that interest rates are zero, we can write the most
general system of SDEs describing the risk-neutral local stochastic
volatility (LSV) dynamics in the form%
\begin{equation}
\begin{array}{ll}
dF_{t}=\mathsf{\sigma }\left( t,F_{t},A_{t}\right) dW_{t}, & F_{0}=F, \\ 
dA_{t}=f\left( t,A_{t}\right) dt+g\left( t,A_{t}\right) dZ_{t}, & A_{0}=A,
\\ 
dW_{t}dZ_{t}=\rho \left( t,F_{t},A_{t}\right) dt, & 
\end{array}
\label{lsv_sde}
\end{equation}%
Here $F_{t}$ is an observable price of the underlying, $A_{t}$ is an
unobservable auxiliary variable, and $W_{t},Z_{t}$ are two correlated
Brownian motions with correlation $\rho $, $\left\vert \rho \right\vert <1$.
We emphasize that here and below $A_{t}$ is a hidden variable which is not
directly observable, but can (potentially) be filtered by using statistical
methods. The corresponding pricing PDE has the form%
\begin{equation}
\begin{array}{l}
V_{t}+\QTOVERD. . {1}{2}\mathsf{\sigma }^{2}\left( t,F,A\right) V_{FF}+\rho
\left( t,F,A\right) g\left( t,A\right) \mathsf{\sigma }\left( t,A,F\right)
V_{FA}+\QTOVERD. . {1}{2}g^{2}\left( t,A\right) V_{AA} \\ 
+f\left( t,A\right) V_{A}=0.%
\end{array}
\label{lsv}
\end{equation}%
This equation should be augmented with proper boundary and final conditions
which depend on the derivative instrument under consideration. Analytical or
semi-analytical solution of the pricing problem with this degree of
generality is not possible, while its numerical solution, which is formally
relatively straightforward (see below), might require substantial
computational efforts.

Below we wish to be more specific and assume that 
\begin{equation}
\mathsf{\sigma }\left( t,F_{t},v_{t}\right) =\sqrt{v_{t}}\mathsf{\sigma }%
\left( F_{t}\right) ,  \label{lsv_vol}
\end{equation}%
where $v_{t}\left( \equiv A_{t}\right) $ is a (still unobservable) scaling
factor, which follows the standard Feller square-root process, \cite{feller}%
, so that%
\begin{equation}
\begin{array}{ll}
dF_{t}=\sqrt{v_{t}}\mathsf{\sigma }\left( F_{t}\right) dW_{t}, & F_{0}=F, \\ 
dv_{t}=\kappa \left( \theta -v_{t}\right) dt+\varepsilon \sqrt{v_{t}}dZ_{t},
& v_{0}=v, \\ 
dW_{t}dZ_{t}=\rho dt. & 
\end{array}
\label{quad_vol_sde}
\end{equation}%
The corresponding PDE reads%
\begin{equation}
V_{t}+\QTOVERD. . {1}{2}v\mathsf{\sigma }^{2}\left( F\right) V_{FF}+\rho
\varepsilon v\mathsf{\sigma }\left( F\right) V_{Fv}+\QTOVERD. .
{1}{2}\varepsilon ^{2}vV_{vv}+\kappa \left( \theta -v\right) V_{v}=0.
\label{quad_vol_pde1}
\end{equation}%
A properly normalized system of SDEs can be written as follows%
\begin{equation}
\begin{array}{ll}
d\bar{F}_{\bar{t}}=\sqrt{\bar{v}_{\bar{t}}}\mathsf{\bar{\sigma}}\left( \bar{F%
}_{\bar{t}}\right) dW_{\bar{t}}, & \bar{F}_{0}=1, \\ 
d\bar{v}_{\bar{t}}=\bar{\kappa}\left( 1-\bar{v}_{\bar{t}}\right) d\bar{t}+%
\bar{\varepsilon}\sqrt{\bar{v}_{\bar{t}}}dZ_{\bar{t}}, & \bar{v}_{0}=\bar{v},
\\ 
dW_{\bar{t}}dZ_{\bar{t}}=\rho d\bar{t}, & 
\end{array}
\label{quad_vol_sde_norm1}
\end{equation}%
where 
\begin{equation}
\begin{array}{l}
\bar{t}=\Sigma ^{2}t,\ \ \ dW_{\bar{t}}=\Sigma dW_{t},\ \ \ dZ_{\bar{t}%
}=\Sigma dZ_{t},\ \ \ \bar{F}_{\bar{t}}=\frac{F_{t}}{F},\ \ \ \bar{v}_{\bar{t%
}}=\frac{v_{t}}{\theta }, \\ 
\mathsf{\bar{\sigma}}\left( \bar{F}_{t}\right) =\frac{\mathsf{\sigma }\left(
F\bar{F}_{\bar{t}}\right) }{\mathsf{\sigma }\left( F\right) },\ \ \ \bar{%
\kappa}=\frac{\kappa }{\Sigma ^{2}},\ \ \ \bar{\varepsilon}=\frac{%
\varepsilon }{\sqrt{\theta }\Sigma },\ \ \ \bar{v}=\frac{v}{\theta },%
\end{array}
\label{nondim_param}
\end{equation}%
are non-dimensional quantities. Here 
\begin{equation}
\Sigma =\sqrt{\theta }\mathsf{\sigma }\left( F\right) /F.  \label{sigma}
\end{equation}%
Below we omit bars and write%
\begin{equation}
\begin{array}{ll}
dF_{t}=\sqrt{v_{t}}\mathsf{\sigma }\left( F_{t}\right) dW_{t}, & F_{0}=1, \\ 
dv_{t}=\kappa \left( 1-v_{t}\right) dt+\varepsilon \sqrt{v_{t}}dZ_{t}, & 
v_{0}=v, \\ 
dW_{t}dZ_{t}=\rho dt. & 
\end{array}
\label{quad_vol_sde_norm2}
\end{equation}%
The corresponding normalized PDE reads,%
\begin{equation}
V_{t}+\QTOVERD. . {1}{2}v\sigma ^{2}\left( F\right) V_{FF}+\rho \varepsilon v%
\mathsf{\sigma }\left( F\right) V_{Fv}+\QTOVERD. . {1}{2}\varepsilon
^{2}vV_{vv}+\kappa \left( 1-v\right) V_{v}=0.  \label{quad_vol_pde2}
\end{equation}%
Since the coefficients of Eq. (\ref{quad_vol_pde2}) are time-independent, it
is convenient to introduce $\tau =T-t$ and rewrite it as a forward equation
of the form%
\begin{equation}
V_{\tau }-\QTOVERD. . {1}{2}v\mathsf{\sigma }^{2}\left( F\right) V_{FF}-\rho
\varepsilon v\mathsf{\sigma }\left( F\right) V_{Fv}-\QTOVERD. .
{1}{2}\varepsilon ^{2}vV_{vv}-\kappa \left( 1-v\right) V_{v}=0.
\label{quad_vol_pde0}
\end{equation}

We are particularly interested in the following concrete and popular choice
of $\mathsf{\sigma }$:%
\begin{equation}
\mathsf{\sigma }\left( F_{t}\right) =\QTOVERD. . {1}{2}\alpha
F_{t}^{2}+\beta F_{t}+\gamma ,  \label{quad_vol}
\end{equation}%
where $\mathsf{\sigma }\left( F\right) $ is a quadratic polynomial which
does not vanish on the positive semi-axis, including the degenerate case
when $\mathsf{\sigma }\left( F\right) $, is a linear polynomial which is
positive on the positive semi-axis, 
\begin{equation}
\mathsf{\sigma }\left( F_{t}\right) =\beta F_{t}+\gamma ,  \label{lin_vol}
\end{equation}%
and the classical Heston model, $\left( \alpha =0,~\beta =1,~\gamma
=0\right) $,%
\begin{equation}
\mathsf{\sigma }\left( F_{t}\right) =F_{t}.  \label{sigma_hest}
\end{equation}%
This model was introduced in \cite{lipton-2}; since then it has become
popular among both practitioners and academics. (For example, it is offered
commercially by a well-known software provider.) In the LV context,
quadratic volatility is discussed in \cite{rady}, \cite{zuhlsdorff}, \cite%
{lipton-book}, and \cite{andersen2}, among others.

Another popular choice of $\mathsf{\sigma }\left( F\right) $ is
SABR-inspired, see \cite{hagan et al}, 
\begin{equation}
\mathsf{\sigma }\left( F\right) =\alpha F_{t}^{\iota }.  \label{CEV}
\end{equation}%
While most of our result can be extended verbatim to this case, we do not
discuss it in detail for the sake of brevity.

When dimensional $\mathsf{\sigma }\left( F_{t}\right) $ has the form (\ref%
{quad_vol}), the corresponding non-dimensional $\mathsf{\bar{\sigma}}\left( 
\bar{F}_{t}\right) $ can be written as follows%
\begin{equation}
\begin{array}{l}
\mathsf{\bar{\sigma}}\left( \bar{F}_{t}\right) =\QTOVERD. . {1}{2}\bar{\alpha%
}\left( \bar{F}_{\bar{t}}-1\right) ^{2}+\bar{\beta}\left( \bar{F}_{\bar{t}%
}-1\right) +\bar{\gamma}, \\ 
\bar{\alpha}=\frac{\alpha F^{2}}{\sigma \left( F\right) },\ \ \ \bar{\beta}=%
\frac{\alpha F^{2}+\beta F}{\sigma \left( F\right) },\ \ \ \bar{\gamma}=1,%
\end{array}
\label{quad_vol2}
\end{equation}%
or, with bars omitted,%
\begin{equation}
\mathsf{\sigma }\left( F\right) =\QTOVERD. . {1}{2}\alpha \left( F-1\right)
^{2}+\beta \left( F-1\right) +1.  \label{quad_vol1}
\end{equation}

We wish to simplify Eq. (\ref{quad_vol_pde0}). To this end we follow \cite%
{lipton-book}, \cite{albanese2}, and \cite{carrlipton}, apply the Liouville
transform $\left( F,V\right) \Rightarrow \left( X,U\right) $, where%
\begin{equation}
\frac{dF}{\mathsf{\sigma }\left( F\right) }=dX,\ \ \ \ \ X=\int_{1}^{F}\frac{%
dF}{\mathsf{\sigma }\left( F\right) },\ \ \ \ \ V=\sqrt{\mathsf{\sigma }}U,
\label{Liouville}
\end{equation}%
and write the transformed pricing PDE in the form%
\begin{equation}
\begin{array}{l}
U_{\tau }-\QTOVERD. . {1}{2}v\left( U_{XX}+\QTOVERD. . {1}{4}\left( 2\mathsf{%
\sigma \sigma }^{\prime \prime }-\left( \mathsf{\sigma }^{\prime }\right)
^{2}\right) \right) U-\rho \varepsilon vU_{Xv} \\ 
-\QTOVERD. . {1}{2}\varepsilon ^{2}vU_{vv}-\left( \kappa -\left( \kappa
-\QTOVERD. . {1}{2}\rho \varepsilon \mathsf{\sigma }^{\prime }\right)
v\right) U_{v}=0,%
\end{array}
\label{quad_vol_pde3}
\end{equation}%
where $^{\prime }=d/dF$.

Assuming that $\mathsf{\sigma }\left( F\right) $ is a quadratic polynomial (%
\ref{quad_vol1}), the corresponding PDE can be written in the form:%
\begin{equation}
\begin{array}{l}
U_{\tau }-\QTOVERD. . {1}{2}v\left( U_{XX}-\omega U\right) -\rho \varepsilon
vU_{Xv} \\ 
-\QTOVERD. . {1}{2}\varepsilon ^{2}vU_{vv}-\left( \kappa -\left( \kappa
-\QTOVERD. . {1}{2}\rho \varepsilon \left( \alpha \left( F-1\right) +\beta
\right) \right) v\right) U_{v}=0,%
\end{array}
\label{quad_vol_pde4}
\end{equation}%
where%
\begin{equation}
\omega =\QTOVERD. . {1}{4}\left( \beta ^{2}-2\alpha \right) .
\label{varomega}
\end{equation}

When $\alpha =0$, we end up with a pricing equation whose coefficients are $%
X $ independent. For the standard Heston model we have%
\begin{equation}
U_{\tau }-\QTOVERD. . {1}{2}v\left( U_{XX}-\QTOVERD. . {1}{4}U\right) -\rho
\varepsilon vU_{Xv}-\QTOVERD. . {1}{2}\varepsilon ^{2}vU_{vv}-\left( \kappa
-\left( \kappa -\QTOVERD. . {1}{2}\rho \varepsilon \right) v\right) U_{v}=0,
\label{Heston_pde1}
\end{equation}%
where%
\begin{equation}
X=\ln \left( F\right) ,\ \ \ X\in \left[ X_{0}^{H},X_{\infty }^{H}\right] =%
\left[ -\infty ,\infty \right] ,\ \ \ v\in \left[ 0,\infty \right] .
\label{Heston_dom}
\end{equation}%
When $0\leq \beta <1$ we deal with the so-called displaced Heston model. The
corresponding pricing equation has the form%
\begin{equation}
U_{\tau }-\QTOVERD. . {1}{2}v\left( U_{XX}-\QTOVERD. . {1}{4}\beta
^{2}U\right) -\rho \varepsilon vU_{Xv}-\QTOVERD. . {1}{2}\varepsilon
^{2}vU_{vv}-\left( \kappa -\left( \kappa -\QTOVERD. . {1}{2}\rho \varepsilon
\beta \right) v\right) U_{v}=0,  \label{shifted_pde1}
\end{equation}%
where%
\begin{equation}
X=\frac{1}{\beta }\ln \left( \beta \left( F-1\right) +1\right) .
\label{shifted_X}
\end{equation}%
The natural domain for the independent variables $\left( X,v\right) $, has
the form%
\begin{equation}
\ X\in \left[ X_{0}^{DH},X_{\infty }^{DH}\right] ,\ \ \ X_{0}^{DH}=\frac{1}{%
\beta }\ln \left( 1-\beta \right) ,\ \ \ X_{\infty }^{DH}=\infty ,\ \ \ v\in %
\left[ 0,\infty \right] .  \label{shifted_dom}
\end{equation}%
We discuss the proper boundary and initial conditions for the above
equations later.

When $\alpha >0$, the situation is more complex. The roots of the quadratic
equation 
\begin{equation}
\mathsf{\sigma }\left( F\right) =0,  \label{quad_equation}
\end{equation}%
are given by%
\begin{equation}
R_{\pm }=\frac{\alpha -\beta \pm 2\sqrt{\omega }}{\alpha },  \label{roots}
\end{equation}%
so that%
\begin{equation}
\omega =\QTOVERD. . {1}{16}\alpha ^{2}\left( R_{+}-R_{-}\right) ^{2}.
\label{varomega1}
\end{equation}%
Since we wish $\sigma $ to be positive on the positive semi-axis $\left[
0,\infty \right) $, we have to restrict ourselves to two possibilities: (A)\
two complex roots, $R_{\pm }=\mathsf{m}\pm i\mathsf{n}$,%
\begin{equation}
\mathsf{\sigma }\left( F\right) =\QTOVERD. . {1}{2}\alpha \left( \left( F-%
\mathsf{m}\right) ^{2}+\mathsf{n}^{2}\right) ,\ \ \ \mathsf{n}>0,\ \ \
\omega =\omega ^{I}=-\QTOVERD. . {1}{4}\alpha ^{2}\mathsf{n}^{2}<0;
\label{sigma_im}
\end{equation}%
(B) two negative roots, $R_{-}=\mathsf{p}$, $R_{+}=\mathsf{q}$,%
\begin{equation}
\mathsf{\sigma }\left( F\right) =\QTOVERD. . {1}{2}\alpha \left( F-\mathsf{p}%
\right) \left( F-\mathsf{q}\right) ,\ \ \ \mathsf{p}<\mathsf{q}<0,\ \ \
\omega =\omega ^{R}=\QTOVERD. . {1}{16}\alpha ^{2}\left( \mathsf{q}-\mathsf{p%
}\right) ^{2}>0.  \label{sigma_re}
\end{equation}%
In case (A) we have $\left\vert \beta \right\vert <\sqrt{2\alpha }$, and%
\begin{equation}
\mathsf{m}=\frac{\alpha -\beta }{\alpha }=\func{Re}\left[ R_{+}\right] ,\ \
\ \ \ \mathsf{n}=\frac{2\sqrt{\left\vert \omega ^{I}\right\vert }}{\alpha }=%
\func{Im}\left[ R_{+}\right] .  \label{roots_im}
\end{equation}%
In case (B) we have $\max \left\{ \alpha ,\sqrt{2\alpha }\right\} <\beta
<1+\alpha /2$, and 
\begin{equation}
\mathsf{p}=\frac{\alpha -\beta -2\sqrt{\omega ^{R}}}{\alpha }=R_{-},\ \ \ \
\ \mathsf{q}=\frac{\alpha -\beta +2\sqrt{\omega ^{R}}}{\alpha }=R_{+}.
\label{roots_re}
\end{equation}

We start with case (A). Straightforward evaluation of the Liouville integral
(\ref{Liouville}) yields%
\begin{equation}
F\rightarrow X=\frac{1}{\sqrt{\left\vert \omega ^{I}\right\vert }}\left(
\arctan \left( \frac{F-\mathsf{m}}{\mathsf{n}}\right) -\arctan \left( \frac{%
1-\mathsf{m}}{\mathsf{n}}\right) \right) .  \label{map_im}
\end{equation}%
The Liouville transform compactifies the positive semi-axis and maps it into
a finite interval: 
\begin{equation}
\left[ X_{0}^{I},X_{\infty }^{I}\right] =\frac{1}{\sqrt{\left\vert \omega
^{I}\right\vert }}\left[ -\arctan \left( \frac{\mathsf{m}}{\mathsf{n}}%
\right) -\arctan \left( \frac{1-\mathsf{m}}{\mathsf{n}}\right) ,\frac{\pi }{2%
}-\arctan \left( \frac{1-\mathsf{m}}{\mathsf{n}}\right) \right] .
\label{dom_im}
\end{equation}%
The length of this interval is denoted by $\Delta ^{I}=X_{\infty
}^{I}-X_{0}^{I}$. The inverse mappings $X\rightarrow F$ and $\sqrt{\sigma
\left( F\right) }$ have the form:%
\begin{equation}
\begin{array}{lll}
F & = & \frac{\sqrt{\mathsf{m}^{2}+\mathsf{n}^{2}}\sin \left( \sqrt{%
\left\vert \omega ^{I}\right\vert }\left( X-X_{0}^{I}\right) \right) }{\sin
\left( \sqrt{\left\vert \omega ^{I}\right\vert }\left( X_{\infty
}^{I}-X\right) \right) }, \\ 
\sqrt{\sigma \left( F\right) } & = & \frac{\sqrt{\left\vert \omega
^{I}\right\vert }}{\sqrt{\frac{\alpha }{2}}\sin \left( \sqrt{\left\vert
\omega ^{I}\right\vert }\left( X_{\infty }^{I}-X\right) \right) }.%
\end{array}
\label{spot_im}
\end{equation}%
Eq. (\ref{quad_vol_pde4}) has the form%
\begin{equation}
\begin{array}{l}
U_{\tau }-\QTOVERD. . {1}{2}v\left( U_{XX}-\omega ^{I}U\right) -\rho
\varepsilon vU_{Xv} \\ 
-\QTOVERD. . {1}{2}\varepsilon ^{2}vU_{vv}-\left( \kappa -\left( \kappa
-\QTOVERD. . {1}{2}\rho \varepsilon \alpha \left( F-\mathsf{m}\right)
\right) v\right) U_{v}=0,%
\end{array}
\label{pde_im1}
\end{equation}%
or, expressing $F$ in terms of $X$ and rearranging terms,%
\begin{equation}
\begin{array}{l}
U_{\tau }-\QTOVERD. . {1}{2}v\left( U_{XX}-\omega ^{I}U\right) -\rho
\varepsilon vU_{Xv} \\ 
-\QTOVERD. . {1}{2}\varepsilon ^{2}vU_{vv}-\left( \kappa -\left( \kappa
-\rho \varepsilon \sqrt{\left\vert \omega ^{I}\right\vert }\cot \left( \sqrt{%
\left\vert \omega ^{I}\right\vert }\left( X_{\infty }^{I}-X\right) \right)
\right) v\right) U_{v}=0.%
\end{array}
\label{pde_im2}
\end{equation}%
Here $X\in \left[ X_{0}^{I},X_{\infty }^{I}\right] ,\ v\in \left[ 0,\infty %
\right] $.

In case (B) we have%
\begin{equation}
X=\frac{1}{2\sqrt{\omega ^{R}}}\ln \left( \frac{\left( 1-\mathsf{p}\right)
\left( F-\mathsf{q}\right) }{\left( 1-\mathsf{q}\right) \left( F-\mathsf{p}%
\right) }\right) .  \label{map_re}
\end{equation}%
The positive semi-axis is compactified and mapped into a finite interval 
\begin{equation}
\left[ X_{0}^{R},X_{\infty }^{R}\right] =\frac{1}{2\sqrt{\omega ^{R}}}\left[
\ln \left( \frac{\left( 1-\mathsf{p}\right) \mathsf{q}}{\left( 1-\mathsf{q}%
\right) \mathsf{p}}\right) ,\ln \left( \frac{\left( 1-\mathsf{p}\right) }{%
\left( 1-\mathsf{q}\right) }\right) \right] .  \label{dom_re}
\end{equation}%
The length of this interval is denoted by $\Delta ^{R}=X_{\infty
}^{R}-X_{0}^{R}$. The inverse mappings $X\rightarrow F$ and $\sqrt{\sigma
\left( F\right) }$ have the form:%
\begin{equation}
\begin{array}{lll}
F & = & \frac{\sqrt{\mathsf{pq}}\sinh \left( \sqrt{\omega ^{R}}\left(
X-X_{0}^{R}\right) \right) }{\sinh \left( \sqrt{\omega ^{R}}\left( X_{\infty
}^{R}-X\right) \right) }, \\ 
\sqrt{\sigma \left( F\right) } & = & \frac{\sqrt{\omega ^{R}}}{\sqrt{\frac{%
\alpha }{2}}\sinh \left( \sqrt{\omega ^{R}}\left( X_{\infty }^{R}-X\right)
\right) }.%
\end{array}
\label{spot_re}
\end{equation}%
Equation (\ref{quad_vol_pde4}) has the form%
\begin{equation}
\begin{array}{l}
U_{\tau }-\QTOVERD. . {1}{2}v\left( U_{XX}-\omega ^{R}U\right) -\rho
\varepsilon vU_{Xv} \\ 
-\QTOVERD. . {1}{2}\varepsilon ^{2}vU_{vv}-\left( \kappa -\left( \kappa
-\QTOVERD. . {1}{2}\rho \varepsilon \alpha \left( F-\frac{\mathsf{p}+\mathsf{%
q}}{2}\right) \right) v\right) U_{v}=0,%
\end{array}
\label{pde_re1}
\end{equation}%
or, expressing $F$ in terms of $X$ and rearranging terms,%
\begin{equation}
\begin{array}{l}
U_{\tau }-\QTOVERD. . {1}{2}v\left( U_{XX}-\omega ^{R}U\right) -\rho
\varepsilon vU_{Xv} \\ 
-\QTOVERD. . {1}{2}\varepsilon ^{2}vU_{vv}-\left( \kappa -\left( \kappa
-\rho \varepsilon \sqrt{\omega ^{R}}\coth \left( \sqrt{\omega ^{R}}\left(
X_{\infty }^{R}-X\right) \right) \right) v\right) U_{v}=0,%
\end{array}
\label{pde_re2}
\end{equation}%
where $X\in \left[ X_{0}^{R},X_{\infty }^{R}\right] ,\ v\in \left[ 0,\infty %
\right] $.

In order to simplify subsequent developments, it is useful to rewrite the
corresponding pricing equations in a unified form. To this end we introduce
new variables $x_{1}\triangleq X,x_{2}\triangleq v$, and obtain%
\begin{equation}
\begin{array}{l}
U_{\tau }-\QTOVERD. . {1}{2}\mathsf{a}_{11}\left( x_{2}\right) \left(
U_{x_{1}x_{1}}-\omega ^{s}U\right) -\mathsf{a}_{12}\left( x_{2}\right)
U_{x_{1}x_{2}} \\ 
-\QTOVERD. . {1}{2}\mathsf{a}_{22}\left( x_{2}\right) U_{x_{2}x_{2}}-\mathsf{%
b}_{2}^{s}\left( x_{1},x_{2}\right) U_{x_{2}}=0,%
\end{array}
\label{pde_univ}
\end{equation}%
where%
\begin{equation}
\mathsf{a}_{11}\left( x_{2}\right) =x_{2},\ \ \ \mathsf{a}_{12}\left(
x_{2}\right) =\rho \varepsilon x_{2},\ \ \ \mathsf{a}_{22}\left(
x_{2}\right) =\varepsilon ^{2}x_{2},  \label{pde_aij}
\end{equation}%
\begin{equation}
\mathsf{b}_{2}^{s}\left( x_{1},x_{2}\right) =\left\{ 
\begin{array}{ll}
\kappa -\left( \kappa -\QTOVERD. . {1}{2}\rho \varepsilon \right) x_{2}, & 
s=H \\ 
\kappa -\left( \kappa -\QTOVERD. . {1}{2}\rho \varepsilon \beta \right)
x_{2}, & s=DH, \\ 
\kappa -\left( \kappa -\rho \varepsilon \sqrt{\left\vert \omega
^{I}\right\vert }\cot \left( \sqrt{\left\vert \omega ^{I}\right\vert }\left(
X_{\infty }^{I}-x_{1}\right) \right) \right) x_{2}, & s=I, \\ 
\kappa -\left( \kappa -\rho \varepsilon \sqrt{\omega ^{R}}\coth \left( \sqrt{%
\omega ^{R}}\left( X_{\infty }^{R}-x_{1}\right) \right) \right) x_{2}, & s=R,%
\end{array}%
\right.  \label{pde_b2}
\end{equation}%
The natural domain for $x_{1}$ in Eq. (\ref{pde_univ}) is the interval $%
\left[ X_{0}^{s},X_{\infty }^{s}\right] $, which might be bounded or
unbounded depending on $s$.

The choice of the proper initial and boundary conditions augmenting Eq. (\ref%
{pde_univ}) depends on the actual derivative product under consideration. We
are interested in vanillas and first generation exotics, such as barrier
calls and puts, single and double no-touch options and the like. For such
options the domain of $x_{1}$ has the form $\left[ X_{L}^{s},X_{U}^{s}\right]
\subset \left[ X_{0}^{s},X_{\infty }^{s}\right] $. The corresponding initial
condition can be written as 
\begin{equation}
U\left( 0,x_{1},x_{2}\right) =u^{s}\left( x_{1}\right) ,  \label{ic_univ}
\end{equation}%
where $u^{s}$ reflects the payoff of the instrument in question. For
instance, for a covered call option $\left[ X_{L}^{s},X_{U}^{s}\right] =%
\left[ X_{0}^{s},X_{\infty }^{s}\right] $, and $u^{s}$ has the form (\ref%
{call_payoff}), while for a DNT option $X_{0}^{s}<X_{L}^{s}<X_{U}^{s}<X_{%
\infty }^{s}$, and $u^{s}$ has the form (\ref{dnt_payoff}).

The boundary conditions in the $x_{1}$ direction are simple%
\begin{equation}
U\left( \tau ,X_{L}^{s},x_{2}\right) =r_{L}\left( \tau \right) ,\ \ \ \ \
U^{s}\left( \tau ,X_{U}^{s},x_{2}\right) =r_{U}\left( \tau \right) ,
\end{equation}%
where $r_{L},r_{U}$ represent the corresponding rebates at barriers. These
equations are understood in the limiting sense when $\left\vert X_{\left\{
L,U\right\} }^{s}\right\vert =\infty $. At the same time, the exact form of
the boundary conditions in the $x_{2}$ direction are somewhat difficult to
formulate. We shall see later that for our purposes it is not necessary,
since we can use the pricing equation itself as a boundary condition.

\section{Numerical solution of the generic pricing problem\label{NumSol}}

Our inability to find an analytical solution for the LSV pricing problem
with nonzero correlation makes it necessary to develop appropriate numerical
methods for its solution. In this section we discuss such methods. In
Section \ref{Discretization} we show how to discretize the pricing problem
in time and in space on a non-uniform grid. While inside the computational
domain this operation is completely standard, we do exploit somewhat
non-standard approach to the discretization of the boundary conditions, and,
by implication, to the closure of the problem. Namely, we distinguish two
cases: (A) the case of endogenous boundary condition, when the equation
itself provides a boundary condition; (B) the case of exogenous boundary
condition, when we simply impose the usual Dirichlet boundary conditions. In
the one-dimensional case, endogenous discretization has been successfully
used by several researchers, see, e.g., \cite{ekstrom1}, \cite{ekstrom2}.
Once the pricing problem (with appropriate boundary conditions) is
discretized, we have several avenues of attack, which we discuss in turn. In
Section \ref{Explicit} we introduce the explicit method.\ While seldom used
in practice due to its unfavorable stability properties, we discuss it
nevertheless, first, to gain an extra data point for comparison of different
numerical results, and, second, to illustrate a practically viable way of
implementing such a method by virtue of the so-called Fast Exponentiation,
which was recently popularized by Albanese and his co-workers, see, e.g., 
\cite{albanese}. Also, recently O'Sullivan-O'Sullivan, \cite{osullivan},
proposed a version of the EFD scheme, which is more efficient that the basic
one. In Section \ref{ADI} we introduce several ADI methods for solving the
pricing problem, including the original Douglas (Do) method, \cite{douglas},
its improvement due to Craig-Sneyd (CS), \cite{craig}, as well as two
modified CS-type methods due to Hunsdorfer and Verwer (HV), \cite{hunsdorfer}%
, and in 't Hout and Welfert (HW), \cite{inthout2}. ADI methods have been
successfully used to solve the Heston pricing problem by \cite{kluge}, \cite%
{ikonen}, \cite{inthout1}, among several others. They have also been used to
price cross-currency swaps, see, e.g., \cite{dempster}, and to solve many
other problems in the field of financial engineering. The next method, which
we introduce in Section \ref{Galerkin} is much less standard than the ones
which were mentioned earlier, in fact, to the best of our knowledge, it had
not been applied before in the context we are interested in. This method,
which is inspired by the classical Galerkin-Ritz ideas, \cite{galerkin}, 
\cite{ritz}, judiciously exploits the structure of the two-dimensional
pricing equation in the spot and variance directions and reduces it to a
coupled system of one-dimensional equations in the variance direction alone.
The corresponding system is solved by treating the mixing terms fully
explicitly. We emphasize that when the correlation $\rho $ between
stochastic drivers is zero, the corresponding system becomes uncoupled and
can be solved \emph{exactly}, as was pointed out by Lipton \cite{lipton-book}%
. This observation is a starting point of Section \ref{Expansion}, where an
expansion in powers of $\rho $ is presented in a semi-explicit fashion. We
emphasize that the idea of using $\rho $ as a small parameter is not new,
see, e.g., \cite{antonelli}. However, we improve the known results
significantly, as well as emphasize the links between the Galerkin and the
expansion methods. Finally, in Section \ref{MonteCarlo} we briefly discuss
pricing vanilla and first generation exotics via a version of the MC method.

\subsection{Discretization of a differential operator\label{Discretization}}

In view of the previous discussion, it is clear that the pricing problem can
be written in the form%
\begin{equation}
\frac{\partial U}{\partial \tau }\left( \tau ,x_{1},x_{2}\right) -\mathcal{L}%
^{s}U\left( \tau ,x_{1},x_{2}\right) =0,  \label{pde_symb}
\end{equation}%
\begin{equation}
U\left( 0,x_{1},x_{2}\right) =u\left( x_{1}\right) ,  \label{ic_sym}
\end{equation}%
\begin{equation}
U\left( \tau ,X_{L},x_{2}\right) =0,\ \ \ \ \ U\left( \tau
,X_{U},x_{2}\right) =0.  \label{bc_sym}
\end{equation}%
Here the operator of interest can be represented as follows:%
\begin{equation}
\mathcal{L}^{s}=\mathcal{L}_{\left( 11\right) }^{s}+\mathcal{L}_{\left(
12\right) }+\mathcal{L}_{\left( 22\right) }^{s},  \label{diff_op_split}
\end{equation}%
\begin{equation}
\mathcal{L}_{\left( 11\right) }^{s}U=\QTOVERD. . {1}{2}\mathsf{a}_{11}\left(
x_{2}\right) \left( U_{x_{1}x_{1}}-\omega ^{s}U\right) ,  \label{diff_op_11}
\end{equation}%
\begin{equation}
\mathcal{L}_{\left( 12\right) }U=\mathsf{a}_{12}\left( x_{2}\right)
U_{x_{1}x_{2}},  \label{diff_op_12}
\end{equation}%
\begin{equation}
\mathcal{L}_{\left( 22\right) }^{s}=\QTOVERD. . {1}{2}\mathsf{a}_{22}\left(
x_{2}\right) U_{x_{2}x_{2}}+\mathsf{b}_{2}^{s}\left( x_{1},x_{2}\right)
U_{x_{2}},  \label{diff_op_22}
\end{equation}%
where the coefficients are given by Eqs (\ref{pde_aij}), (\ref{pde_b2}).

First, we discretize Eq. (\ref{pde_symb}) in the $\tau $ direction. This
procedure is straightforward. We choose a grid $\mathfrak{T}=\left\{ \tau
_{0}=0,\tau _{1},...,\tau _{n},...,\tau _{N-1},\tau _{N}=T\right\} $ with $%
N+1$ points, and write the dynamic equation as follows%
\begin{equation}
\frac{U_{n+1}\left( x_{1},x_{2}\right) -U_{n}\left( x_{1},x_{2}\right) }{%
\Delta \tau _{n+1,n}}-\mathcal{L}^{s}\left( \varsigma U_{n}\left(
x_{1},x_{2}\right) +\left( 1-\varsigma \right) U_{n+1}\left(
x_{1},x_{2}\right) \right) =0,  \label{pde_disc1}
\end{equation}%
where $\Delta \tau _{n+1,n}=\tau _{n+1}-\tau _{n}$, and $U_{n}\left(
x_{1},x_{2}\right) =U\left( \tau _{n},x_{1},x_{2}\right) $. It is clear that%
\begin{equation}
U_{0}\left( x_{1},x_{2}\right) =u\left( x_{1},x_{2}\right) .  \label{ic_sym1}
\end{equation}%
Here $\varsigma \in \left[ 0,1\right] $ is a mixing parameter, which defines
the degree of explicitness of the scheme under consideration. In most cases
we use a uniform grid in time, so that $\Delta \tau _{n+1,n}=\Delta \tau $.
We emphasize that this is the most common but by no means the only way of
discretizing Eq. (\ref{pde_symb}) in the $\tau $ direction. In some cases
three-level discretization is more accurate.

Discretization of a differential operator on a non-uniform grid is a common
procedure, see, e.g., \cite{tavella}. While the corresponding formulas are
ubiquitous, we present the ones which we actually use in our calculation for
the reader's convenience. We consider a non-uniform grid $\mathfrak{X}%
=\left\{ x_{0},x_{1},...,x_{i},...,x_{I-1},x_{I}\right\} $ with $I+1$
points, and write second order accurate FD expressions for the following
operators $\mathcal{D}_{1}=d/dx,\mathcal{D}_{2}=d^{2}/dx^{2}$. The
differences between the grid points are denoted by $\Delta
x_{i,j}=x_{i}-x_{j}$, $i=1,...,I+1,j=0,...,I$. For internal points $\left\{
x_{1},...,x_{i},...,x_{I-1}\right\} $ we use central differences:%
\begin{equation}
\left. \frac{df}{dx}\right\vert _{x_{i}}=\xi _{i,-1}^{c}f_{i-1}+\xi
_{i,0}^{c}f_{i}+\xi _{i,1}^{c}f_{i+1},\ \ \ \ \ 1\leq i\leq I-1,
\label{D1_c_dis}
\end{equation}%
where%
\begin{equation}
\begin{array}{lll}
\xi _{i,-1}^{c} & = & -\frac{\Delta x_{i+1,i}}{\Delta x_{i,i-1}\Delta
x_{i+1,i-1}}, \\ 
\xi _{i,0}^{c} & = & \frac{\Delta x_{i+1,i}-\Delta x_{i,i-1}}{\Delta
x_{i,i-1}\Delta x_{i+1,i}}, \\ 
\xi _{i,1}^{c} & = & \frac{\Delta x_{i,i-1}}{\Delta x_{i+1,i}\Delta
x_{i+1,i-1}},%
\end{array}
\label{xi_c_dis}
\end{equation}%
\begin{equation}
\xi _{i,-1}^{c}+\xi _{i,0}^{c}+\xi _{i,1}^{c}=0.
\end{equation}%
For the left and right end-points we use forward and backward differences:%
\begin{equation}
\left. \frac{df}{dx}\right\vert _{x_{0}}=\xi _{0,0}^{f}f_{0}+\xi
_{0,1}^{f}f_{1}+\xi _{0,2}^{f}f_{2},  \label{D1_f_dis}
\end{equation}%
\begin{equation}
\left. \frac{df}{dx}\right\vert _{x_{I}}=\xi _{I,0}^{b}f_{I}+\xi
_{I,-1}^{b}f_{I-1}+\xi _{I,-2}^{b}f_{I-2},  \label{D1_b_dis}
\end{equation}%
where%
\begin{equation}
\begin{array}{lll}
\xi _{0,0}^{f} & = & -\frac{\Delta x_{1,0}+\Delta x_{2,0}}{\Delta
x_{1,0}\Delta x_{2,0}}, \\ 
\xi _{0,1}^{f} & = & \frac{\Delta x_{2,0}}{\Delta x_{1,0}\Delta x_{2,1}}, \\ 
\xi _{0,2}^{f} & = & -\frac{\Delta x_{1,0}}{\Delta x_{2,1}\Delta x_{2,0}},%
\end{array}
\label{xi_f_dis}
\end{equation}%
\begin{equation}
\xi _{0,0}^{f}+\xi _{0,1}^{f}+\xi _{0,2}^{f}=0,
\end{equation}%
\begin{equation}
\begin{array}{lll}
\xi _{I,0}^{b} & = & \frac{\Delta x_{I,I-2}+\Delta x_{I,I-1}}{\Delta
x_{I,I-1}\Delta x_{I,I-2}}, \\ 
\xi _{I,-1}^{b} & = & -\frac{\Delta x_{I,I-2}}{\Delta x_{I-1,I-2}\Delta
x_{I,I-1}}, \\ 
\xi _{I,-2}^{b} & = & \frac{\Delta x_{I,I-1}}{\Delta x_{I-1,I-2}\Delta
x_{I,I-2}},%
\end{array}
\label{xi_b_dis}
\end{equation}%
\begin{equation}
\xi _{I,0}^{b}+\xi _{I,-1}^{b}+\xi _{I,-2}^{b}=0.
\end{equation}%
Similarly, we write%
\begin{equation}
\QTOVERD. . {1}{2}\left. \frac{d^{2}f}{dx^{2}}\right\vert _{x_{i}}=\eta
_{i,-1}^{c}f_{i-1}+\eta _{i,0}^{c}f_{i}+\eta _{i,1}^{c}f_{i+1},\ \ \ \ \
1\leq i\leq I-1,  \label{D2_c_dis}
\end{equation}%
where%
\begin{equation}
\begin{array}{lll}
\eta _{i,-1}^{c} & = & \frac{1}{\Delta x_{i,i-1}\Delta x_{i+1,i-1}}, \\ 
\eta _{i,0}^{c} & = & -\frac{1}{\Delta x_{i,i-1}\Delta x_{i+1,i}}, \\ 
\eta _{i,1}^{c} & = & \frac{1}{\Delta x_{i+1,i}\Delta x_{i+1,i-1}},%
\end{array}
\label{eta_c_dis}
\end{equation}%
\begin{equation}
\eta _{i,-1}^{c}+\eta _{i,0}^{c}+\eta _{i,1}^{c}=0,
\end{equation}%
and%
\begin{equation}
\left. \frac{d^{2}f}{dx^{2}}\right\vert _{x_{0}}=\eta _{0,0}^{f}f_{0}+\eta
_{0,1}^{f}f_{1}+\eta _{0,2}^{f}f_{2}+\eta _{0,3}^{f}f_{3},  \label{D2_f_dis}
\end{equation}%
\begin{equation}
\left. \frac{d^{2}f}{dx^{2}}\right\vert _{x_{I}}=\eta _{I,0}^{b}f_{I}+\eta
_{I,-1}^{b}f_{I-1}+\eta _{I,-2}^{b}f_{I-2}+\eta _{I,-3}^{b}f_{I-3},
\label{D2_b_dis}
\end{equation}%
where%
\begin{equation}
\eta _{0,0}^{f}=\varkappa _{1}^{f}+\varkappa _{2}^{f}+\varkappa _{3}^{f},\ \
\eta _{0,i}^{f}=-\varkappa _{i}^{f},\ \ i=1,2,3,  \label{eta_f_dis}
\end{equation}%
\begin{equation}
\begin{array}{lll}
\varkappa _{1}^{f} & = & \frac{\Delta x_{2,0}+\Delta x_{3,0}}{\Delta
x_{1,0}\Delta x_{2,1}\Delta x_{3,1}}, \\ 
\varkappa _{2}^{f} & = & -\frac{\Delta x_{1,0}+\Delta x_{3,0}}{\Delta
x_{2,0}\Delta x_{2,1}\Delta x_{3,2}}, \\ 
\varkappa _{3}^{f} & = & \frac{\Delta x_{1,0}+\Delta x_{2,0}}{\Delta
x_{3,0}\Delta x_{3,1}\Delta x_{3,2}},%
\end{array}%
\end{equation}%
\begin{equation}
\eta _{I,0}^{b}=\varkappa _{-1}^{b}+\varkappa _{-2}^{b}+\varkappa
_{-3}^{b},\ \ \eta _{I,-i}^{b}=-\varkappa _{-i}^{b},\ \ i=1,2,3,
\label{eta_b_dis}
\end{equation}%
\begin{equation}
\begin{array}{ccc}
\varkappa _{-1}^{b} & = & \frac{\Delta x_{I,I-2}+\Delta x_{I,I-3}}{\Delta
x_{I,I-1}\Delta x_{I-1,I-2}\Delta x_{I-1,I-3}}, \\ 
\varkappa _{-2}^{b} & = & -\frac{\Delta x_{I,I-1}+\Delta x_{I,I-3}}{\Delta
x_{I,I-2}\Delta x_{I-1,I-2}\Delta x_{I-2,I-3}}, \\ 
\varkappa _{-3}^{b} & = & \frac{\Delta x_{I,I-1}+\Delta x_{I,I-2}}{\Delta
x_{I,I-3}\Delta x_{I-1,I-3}\Delta x_{I-2,I-3}}.%
\end{array}%
\end{equation}

We are now prepared to discretize a one-dimensional second-order
differential operator $\mathcal{L}$ of the form%
\begin{equation}
\mathcal{L}=\QTOVERD. . {1}{2}\mathsf{a}\left( x\right) \frac{d^{2}}{dx^{2}}+%
\mathsf{b}\left( x\right) \frac{d}{dx}-\mathsf{c}\left( x\right) ,
\label{diff_op}
\end{equation}%
on a finite grid $\mathfrak{X}=\left\{
x_{0},x_{1},...,x_{i},...,x_{I-1},x_{I}\right\} $. By using the above
formulas, we represent it as a penta-diagonal matrix $\mathfrak{L}$ of the
form%
\begin{equation}
\mathfrak{L=}\left[ 
\begin{array}{ccccccccccc}
d_{0} & d_{1} & d_{2} & d_{3} &  &  &  &  &  &  &  \\ 
a_{1} & b_{1} & c_{1} &  &  &  &  &  &  &  &  \\ 
& \ast & \ast & \ast &  &  &  &  &  &  &  \\ 
&  & \ast & \ast & \ast &  &  &  &  &  &  \\ 
&  &  & \ast & \ast & \ast &  &  &  &  &  \\ 
&  &  &  & a_{i} & b_{i} & c_{i} &  &  &  &  \\ 
&  &  &  &  & \ast & \ast & \ast &  &  &  \\ 
&  &  &  &  &  & \ast & \ast & \ast &  &  \\ 
&  &  &  &  &  &  & \ast & \ast & \ast &  \\ 
&  &  &  &  &  &  &  & a_{I-1} & b_{I-1} & c_{I-1} \\ 
&  &  &  &  &  &  & d_{I-3} & d_{I-2} & d_{I-1} & d_{I}%
\end{array}%
\right] ,  \label{matrix_op}
\end{equation}%
where%
\begin{equation}
\begin{array}{lll}
a_{i} & = & \mathsf{a}\left( x_{i}\right) \eta _{i,-1}^{c}+\mathsf{b}\left(
x_{i}\right) \xi _{i,-1}^{c}, \\ 
b_{i} & = & \mathsf{a}\left( x_{i}\right) \eta _{i,0}^{c}+\mathsf{b}\left(
x_{i}\right) \xi _{i,0}^{c}-\mathsf{c}\left( x_{i}\right) , \\ 
c_{i} & = & \mathsf{a}\left( x_{i}\right) \eta _{i,+1}^{c}+\mathsf{b}\left(
x_{i}\right) \xi _{i,+1}^{c}.%
\end{array}
\label{aibici}
\end{equation}%
We consider two possibilities: (A) boundary conditions are endogenous and
determined by the operator itself; (B) boundary conditions are exogenous and
determined by the nature of the derivative product in question; for brevity,
in the latter case we only consider exogenous Dirichlet boundary conditions.
In case (A) we have%
\begin{equation}
\begin{array}{lll}
d_{0} & = & \mathsf{a}\left( x_{0}\right) \eta _{0,0}^{f}+\mathsf{b}\left(
x_{0}\right) \xi _{0,0}^{f}-\mathsf{c}\left( x_{0}\right) , \\ 
d_{1} & = & \mathsf{a}\left( x_{0}\right) \eta _{0,1}^{f}+\mathsf{b}\left(
x_{0}\right) \xi _{0,1}^{f}, \\ 
d_{2} & = & \mathsf{a}\left( x_{0}\right) \eta _{0,2}^{f}+\mathsf{b}\left(
x_{0}\right) \xi _{0,2}^{f}, \\ 
d_{3} & = & \mathsf{a}\left( x_{0}\right) \eta _{0,3}^{f},%
\end{array}
\label{d0123}
\end{equation}%
\begin{equation}
\begin{array}{lll}
d_{I} & = & \mathsf{a}\left( x_{I}\right) \eta _{I,0}^{b}+\mathsf{b}\left(
x_{I}\right) \xi _{I,0}^{b}-\mathsf{c}\left( x_{I}\right) , \\ 
d_{I-1} & = & \mathsf{a}\left( x_{I}\right) \eta _{I,-1}^{f}+\mathsf{b}%
\left( x_{I}\right) \xi _{I,-1}^{f}, \\ 
d_{I-2} & = & \mathsf{a}\left( x_{I}\right) \eta _{I,-2}^{b}+\mathsf{b}%
\left( x_{I}\right) \xi _{I,-2}^{b}, \\ 
d_{I-3} & = & \mathsf{a}\left( x_{I}\right) \eta _{I,-3}^{b}.%
\end{array}
\label{dI0123}
\end{equation}%
In case (B) we have%
\begin{equation}
d_{0}=1,\ \ \ d_{i}=0,\ \ \ d_{I}=1,\ \ \ d_{I-i}=0,\ \ \ \ \ i=1,...,3.
\label{d_Dirichlet}
\end{equation}%
Endogenous boundary conditions have been used in the past for the single
factor term structure problems, see, e.g., \cite{ekstrom1}, \cite{ekstrom2}.

It is natural to represent the discretized operator $\mathfrak{L}$ in the
form

\begin{equation}
\mathfrak{L}^{s}=\mathfrak{L}_{\left( 11\right) }^{s}+\mathfrak{L}_{\left(
12\right) }+\mathfrak{L}_{\left( 22\right) }^{s}.  \label{matrix_split}
\end{equation}%
We use the above formulae in order to obtain the discretized versions $%
\mathfrak{L}_{\left( 11\right) }^{s},\mathfrak{L}_{\left( 22\right) }^{s}$
of the differential operators $\mathcal{L}_{\left( 11\right) }^{s},\mathcal{L%
}_{\left( 22\right) }^{s}$on one-dimensional grids $\mathfrak{X}_{1},%
\mathfrak{X}_{2}$. Below we denote matrix elements of the penta-diagonal $%
\left( I_{\iota }+1\right) \times \left( I_{\iota }+1\right) $ matrices $%
\mathfrak{L}_{\left( \iota \iota \right) }$ by $\mathfrak{l}_{\left( \iota
\iota \right) i_{\iota },i_{\iota }^{\prime }}$, $\iota =1,2$.

In order to obtain the discretized version $\mathfrak{L}_{\left( 12\right) }$
of $\mathcal{L}_{\left( 12\right) }$ on a rectangular grid $\mathfrak{X}%
_{1}\otimes \mathfrak{X}_{2}$, we use formula (\ref{D1_c_dis}) twice and get
a nine-point stencil representation for the cross derivative%
\begin{equation}
\left. \frac{\partial ^{2}f}{\partial x_{1}\partial x_{2}}\right\vert
_{x_{1,i_{1}},x_{2,i_{2}}}=\dsum\limits_{\alpha _{1},\alpha _{2}\in \aleph
}\xi _{i_{1},\alpha _{1},i_{2},\alpha _{2}}^{c}f_{i_{1}+\alpha
_{1},i_{2}+\alpha _{2}},  \label{dx1dx2}
\end{equation}%
where%
\begin{equation}
\xi _{i_{1},\alpha _{1},i_{2},\alpha _{2}}^{c}=\xi _{i_{1},\alpha
_{1}}^{c}\xi _{i_{2},\alpha _{2}}^{c},\ \ \ \ \ \alpha _{\iota }\in \aleph
\equiv \left\{ -1,0,1\right\} .  \label{xi_c_12}
\end{equation}%
Here $1\leq i_{\iota }\leq I_{\iota }-1$. For the end points of the grid $%
\mathfrak{X}_{1}\otimes \mathfrak{X}_{2}$ the corresponding expressions are
slightly different and are left for the reader to derive. Accordingly,%
\begin{equation}
\begin{array}{ll}
\left. \mathfrak{L}_{\left( 12\right) }f\right\vert _{i_{1},i_{2}} & =\rho
\varepsilon x_{i_{2}}\dsum\limits_{\alpha _{1},\alpha _{2}\in \aleph }\xi
_{i_{1},\alpha _{1},i_{2},\alpha _{2}}^{c}f_{i_{1}+\alpha _{1},i_{2}+\alpha
_{2}} \\ 
& =\dsum\limits_{\alpha _{1},\alpha _{2}\in \aleph }\mathfrak{l}_{\left(
12\right) i_{1},\alpha _{1},i_{2},\alpha _{2}}f_{i_{1}+\alpha
_{1},i_{2}+\alpha _{2}}.%
\end{array}
\label{Mat_12}
\end{equation}

\subsection{Explicit method\label{Explicit}}

The fully explicit scheme is straightforward and can be presented by a
single step%
\begin{equation}
U_{n}\Longrightarrow U_{n+1}=\mathcal{F}_{E}\left( U_{n}\right) ,
\label{Map_explicit}
\end{equation}%
where%
\begin{equation}
U_{n+1}=\mathfrak{P}U_{n},\ \ \ \ \ \mathfrak{P}\mathcal{=}\mathfrak{I}%
+\Delta \tau \mathfrak{L}^{s}\mathfrak{=I}+\Delta \tau \left( \mathfrak{L}%
_{\left( 11\right) }^{s}+\mathfrak{L}_{\left( 12\right) }+\mathfrak{L}%
_{\left( 22\right) }^{s}\right) ,  \label{Step_explicit}
\end{equation}%
and $\mathfrak{I}$ is the identity operator. Given the fact that $%
U_{n}=U_{n,i_{1},i_{2}}$ is a \emph{matrix} rather than a \emph{vector}, we
have to define $\mathfrak{P}$ as a four index \emph{tensor}, $\mathfrak{P=P}%
_{i_{1},i_{2},j_{1},j_{2}}$, and represent the mapping (\ref{Map_explicit})
as follows%
\begin{equation*}
U_{n+1,i_{1},i_{2}}=\dsum\limits_{0\leq j_{\iota }\leq I_{\iota }}\mathfrak{P%
}_{i_{1},i_{2},j_{1},j_{2}}U_{n,j_{1},j_{2}}.
\end{equation*}%
Tensor elements $\mathfrak{p}_{i_{1},i_{2},j_{1},j_{2}}$ for $1\leq i_{\iota
}\leq I_{\iota }-1$ have the form%
\begin{equation}
\begin{array}{l}
\mathfrak{p}_{i_{1},i_{2},j_{1},j_{2}}=\delta _{i_{1},j_{1}}\delta
_{i_{2},j_{2}} \\ 
+\Delta \tau \left( \mathfrak{l}_{\left( 11\right) i_{1},j_{1}}\delta
_{i_{2},j_{2}}+\dsum\limits_{\alpha _{1},\alpha _{2}\in \aleph }\mathfrak{l}%
_{\left( 12\right) i_{1},\alpha _{1},i_{2},\alpha _{2}}\delta _{i_{1}+\alpha
_{1},j_{1}}\delta _{i_{2}+\alpha _{2},j_{2}}+\delta _{i_{1},j_{1}}\mathfrak{l%
}_{\left( 22\right) i_{2},j_{2}}^{s}\right) ,%
\end{array}
\label{mat_elem}
\end{equation}%
where $\delta _{i,j}$ is the Kronecker delta. It is clear that the
corresponding tensor is very sparse. \footnote{%
We note in passing that we can uniquely map a matrix $U_{n,i_{1},i_{2}}$
into a vector $\tilde{U}_{n,I}$, where $I\left( i_{1},i_{2}\right)
=i_{1}+i_{2}\left( I_{1}+1\right) ,\ \ \ \ \ 0\leq I\leq
I_{1}I_{2}+I_{1}+I_{2}$. By doing so, we can define a matrix $\widetilde{%
\mathfrak{P}}$ and avoid using tensors altogether.}

In spite of its simplicity, this scheme is seldom used in practice because
it is unstable unless the corresponding time step is prohibitively small,
say one hour for an option with maturity of one year. Thus, in order to
compute%
\begin{equation}
U_{N}=\mathfrak{P}^{N}U_{0},  \label{Map_symbolic}
\end{equation}%
one has to perform $N\gg 1$ matrix multiplications, which is extraordinary
costly. However, recently this scheme won new lease of life, by using the
Fast Exponentiation, see, e.g., Albanese \textit{et al.} \cite{albanese}.
Provided that $N=2^{N^{\prime }}$, one can calculate $\mathfrak{P}^{N}$ in $%
N^{\prime }$ steps via the following recursion%
\begin{equation}
\mathfrak{P}_{1}=\mathfrak{P},\ \ \ \mathfrak{P}_{2}=\mathfrak{P}_{1}^{2},\
\ \ \mathfrak{P}_{N^{\prime }}=\mathfrak{P}_{N^{\prime }-1}^{2},
\label{Russian_doll1}
\end{equation}%
since%
\begin{equation}
\mathfrak{P}_{N^{\prime }}=\mathfrak{P}^{N}.  \label{Russian_doll2}
\end{equation}%
While, in our experience this method is still too cumbersome to be viable,
(at least if GPUs are not used,) it can be used for comparison purposes.

\subsection{ADI methods\label{ADI}}

It is natural to use Eq. (\ref{matrix_split}) in order to construct the
so-called ADI schemes for solving the discretized pricing problem. Here we
discretize the differential operators $\mathfrak{L}_{\left( 11\right) }^{s}$
and $\mathfrak{L}_{\left( 22\right) }^{s}$ via an implicit--explicit FD
scheme parametrized by $\varsigma $, and treat the operator $\mathcal{L}%
_{\left( 12\right) }$ in an explicit manner.

We start with the Do scheme, which consists of a single predictor step and
two correction steps, and can be symbolically written as follows%
\begin{equation}
U_{n}\Longrightarrow Y_{0}\Longrightarrow Y_{1}\Longrightarrow
Y_{2}\Longrightarrow U_{n+1}=\mathcal{F}_{D}\left( U_{n}\right) ,
\label{Do_sequence}
\end{equation}%
where%
\begin{equation}
\begin{array}{l}
Y_{0}=U_{n}+\Delta \tau \mathfrak{L}^{s}U_{n}, \\ 
Y_{1}=Y_{0}+\varsigma \Delta \tau \left( \mathfrak{L}_{\left( 11\right)
}^{s}Y_{1}-\mathfrak{L}_{\left( 11\right) }^{s}U_{n}\right) , \\ 
Y_{2}=Y_{1}+\varsigma \Delta \tau \left( \mathfrak{L}_{\left( 22\right)
}^{s}Y_{2}-\mathfrak{L}_{\left( 22\right) }^{s}U_{n}\right) , \\ 
U_{n+1}=Y_{2}.%
\end{array}
\label{Do_details}
\end{equation}%
It is first order accurate in time.

More accurate schemes repeat the Do scheme twice, once for prediction, once
for correction. We consider the following three: CS scheme:%
\begin{equation}
U_{n}\Longrightarrow Y_{0}\Longrightarrow Y_{1}\Longrightarrow
Y_{2}\Longrightarrow \tilde{Y}_{0}\Longrightarrow \tilde{Y}%
_{1}\Longrightarrow \tilde{Y}_{2}\Longrightarrow U_{n+1}=\mathcal{F}%
_{CS}\left( U_{n}\right) ,  \label{CS_sequence}
\end{equation}%
where%
\begin{equation}
\begin{array}{l}
Y_{0}=U_{n}+\Delta \tau \mathfrak{L}^{s}U_{n}, \\ 
Y_{1}=Y_{0}+\varsigma \Delta \tau \left( \mathfrak{L}_{\left( 11\right)
}^{s}Y_{1}-\mathfrak{L}_{\left( 11\right) }^{s}U_{n}\right) , \\ 
Y_{2}=Y_{1}+\varsigma \Delta \tau \left( \mathfrak{L}_{\left( 22\right)
}^{s}Y_{2}-\mathfrak{L}_{\left( 22\right) }^{s}U_{n}\right) , \\ 
\tilde{Y}_{0}=Y_{0}+\QTOVERD. . {1}{2}\Delta \tau \mathfrak{L}_{\left(
12\right) }\left( Y_{2}-U_{n}\right) , \\ 
\tilde{Y}_{1}=\tilde{Y}_{0}+\varsigma \Delta \tau \left( \mathfrak{L}%
_{\left( 11\right) }^{s}\tilde{Y}_{1}-\mathfrak{L}_{\left( 11\right)
}^{s}U_{n}\right) , \\ 
\tilde{Y}_{2}=\tilde{Y}_{1}+\varsigma \Delta \tau \left( \mathfrak{L}%
_{\left( 22\right) }^{s}\tilde{Y}_{2}-\mathfrak{L}_{\left( 22\right)
}^{s}U_{n}\right) , \\ 
U_{n+1}=\tilde{Y}_{2}.%
\end{array}
\label{CS_details}
\end{equation}%
In HW scheme:%
\begin{equation}
U_{n}\Longrightarrow Y_{0}\Longrightarrow Y_{1}\Longrightarrow
Y_{2}\Longrightarrow \tilde{Y}_{0}\Longrightarrow \tilde{Y}%
_{1}\Longrightarrow \tilde{Y}_{2}\Longrightarrow U_{n+1}=\mathcal{F}%
_{IW}\left( U_{n}\right) ,  \label{IW_sequence}
\end{equation}%
where the fourth step in sequence (\ref{CS_details}) is replaced by the
following one%
\begin{equation*}
\tilde{Y}_{0}=Y_{0}+\left( \left( \QTOVERD. . {1}{2}-\varsigma \right)
\Delta \tau \left( \mathfrak{L}_{\left( 11\right) }^{s}+\mathfrak{L}_{\left(
22\right) }^{s}\right) +\QTOVERD. . {1}{2}\Delta \tau \mathfrak{L}_{\left(
12\right) }\right) \left( Y_{2}-U_{n}\right) .
\end{equation*}%
HV scheme:%
\begin{equation}
U_{n}\Longrightarrow Y_{0}\Longrightarrow Y_{1}\Longrightarrow
Y_{2}\Longrightarrow \tilde{Y}_{0}\Longrightarrow \tilde{Y}%
_{1}\Longrightarrow \tilde{Y}_{2}\Longrightarrow U_{n+1}=\mathcal{F}%
_{HV}\left( U_{n}\right) ,  \label{HV_sequence}
\end{equation}%
where the fourth step in sequence (\ref{CS_details}) is replaced by the
following one%
\begin{equation}
\tilde{Y}_{0}=Y_{0}+\QTOVERD. . {1}{2}\Delta \tau \mathfrak{L}^{s}\left(
Y_{2}-U_{n}\right) .  \label{HV_details}
\end{equation}

The Do scheme is always first order accurate, CS is second order accurate
when $\varsigma =1/2$, whilst IW and HV schemes are second order accurate
for any $\varsigma $. The Do and CS schemes are unconditionally stable when $%
\varsigma \geq 1/2$ (hence the only practical choice for CS scheme is $%
\varsigma =1/2$). IW and HV schemes (without convection terms) are stable
when $\varsigma \geq 1/3$ and $\varsigma \geq 1-\sqrt{1/2}$, respectively;
it is conjectured that HV is stable in the convection-diffusion setup when $%
\varsigma \geq (1+\sqrt{1/3})/2$. Following \cite{inthout1} we choose $%
\varsigma =1/3$ and $\varsigma =(1+\sqrt{1/3})/2$ for IW and HV schemes,
respectively.

\subsection{Galerkin method\label{Galerkin}}

We now describe the Galerkin method for solving the problem (\ref{pde_univ}%
), (\ref{ic_univ}). Depending on the instrument under consideration, the
problem can be defined on the whole axis $\left( -\infty ,\infty \right) $,
a semi-axis, or on a finite interval. To be concrete, we assume that the
problem is defined on a finite interval $\left[ X_{L}^{s},X_{U}^{s}\right] $%
. (Of course, when $s=I,R$, it is possible that $X_{L}^{s}=X_{0}^{s}$, $%
X_{R}^{s}=X_{\infty }^{s}$.) As usual, we can choose a convenient basis in
the $x_{1}$ direction and represent $U\left( \tau ,x_{1},x_{2}\right) $ in
the form%
\begin{equation}
U\left( \tau ,x_{1},x_{2}\right) \boldsymbol{=}\dsum\limits_{k=1}^{\infty
}U_{k}\left( \tau ,x_{2}\right) e_{k}\left( x_{1}\right) .
\label{expansion_vector}
\end{equation}%
Here $e_{k}$ are appropriately chosen basis functions of $x_{1}$, $%
X_{L}^{s}\leq x_{1}\leq X_{U}^{s}$. In the case in question, it is
convenient to use an orthogonal (but not an ortho-normal) basis of the form%
\begin{equation}
e_{k}\left( x_{1}\right) =\sin \left( \zeta _{k}\left(
x_{1}-X_{0}^{s}\right) \right) ,  \label{e_k}
\end{equation}%
where $\zeta _{k}=\pi k/\Delta ^{s}$. Naturally, we consider a truncated
series%
\begin{equation}
U\left( \tau ,x_{1},x_{2}\right) \boldsymbol{=}\dsum\limits_{k=1}^{M}U_{k}%
\left( \tau ,x_{2}\right) e_{k}\left( x_{1}\right) ,
\label{expansion_vector_M}
\end{equation}%
where $M$ is suitably large. We can now think of $U\left( \tau
,x_{1},x_{2}\right) $ as a \emph{vector function} of two variables $\left(
\tau ,x_{2}\right) $ with vector components parametrized by the index $k$.%
\begin{equation*}
U\left( \tau ,x_{1},x_{2}\right) \Rightarrow \overrightarrow{U}\left( \tau
,x_{2}\right) =\left\{ U_{k}\left( \tau ,x_{2}\right) \right\} ,
\end{equation*}%
and rewrite the problem (\ref{pde_univ}), (\ref{ic_univ}) in the form%
\begin{equation}
\partial _{\tau }\overrightarrow{U}-\mathcal{\tilde{L}}_{\left( 22\right) }%
\overrightarrow{U}-\rho \varepsilon x_{2}\partial _{x_{2}}\mathbb{B}^{s}%
\overrightarrow{U}-x_{2}\mathbb{C}^{s}\overrightarrow{U}=0,
\label{PDE_vector}
\end{equation}%
\begin{equation}
\overrightarrow{U}\left( 0\right) =\overrightarrow{u},  \label{ic_vector}
\end{equation}%
where%
\begin{equation}
\mathcal{\tilde{L}}_{\left( 22\right) }\overrightarrow{U}=\QTOVERD. .
{1}{2}\varepsilon ^{2}x_{2}\overrightarrow{U}_{x_{2}x_{2}}+\kappa \left(
1-x_{2}\right) \overrightarrow{U}_{x_{2}},  \label{L22_scalar}
\end{equation}%
\begin{equation}
\begin{array}{c}
\mathbb{B}^{s}\overrightarrow{U}=\overrightarrow{U}_{x_{1}}+\mathsf{\tilde{b}%
}_{2}^{s}\left( x_{1}\right) \overrightarrow{U}, \\ 
\mathbb{C}^{s}\overrightarrow{U}=\QTOVERD. . {1}{2}\overrightarrow{U}%
_{x_{1}x_{1}}-\QTOVERD. . {1}{2}\omega ^{s}\overrightarrow{U},%
\end{array}
\label{B_and_C}
\end{equation}%
and%
\begin{equation}
\mathsf{\tilde{b}}_{2}^{s}\left( x_{1}\right) =\left\{ 
\begin{array}{ll}
\QTOVERD. . {1}{2}, & s=H, \\ 
\QTOVERD. . {1}{2}\beta , & s=DH, \\ 
\sqrt{\left\vert \omega ^{I}\right\vert }\cot \left( \sqrt{\left\vert \omega
^{I}\right\vert }\left( X_{\infty }^{I}-x_{1}\right) \right) , & s=I, \\ 
\sqrt{\omega ^{R}}\coth \left( \sqrt{\omega ^{R}}\left( X_{\infty
}^{R}-x_{1}\right) \right) , & s=R.%
\end{array}%
\right.  \label{b2_tilde}
\end{equation}

It is clear that%
\begin{equation}
\mathbb{B}^{s}e_{k}=\dsum\limits_{l=1}^{M}\left( \hat{\mu}_{kl}^{s}+\bar{\mu}%
_{kl}^{s}\right) e_{l}\equiv \dsum\limits_{l=1}^{M}\mu _{kl}^{s}e_{l},
\label{B_matrix}
\end{equation}%
\begin{equation}
\mathbb{C}^{s}e_{k}=-\QTOVERD. . {1}{2}\left( \zeta _{k}^{2}+\omega
^{s}\right) e_{k}\equiv -\lambda _{k}^{s}e_{k},  \label{C_matrix}
\end{equation}%
\begin{equation}
u=\dsum\limits_{k=1}^{M}\nu _{k}^{s}e_{k},  \label{u_vector}
\end{equation}%
where%
\begin{equation}
\hat{\mu}_{kl}^{s}=\left\{ 
\begin{array}{ll}
0, & l=k, \\ 
\frac{2kl\left( \left( -1\right) ^{k-l}-1\right) }{\left( k^{2}-l^{2}\right)
\left( X_{U}^{s}-X_{L}^{s}\right) }, & l\neq k,%
\end{array}%
\right.
\end{equation}%
\begin{equation}
\bar{\mu}_{kl}^{s}=\left\{ 
\begin{array}{ll}
\QTOVERD. . {1}{2}\delta _{kl}, & s=H, \\ 
\QTOVERD. . {1}{2}\beta \delta _{kl}, & s=DH, \\ 
\begin{array}{l}
\frac{2\sqrt{\left\vert \omega ^{I}\right\vert }}{\left(
X_{U}^{I}-X_{L}^{I}\right) }\int_{X_{L}^{I}}^{X_{U}^{I}}\cot \left( \sqrt{%
\left\vert \omega ^{I}\right\vert }\left( X_{\infty }^{I}-x_{1}\right)
\right) \\ 
\times \sin \left( \zeta _{k}\left( x_{1}-X_{0}^{I}\right) \right) \sin
\left( \zeta _{l}\left( x_{1}-X_{0}^{I}\right) \right) dx_{1},%
\end{array}
& s=I, \\ 
\begin{array}{l}
\frac{2\sqrt{\omega ^{R}}}{\left( X_{U}^{R}-X_{L}^{R}\right) }%
\int_{X_{L}^{R}}^{X_{U}^{R}}\coth \left( \sqrt{\omega ^{R}}\left( X_{\infty
}^{R}-x_{1}\right) \right) \\ 
\times \sin \left( \zeta _{k}\left( x_{1}-X_{0}^{R}\right) \right) \sin
\left( \zeta _{l}\left( x_{1}-X_{0}^{R}\right) \right) dx_{1},%
\end{array}
& s=R,%
\end{array}%
\right.  \label{mu_kl}
\end{equation}%
\begin{equation}
\nu _{k}^{s}=\frac{2}{\left( X_{U}^{I}-X_{L}^{I}\right) }%
\int_{X_{L}^{I}}^{X_{U}^{I}}u\left( x_{1}\right) \sin \left( \zeta
_{k}\left( x_{1}-X_{0}^{I}\right) \right) dx_{1},  \label{nu_k1}
\end{equation}%
and $\zeta _{l}=\pi l/\left( X_{U}^{s}-X_{L}^{s}\right) $. We notice that
the corresponding integrands are singular at $X=X_{\infty }^{s}$, provided,
of course, that $X_{U}^{s}=X_{\infty }^{s}$, but the integrals are well
defined anyway. While it is possible to express $\bar{\mu}_{kl}^{s}$, $s=I,R$%
, in terms of hypergeometric functions, it is easier to compute them
numerically, which is what we do. Substitution of the above formulas in the
pricing equation and initial condition yields%
\begin{equation}
\partial _{\tau }U_{k}-\mathcal{\tilde{L}}_{\left( 2,2\right)
k}^{s}U_{k}-\rho \varepsilon x_{2}\partial _{x_{2}}\dsum\limits_{l=1}^{M}\mu
_{kl}^{s}U_{l}=0,  \label{pde_scalar}
\end{equation}%
\begin{equation}
U_{k}\left( 0\right) =\nu _{k},  \label{ic_scalar}
\end{equation}%
where%
\begin{equation}
\mathcal{\tilde{L}}_{\left( 2,2\right) k}=\QTOVERD. . {1}{2}\varepsilon
^{2}x_{2}\partial _{x_{2}}^{2}+\kappa \left( 1-x_{2}\right) \partial
_{x_{2}}-\QTOVERD. . {1}{2}\lambda _{k}^{s}x_{2}.  \label{L22_scalar_k}
\end{equation}

In words, we replace a two-factor parabolic PDE with a coupled system of
one-factor parabolic PDEs. We solve this system of equations by treating the
cross term fully explicitly, which allows us to use the standard technique
for solving scalar one-factor PDEs with nonzero source terms. We emphasize
that this system becomes uncoupled when $\rho =0$. In the latter case it can
be solved analytically.

When applicable, the Galerkin method generally beats the standard ADI
methods due to the fact that it is treating the problem in the $x_{1}$%
-direction in a natural way. In general, computational savings are of order $%
I_{1}/M$.

\subsection{Small $\protect\rho $ expansion\label{Expansion}}

Consider Eq. (\ref{PDE_vector}). If we assume that $\rho $ is small, we can
use it as an expansion parameter and write $\overrightarrow{U}$ in the form%
\begin{equation}
\overrightarrow{U}=\overrightarrow{U}^{\left( 0\right) }+\rho \varepsilon 
\overrightarrow{U}^{\left( 1\right) }+\left( \rho \varepsilon \right) ^{2}%
\overrightarrow{U}^{\left( 2\right) }+\left( \rho \varepsilon \right) ^{3}%
\overrightarrow{U}^{\left( 3\right) }...=\sum_{n=0}^{\infty }\left( \rho
\varepsilon \right) ^{n}\overrightarrow{U}^{\left( n\right) },
\label{expansion}
\end{equation}%
where%
\begin{equation}
\partial _{\tau }\overrightarrow{U}^{\left( 0\right) }-\mathcal{\tilde{L}}%
_{\left( 22\right) }\overrightarrow{U}^{\left( 0\right) }-x_{2}\mathbb{C}^{s}%
\overrightarrow{U}^{\left( 0\right) }=0,\ \ \ \ \ \overrightarrow{U}^{\left(
0\right) }\left( 0\right) =\overrightarrow{u},  \label{expansion_0}
\end{equation}%
\begin{equation}
\partial _{\tau }\overrightarrow{U}^{\left( 1\right) }-\mathcal{\tilde{L}}%
_{\left( 22\right) }\overrightarrow{U}^{\left( 1\right) }-x_{2}\mathbb{C}^{s}%
\overrightarrow{U}^{\left( 1\right) }=x_{2}\partial _{x_{2}}\mathbb{B}^{s}%
\overrightarrow{U}^{\left( 0\right) },\ \ \ \ \ \overrightarrow{U}^{\left(
1\right) }\left( 0\right) =0,  \label{expansion_1}
\end{equation}%
\begin{equation}
\partial _{\tau }\overrightarrow{U}^{\left( 2\right) }-\mathcal{\tilde{L}}%
_{\left( 22\right) }\overrightarrow{U}^{\left( 2\right) }-x_{2}\mathbb{C}^{s}%
\overrightarrow{U}^{\left( 2\right) }=x_{2}\partial _{x_{2}}\mathbb{B}^{s}%
\overrightarrow{U}^{\left( 1\right) },\ \ \ \ \ \overrightarrow{U}^{\left(
2\right) }\left( 0\right) =0,  \label{expansion_2}
\end{equation}%
etc. In general,%
\begin{equation}
\partial _{\tau }\overrightarrow{U}^{\left( n\right) }-\mathcal{\tilde{L}}%
_{\left( 22\right) }\overrightarrow{U}^{\left( n\right) }-x_{2}\mathbb{C}^{s}%
\overrightarrow{U}^{\left( n\right) }=x_{2}\partial _{x_{2}}\mathbb{B}^{s}%
\overrightarrow{U}^{\left( n-1\right) },\ \ \ \ \ \overrightarrow{U}^{\left(
n\right) }\left( 0\right) =0,  \label{expansion_n}
\end{equation}

Below we need to solve the following initial value problems%
\begin{equation}
\partial _{\tau }\mathsf{w}_{n}-\mathcal{\tilde{L}}_{\left( 22\right) }%
\mathsf{w}_{n}+\QTOVERD. . {1}{2}\lambda x_{2}\mathsf{w}_{n}=0,\ \ \ \ \ 
\mathsf{w}_{n}\left( 0\right) =x_{2}^{n}e^{\psi x_{2}},  \label{w_n_eq}
\end{equation}%
where $\lambda ,\psi $ are given constants, and $n=0,1,2,...$. The
corresponding solutions $\mathsf{w}_{n}$ can be found by using an affine
ansatz:%
\begin{equation}
\begin{array}{l}
\mathsf{w}_{0}\left( \tau ,x_{2},\lambda ,\psi \right) =D_{0,0}\left( \tau
,\lambda ,\psi \right) E\left( \tau ,x_{2},\lambda ,\psi \right) , \\ 
\mathsf{w}_{1}\left( \tau ,x_{2},\lambda ,\psi \right) =\left( D_{1,0}\left(
\tau ,\lambda ,\psi \right) +D_{1,1}\left( \tau ,\lambda ,\psi \right)
x_{2}\right) E\left( \tau ,x_{2},\lambda ,\psi \right) , \\ 
\mathsf{w}_{2}\left( \tau ,x_{2},\lambda ,\psi \right) =\left( D_{2,0}\left(
\tau ,\lambda ,\psi \right) +D_{2,1}\left( \tau ,\lambda ,\psi \right)
x_{2}+D_{2,2}\left( \tau ,\lambda ,\psi \right) x_{2}^{2}\right) E\left(
\tau ,x_{2},\lambda ,\psi \right) ,%
\end{array}
\label{w012}
\end{equation}%
etc., where, by definition, $D_{0,0}\left( \tau ,\lambda ,\psi \right) =1$,
and%
\begin{equation}
E\left( \tau ,x_{2},\lambda ,\psi \right) =e^{A\left( \tau ,\lambda ,\psi
\right) +B\left( \tau ,\lambda ,\psi \right) x_{2}},  \label{E(tau,x2)}
\end{equation}%
In general,%
\begin{equation}
\mathsf{w}_{n}\left( \tau ,x_{2},\lambda ,\psi \right) =\left(
\sum_{m=0}^{n}D_{n,m}\left( \tau ,\lambda ,\psi \right) x_{2}^{m}\right)
E\left( \tau ,x_{2},\lambda ,\psi \right) .  \label{w_n}
\end{equation}%
All the coefficients can be written explicitly.

Let us calculate the relevant quantities for $n=0,1,2,3$. It is clear that $%
A\left( \tau ,\lambda ,\psi \right) $, $B\left( \tau ,\lambda ,\psi \right) $
satisfy the following system of ODEs%
\begin{equation*}
\begin{array}{l}
A^{\prime }\left( \tau ,\lambda ,\psi \right) -\kappa B\left( \tau ,\lambda
,\psi \right) =0, \\ 
B^{\prime }\left( \tau ,\lambda ,\psi \right) -\QTOVERD. . {1}{2}\varepsilon
^{2}B^{2}\left( \tau ,\lambda ,\psi \right) +\kappa B\left( \tau ,\lambda
,\psi \right) +\QTOVERD. . {1}{2}\lambda =0,%
\end{array}%
\end{equation*}%
supplied with the initial conditions of the form%
\begin{equation}
A\left( 0,\lambda ,\psi \right) =0,\ \ \ \ \ B\left( 0,\lambda ,\psi \right)
=\psi .  \label{affine_ic}
\end{equation}%
The Riccati transform 
\begin{equation}
A\left( \tau ,\lambda ,\psi \right) =-\frac{2\kappa }{\varepsilon ^{2}}\ln
\left( \gamma \left( \tau ,\lambda ,\psi \right) \right) ,\ \ \ B\left( \tau
,\lambda ,\psi \right) =-\frac{2\gamma ^{\prime }\left( \tau ,\lambda ,\psi
\right) }{\varepsilon ^{2}\gamma \left( \tau ,\lambda ,\psi \right) },
\label{Riccati}
\end{equation}%
yields%
\begin{equation}
\gamma ^{\prime \prime }\left( \tau ,\lambda ,\psi \right) +\kappa \gamma
^{\prime }\left( \tau ,\lambda ,\psi \right) -\QTOVERD. . {1}{4}\varepsilon
^{2}\lambda \gamma \left( \tau ,\lambda ,\psi \right) =0,  \label{linear_ODE}
\end{equation}%
supplied with the initial conditions%
\begin{equation}
\gamma \left( 0,\lambda ,\psi \right) =1,\ \ \ \ \ \gamma ^{\prime }\left(
0,\lambda ,\psi \right) =-\QTOVERD. . {1}{2}\varepsilon ^{2}\psi .
\end{equation}%
A simple algebra shows that the corresponding solution can be written in the
form%
\begin{equation}
\gamma \left( \tau ,\lambda ,\psi \right) =\frac{e^{\Xi _{+}\tau /2}\Upsilon
\left( \tau ,\lambda ,\psi \right) }{2\varpi \left( \lambda \right) },
\label{affine_gamma}
\end{equation}%
where%
\begin{equation}
\begin{array}{l}
\Upsilon \left( \tau ,\lambda ,\psi \right) =\left( \Xi _{-}\left( \lambda
\right) -\varepsilon ^{2}\psi \right) +\left( \Xi _{+}\left( \lambda \right)
+\varepsilon ^{2}\psi \right) e^{-\varpi \left( \lambda \right) \tau }, \\ 
\Xi _{\pm }\left( \lambda \right) =\mp \kappa +\varpi \left( \lambda \right)
, \\ 
\varpi \left( \lambda \right) =\sqrt{\kappa ^{2}+\varepsilon ^{2}\lambda }.%
\end{array}
\label{affine_1}
\end{equation}%
Finally,%
\begin{equation}
\begin{array}{l}
A\left( \tau ,\lambda ,\psi \right) =-\frac{\kappa }{\varepsilon ^{2}}\left(
\Xi _{+}\left( \lambda \right) \tau +2\ln \left( \frac{\Upsilon \left( \tau
,\lambda ,\psi \right) }{2\varpi \left( \lambda \right) }\right) \right) ,
\\ 
B\left( \tau ,\lambda ,\psi \right) =-\frac{\left( \Xi _{+}\left( \lambda
\right) \left( \Xi _{-}\left( \lambda \right) -\varepsilon ^{2}\psi \right)
-\Xi _{-}\left( \lambda \right) \left( \Xi _{+}\left( \lambda \right)
+\varepsilon ^{2}\psi \right) e^{-\varpi \left( \lambda \right) \tau
}\right) }{\varepsilon ^{2}\Upsilon \left( \tau ,\lambda ,\psi \right) }.%
\end{array}
\label{affine_2}
\end{equation}%
In order to compute $D_{n,m}$ we differentiate the affine solution with
respect to $\psi $ and obtain%
\begin{equation}
\begin{array}{l}
D_{0,0}=1, \\ 
D_{1,0}+D_{1,1}x_{2}=\dot{A}+\dot{B}x_{2}, \\ 
D_{2,0}+D_{2,1}x_{2}+D_{2,2}x_{2}^{2}=\left( \dot{A}+\dot{B}x_{2}\right)
^{2}+\ddot{A}+\ddot{B}x_{2}, \\ 
D_{3,0}+D_{3,1}x_{2}+D_{3,2}x_{2}^{2}+D_{3,3}x_{2}^{3} \\ 
=\left( \dot{A}+\dot{B}x_{2}\right) ^{3}+3\left( \dot{A}+\dot{B}x_{2}\right)
\left( \ddot{A}+\ddot{B}x_{2}\right) +\dddot{A}+\dddot{B}x_{2},%
\end{array}
\label{D_mn0}
\end{equation}%
so that%
\begin{equation}
\begin{array}{l}
D_{0,0}=1, \\ 
D_{1,0}=\dot{A},\ \ \ \ \ D_{1,1}=\dot{B}, \\ 
D_{2,0}=\dot{A}^{2}+\ddot{A},\ \ \ \ \ D_{2,1}=2\dot{A}\dot{B}+\ddot{B},\ \
\ \ \ D_{2,2}=\dot{B}^{2}, \\ 
D_{3,0}=\dot{A}^{3}+3\dot{A}\ddot{A}+\dddot{A},\ \ \ D_{3,1}=3(\dot{A}^{2}%
\dot{B}+\dot{A}\ddot{B}+\ddot{A}\dot{B})+\dddot{B}, \\ 
D_{3,2}=3\left( \dot{A}\dot{B}^{2}+\dot{B}\ddot{B}\right) ,\ \ \ D_{3,3}=%
\dot{B}^{3}.%
\end{array}
\label{D_mn1}
\end{equation}%
Here%
\begin{eqnarray}
\dot{A} &=&2\kappa \Omega ,\ \ \ \ddot{A}=2\kappa \varepsilon ^{2}\Omega
^{2},\ \ \ \dddot{A}=4\kappa \varepsilon ^{4}\Omega ^{3},  \label{Adot_Bdot}
\\
\dot{B} &=&4\Theta ,\ \ \ \ddot{B}=8\varepsilon ^{2}\Theta \Omega ,\ \ \ 
\dddot{B}=24\varepsilon ^{4}\Theta \Omega ^{2},  \notag
\end{eqnarray}%
with%
\begin{equation}
\Omega =\frac{\left( 1-e^{-\varpi \left( \lambda \right) \tau }\right) }{%
\Upsilon \left( \tau ,\lambda ,\psi \right) },\ \ \ \ \ \Theta =\frac{\varpi
^{2}\left( \lambda \right) e^{-\varpi \left( \lambda \right) \tau }}{%
\Upsilon \left( \tau ,\lambda ,\psi \right) ^{2}}.  \label{Omega_Theta}
\end{equation}

When we consider a perturbation of order $n$ we introduce an ordered set of
times%
\begin{equation}
\boldsymbol{\tau }=\left( \tau _{0}=0<\tau _{1}<...<\tau _{n}<\tau
_{n+1}=\tau \right) .  \label{tau_set}
\end{equation}%
In particular, for $n=0$ (for the leading order term) we have only two
points $\tau _{0}=0<\tau _{1}=\tau $. By using this notation, we can
introduce $\overrightarrow{W}^{\left( 0\right) }\left( \boldsymbol{\tau }%
,x_{2}\right) $ of the form%
\begin{equation}
\overrightarrow{W}^{\left( 0\right) }\left( \boldsymbol{\tau },x_{2}\right)
=\dsum\limits_{k_{1}=1}^{\infty }C_{0,0}e^{A\left( 0,1\right) +B\left(
0,1\right) x_{2}}\nu _{k_{1}}\overrightarrow{e}_{k_{1}}.  \label{W0}
\end{equation}%
where the following notation is used%
\begin{equation}
C_{0,0}=1,\ \ \ A\left( 0,1\right) =A\left( \tau _{1}-\tau _{0},\lambda
_{k_{1}},0\right) ,\ \ \ B\left( 0,1\right) =B\left( \tau _{1}-\tau
_{0},\lambda _{k_{1}},0\right) .  \label{C_00}
\end{equation}%
Since there are no intermediate time points for $n=0$, we can write%
\begin{equation}
\overrightarrow{U}^{\left( 0\right) }\left( \tau ,x_{2}\right) =%
\overrightarrow{W}^{\left( 0\right) }\left( \boldsymbol{\tau },x_{2}\right) .
\label{U0}
\end{equation}%
Generalization of the above construct allows us to introduce $%
\overrightarrow{W}^{\left( n\right) }\left( \tau ,x_{2}\right) $ as follows%
\begin{equation}
\begin{array}{ll}
\overrightarrow{W}^{\left( n\right) }\left( \boldsymbol{\tau },x_{2}\right) =
& \dsum\limits_{k_{1}=1,...,k_{n+1}=1}^{\infty }\left(
\sum_{m=0}^{n}C_{n,m}x_{2}^{m}\right) \\ 
& \times e^{\boldsymbol{A}\left( 0,n+1\right) +B\left( n,n+1\right)
x_{2}}\nu _{k_{1}}\mu _{k_{1}k_{2}}...\mu _{k_{n}k_{n+1}}\overrightarrow{e}%
_{k_{n}+1},%
\end{array}
\label{Wn}
\end{equation}%
where we slightly abuse the notation and write%
\begin{equation}
\begin{array}{ll}
A\left( n,n+1\right) = & A\left( \tau _{n+1}-\tau _{n},\lambda
_{k_{n+1}},B\left( n-1,n\right) \right) , \\ 
B\left( n,n+1\right) = & B\left( \tau _{n+1}-\tau _{n},\lambda
_{k_{n+1}},B\left( n-1,n\right) \right) . \\ 
\boldsymbol{A}\left( 0,n+1\right) = & A\left( 0,1\right) +...+A\left(
n,n+1\right)%
\end{array}
\label{A_n_B_n}
\end{equation}%
This definition is clearly recurrent (telescopic). We claim that $%
\overrightarrow{U}^{\left( n\right) }$ can be expressed in terms of $%
\overrightarrow{W}^{\left( n\right) }$ via a simple integration over
intermediate time steps, i.e., 
\begin{equation}
\overrightarrow{U}^{\left( n\right) }\left( \tau ,x_{2}\right)
=\dint\limits_{0}^{\tau }\dint\limits_{\tau _{1}}^{\tau
}...\dint\limits_{\tau _{n-2}}^{\tau }\dint\limits_{\tau _{n-1}}^{\tau }%
\overrightarrow{W}^{\left( n\right) }\left( \boldsymbol{\tau },x_{2}\right)
d\tau _{1}...d\tau _{n}.  \label{Un}
\end{equation}%
In order to prove this fact, we can use Duhamel's principle and reduce the
corresponding inhomogeneous problems to a family of homogeneous problems.
Carefully accounting for the powers of $x_{2}$, we can derive the following
recurrent relation%
\begin{equation}
C_{n,m}=\sum_{l=1}^{n}\left( B\left( n-1,n\right)
C_{n-1,l-1}+lC_{n-1,l}\right) D_{l,m}\left( n,n+1\right) ,\ \ \ n>0,\ \ \
m=0,...,n,  \label{C_nm}
\end{equation}%
where $C_{0,0}=1$, $C_{n,n^{\prime }}=0$ if $n^{\prime }>n$, and%
\begin{equation}
D_{l,m}\left( n,n+1\right) =D_{l,m}\left( \tau _{n+1}-\tau _{n},\lambda
_{k_{n+1}},B\left( n-1,n\right) \right) .  \label{D_lm}
\end{equation}

By using these formulas, we immediately obtain the following expressions for
the first three expansion terms%
\begin{equation}
\begin{array}{ll}
C_{1,0}= & B\left( 0,1\right) D_{1,0}\left( 1,2\right) , \\ 
C_{1,1}= & B\left( 0,1\right) D_{1,1}\left( 1,2\right) , \\ 
C_{2,0}= & \left( B\left( 1,2\right) C_{1,0}+C_{1,1}\right) D_{1,0}\left(
2,3\right) +B\left( 1,2\right) C_{1,1}D_{2,0}\left( 2,3\right) , \\ 
C_{2,1}= & \left( B\left( 1,2\right) C_{1,0}+C_{1,1}\right) D_{1,1}\left(
2,3\right) +B\left( 1,2\right) C_{1,1}D_{2,1}\left( 2,3\right) , \\ 
C_{2,2}= & B\left( 1,2\right) C_{1,1}D_{2,2}\left( 2,3\right) , \\ 
C_{3,0}= & \left( B\left( 2,3\right) C_{2,0}+C_{2,1}\right) D_{1,0}\left(
3,4\right) +\left( B\left( 2,3\right) C_{2,1}+2C_{2,2}\right) D_{2,0}\left(
3,4\right) \\ 
& +B\left( 2,3\right) C_{2,2}D_{3,0}\left( 3,4\right) , \\ 
C_{3,1}= & \left( B\left( 2,3\right) C_{2,0}+C_{2,1}\right) D_{1,1}\left(
3,4\right) +\left( B\left( 2,3\right) C_{2,1}+2C_{2,2}\right) D_{2,1}\left(
3,4\right) \\ 
& +B\left( 2,3\right) C_{2,2}D_{3,1}\left( 3,4\right) , \\ 
C_{3,2}= & \left( B\left( 2,3\right) C_{2,1}+2C_{2,2}\right) D_{2,2}\left(
3,4\right) +B\left( 2,3\right) C_{2,2}D_{3,2}\left( 3,4\right) , \\ 
C_{3,3}= & B\left( 2,3\right) C_{2,2}D_{3,3}\left( 3,4\right) .%
\end{array}
\label{C3i}
\end{equation}%
Higher order correction can be computed in the same manner.

In order to simplify Eq. (\ref{Un}), we perform a change of variables and
transform the simplex over which the integration is performed into the unit
cube. Specifically, we introduce $\xi _{n}$, $0\leq \xi _{n}\leq 1$, and
write%
\begin{equation}
\begin{array}{l}
\frac{\tau _{1}}{\tau }=\xi _{1}\equiv \eta _{1}, \\ 
\frac{\tau _{2}}{\tau }=\xi _{1}+\left( 1-\xi _{1}\right) \xi _{2}\equiv
\eta _{2}, \\ 
\frac{\tau _{3}}{\tau }=\xi _{1}+\left( 1-\xi _{1}\right) \xi _{2}+\left(
1-\xi _{1}\right) \left( 1-\xi _{2}\right) \xi _{3}\equiv \eta _{3},%
\end{array}
\label{tau->xi}
\end{equation}%
etc. It is clear that%
\begin{equation}
\begin{array}{l}
d\tau _{1}=\tau d\xi _{1}, \\ 
d\tau _{2}=\tau ^{2}\left( 1-\xi _{1}\right) d\xi _{1}d\xi _{2}, \\ 
d\tau _{3}=\tau ^{3}\left( 1-\xi _{1}\right) ^{2}\left( 1-\xi _{2}\right)
d\xi _{1}d\xi _{2}d\xi _{3},%
\end{array}
\label{dtau->dxi}
\end{equation}%
so that,%
\begin{equation}
\begin{array}{l}
\dint\limits_{0}^{\tau }f\left( \tau _{1}\right) d\tau _{1}=\tau
\dint\limits_{0}^{1}f\left( \tau \eta _{1}\right) d\xi _{1}, \\ 
\dint\limits_{0}^{\tau }\dint\limits_{\tau _{1}}^{\tau }f\left( \tau
_{1},\tau _{2}\right) d\tau _{1}d\tau _{2}=\tau
^{2}\dint\limits_{0}^{1}\dint\limits_{0}^{1}f\left( \tau \eta _{1},\tau \eta
_{2}\right) \left( 1-\xi _{1}\right) d\xi _{1}d\xi _{2}, \\ 
\dint\limits_{0}^{\tau }\dint\limits_{\tau _{1}}^{\tau }\dint\limits_{\tau
_{2}}^{\tau }f\left( \tau _{1},\tau _{2},\tau _{3}\right) d\tau _{1}d\tau
_{2}d\tau _{3}=\tau
^{3}\dint\limits_{0}^{1}\dint\limits_{0}^{1}\dint\limits_{0}^{1}f\left( \tau
\eta _{1},\tau \eta _{2},\tau \eta _{3}\right) \left( 1-\xi _{1}\right)
^{2}\left( 1-\xi _{2}\right) d\xi _{1}d\xi _{2}d\xi _{3},%
\end{array}
\label{integrals}
\end{equation}%
etc. Finally, in order to perform integration over the unit interval we use
Bode's rule.

\subsection{Monte Carlo method\label{MonteCarlo}}

Consider the standard Heston SDEs, which we can write as follows%
\begin{equation}
\begin{array}{lll}
dx_{t}= & -\QTOVERD. . {1}{2}v_{t}dt+\sqrt{v_{t}}\left( \rho dZ_{t}+\bar{\rho%
}d\tilde{W}_{t}\right) , & x_{0}=0, \\ 
dv_{t}= & \kappa \left( 1-v_{t}\right) dt+\varepsilon \sqrt{v_{t}}dZ_{t}, & 
v_{0}=v,%
\end{array}
\label{stoch_vol1}
\end{equation}%
where $\bar{\rho}=\sqrt{1-\rho ^{2}}$, $x_{t}=\ln \left( F_{t}/F\right) $,
and $d\tilde{W}_{t}dZ_{t}=0$. A well-known argument (see, e.g., \cite{hull}
for the zero correlation case, and \cite{willard} for the general case),
shows that%
\begin{equation}
x_{T}-x_{t}=-\QTOVERD. . {1}{2}I_{t}^{T}+\rho J_{t}^{T}+\xi _{t}^{T},
\label{x_tT0}
\end{equation}%
where%
\begin{equation}
I_{t}^{T}=\int_{t}^{T}v_{t^{\prime }}dt^{\prime },\ \ \ \ \
J_{t}^{T}=\int_{t}^{T}\sqrt{v_{t^{\prime }}}dZ_{t^{\prime }},
\label{Integrals_I_J}
\end{equation}%
\begin{equation}
\xi _{t}^{T}=N\left( 0,\bar{\rho}^{2}I_{t}^{T}\right) .  \label{xi_tT}
\end{equation}%
Equivalently,%
\begin{equation}
x_{T}-x_{t}=-\QTOVERD. . {1}{2}\rho ^{2}I_{t}^{T}+\rho J_{t}^{T}+\tilde{\xi}%
_{t}^{T},  \label{x_tT1}
\end{equation}%
where%
\begin{equation}
\tilde{\xi}_{t}^{T}=N\left( -\QTOVERD. . {1}{2}\bar{\rho}^{2}I_{t}^{T},\bar{%
\rho}^{2}I_{t}^{T}\right) .  \label{xi_tT_tilde}
\end{equation}%
In particular,%
\begin{equation}
x_{T}=-\QTOVERD. . {1}{2}\rho ^{2}I_{0}^{T}+\rho J_{0}^{T}+\tilde{\xi}%
_{0}^{T}.  \label{x_0T}
\end{equation}%
Thus, conditional on the values of $I_{0}^{T},J_{0}^{T}$ we see that $x_{T}$
is a normal variable. This observation can be used to extend the classical
BSM formula (\ref{BS_formula}) to the case of stochastic volatility. The
corresponding\ formula has the form%
\begin{equation}
\begin{array}{c}
C^{SV}(0,1,v;T,K)=\int_{0}^{\infty }\int_{-\infty }^{\infty
}C^{BS}(0,E_{0}^{T};T,K;\sqrt{\frac{\bar{\rho}^{2}I_{0}^{T}}{T}})\phi
_{T}\left( \left. I_{0}^{T},J_{0}^{T}\right\vert v\right)
dI_{0}^{T}dJ_{0}^{T},%
\end{array}
\label{Willard0}
\end{equation}%
where%
\begin{equation}
E_{t}^{T}=e^{-\QTOVERD. . {1}{2}\rho ^{2}I_{t}^{T}+\rho J_{t}^{T}},
\end{equation}%
$\phi _{\tau }\left( \left. I_{t}^{T},J_{t}^{T}\right\vert v_{t}\right) $ is
the joint p.d.f. of $\left( I_{t}^{T},J_{t}^{T}\right) $ conditional on $%
v_{t}$. It should be noted that, in general, this expression is far too
complex to be of any practical value. Moreover, it cannot be generalized for
pricing first generation exotic options, such as DNT or barrier options,
which is the main topic of this paper.

The above methodology can be extended verbatim to the case of generic SV
dynamics. For the Heston model expression (\ref{x_tT1}) can be simplified.
Namely, the SDE for $v_{t}$ can be integrated%
\begin{equation}
J_{t}^{T}=\frac{1}{\varepsilon }\left( v_{T}-v_{t}-\kappa \tau +\kappa
I_{t}^{T}\right) ,  \label{J_tT}
\end{equation}%
so that%
\begin{equation}
\begin{array}{ll}
x_{T}-x_{t} & =-\QTOVERD. . {1}{2}\rho ^{2}I_{t}^{T}+\frac{\rho }{%
\varepsilon }\left( v_{T}-v_{t}-\kappa \tau +\kappa I_{t}^{T}\right) +\tilde{%
\xi}_{t}^{T} \\ 
& =\frac{\rho }{\varepsilon }\left( v_{T}-v_{t}-\kappa \tau +\hat{\kappa}%
I_{t}^{T}\right) +\tilde{\xi}_{t}^{T}.%
\end{array}
\label{x_tT2}
\end{equation}%
where $\hat{\kappa}=\kappa -\QTOVERD. . {1}{2}\rho \varepsilon $. In
particular,%
\begin{equation}
x_{T}=\frac{\rho }{\varepsilon }\left( v_{T}-v-\kappa T+\hat{\kappa}%
I_{0}^{T}\right) +\tilde{\xi}_{0}^{T}.  \label{x_0T1}
\end{equation}%
In the differential form we have%
\begin{equation}
dx_{t}=\mu _{t}dt+\sigma _{t}d\tilde{W}_{t},  \label{SDE_x}
\end{equation}%
where%
\begin{equation}
\mu _{t}=\frac{\rho }{\varepsilon }\frac{dv_{t}}{dt}+\left( \frac{\rho
\kappa }{\varepsilon }-\QTOVERD. . {1}{2}\right) v_{t}-\frac{\rho \kappa }{%
\varepsilon },\ \ \ \sigma _{t}=\bar{\rho}\sqrt{v_{t}}.  \label{SDE_x_coef1}
\end{equation}%
Accordingly,%
\begin{equation}
\begin{array}{c}
C^{SV}(0,1,v;T,K)=\int_{0}^{\infty }\int_{0}^{\infty }C^{BS}(0,\tilde{E}%
_{0}^{T};T,K;\sqrt{\frac{\bar{\rho}^{2}I_{0}^{T}}{T}})\chi _{T}\left( \left.
I_{0}^{T},v_{T}\right\vert v\right) dI_{0}^{T}dv_{T},%
\end{array}
\label{Willard1}
\end{equation}%
where%
\begin{equation}
\tilde{E}_{t}^{T}=e^{\frac{\rho }{\varepsilon }\left( v_{T}-v_{t}-\kappa
\tau +\hat{\kappa}I_{t}^{T}\right) },
\end{equation}%
while $\chi _{\tau }\left( \left. I_{t}^{T},v_{T}\right\vert v_{t}\right) $
is the joint p.d.f. of $\left( I_{t}^{T},v_{T}\right) $ conditional on $%
v_{t} $. As usual, we can represent $\chi _{\tau }\left( \left.
I_{t}^{T},v_{T}\right\vert v_{t}\right) $ as follows%
\begin{equation}
\chi _{\tau }\left( \left. I_{t}^{T},v_{T}\right\vert v_{t}\right) =\chi
_{\tau }\left( \left. I_{t}^{T}\right\vert v_{t},v_{T}\right) \chi _{\tau
}\left( \left. v_{T}\right\vert v_{t}\right) ,  \label{cond_pdf}
\end{equation}%
where $\chi _{\tau }\left( \left. v_{T}\right\vert v_{t}\right) $ is the
p.d.f. of $v_{T}$ conditional on $v_{t}$, and $\chi _{\tau }\left( \left.
I_{t}^{T}\right\vert v_{t},v_{T}\right) $ is the p.d.f. of $I_{t}^{T}$
conditional on $\left( v_{t},v_{T}\right) $. Accordingly, we can rewrite Eq.
(\ref{Willard1}) as follows%
\begin{equation}
\begin{array}{lll}
C^{SV}(0,1,v;T,K) & = & \int_{0}^{\infty }\int_{0}^{\infty }C^{BS}(0,\tilde{E%
}_{t}^{T};T,K;\sqrt{\frac{\bar{\rho}^{2}I_{0}^{T}}{T}})\chi _{T}\left(
\left. I_{0}^{T}\right\vert v,v_{T}\right) \\ 
&  & \times \chi _{T}\left( \left. v_{T}\right\vert v\right)
dI_{0}^{T}dv_{T}.%
\end{array}
\label{Willard2}
\end{equation}%
Once again, formula (\ref{Willard2}) is too complex to be used in practice,
especially when compared to the Fourier transform based Lewis-Lipton formula 
\cite{lewis-1}, \cite{lewis-2}, \cite{lipton-2}. However, it can give some
useful hint on how to build an accurate (if not practical) MC simulation,
see Appendix \ref{Broadie}.

It is well-known, see, e.g., \cite{feller}, that $\chi _{\tau }\left( \left.
v_{T}\right\vert v_{t}\right) $ is the so-called non-central chi-square
distribution,%
\begin{equation}
\chi _{\tau }\left( \left. v_{T}\right\vert v_{t}\right) =e^{\frac{\kappa
\tau }{2}}\psi \left( \kappa ,\tau \right) \exp \left( -\psi \left( \kappa
,\tau \right) \left( \bar{v}_{t}+\bar{v}_{T}\right) \right) \left( \frac{%
\bar{v}_{T}}{\bar{v}_{t}}\right) ^{\frac{\vartheta }{2}}I_{\vartheta }\left(
2\psi \left( \kappa ,\tau \right) \sqrt{\bar{v}_{t}\bar{v}_{T}}\right) ,
\label{xi_vtvT}
\end{equation}%
where $\vartheta =2\kappa /\varepsilon ^{2}-1$, $I_{\vartheta }\left(
.\right) $ is the modified Bessel function, $\bar{v}_{t}=e^{-\frac{\kappa
\tau }{2}}v_{t}$, $\bar{v}_{T}=e^{\frac{\kappa \tau }{2}}v_{T}$, and%
\begin{equation}
\psi \left( \kappa ,\tau \right) =\frac{\kappa }{\varepsilon ^{2}\sinh
\left( \frac{\kappa \tau }{2}\right) }\underset{\tau \longrightarrow 0}{%
\longrightarrow }\frac{2}{\varepsilon ^{2}\tau }.  \label{psi_alpha_tau1}
\end{equation}%
At the same time, $\chi _{T}\left( \left. I_{0}^{T}\right\vert
v,v_{T}\right) $ cannot be written in closed form; see, however, Eq. (\ref%
{I_cond_dist}) below. The condition $\vartheta >0$, known as the Feller
condition, \cite{feller}, implies that the process $v_{t}$ can never reach
zero; when this condition is violated, the origin is accessible and strongly
reflecting. We shall see below that for realistic FX cases, the Feller
condition is typically violated, which causes numerical complications.

In view of the fact that $\psi \left( \kappa ,\tau \right) $ explodes when $%
\tau \longrightarrow 0$, it is exceedingly difficult to perform direct
sampling of the non-central chi-square random variable when $\tau
\longrightarrow 0$. While for vanilla pricing one large time step is
sufficient, for barrier options very small time-steps are needed, so the
above mentioned obstacle has to be overcome. We considered several MC
schemes, such as \cite{broadie}, \cite{kahl}, \cite{smith}, \cite{andersen1}%
, and concluded that the well-known Andersen Quadratic Exponential (QE)
scheme performs particularly well when time steps are small. We emphasize
that for small time steps it is not necessary to calculate $\chi _{\tau
}\left( \left. I_{t}^{T}\right\vert v_{t},v_{T}\right) $ since $I_{t}^{T}$
can be accurately approximated as follows%
\begin{equation}
I_{t}^{T}\approx \QTOVERD. . {1}{2}\left( v_{t}+v_{T}\right) \tau .
\label{ItT_approx}
\end{equation}%
Accordingly, we can approximate $\mu _{t},\sigma _{t}$ in Eq. (\ref{SDE_x})
as follows:%
\begin{equation}
\begin{array}{ll}
\mu _{t\in \left( t_{i},t_{i+1}\right] }= & \frac{\rho }{\varepsilon }\frac{%
v_{t_{i+1}}-v_{t_{i}}}{t_{i+1}-t_{i}}+\left( \frac{\rho \kappa }{\varepsilon 
}-\QTOVERD. . {1}{2}\right) \frac{\left( v_{t_{i}}+v_{t_{i+1}}\right) }{2}-%
\frac{\rho \kappa }{\varepsilon }, \\ 
\sigma _{t\in \left( t_{i},t_{i+1}\right] }= & \bar{\rho}\sqrt{\frac{\left(
v_{t_{i}}+v_{t_{i+1}}\right) }{2}}.%
\end{array}
\label{SDE_x_coef2}
\end{equation}%
This approximation can be used as a basis for developing a mixed PDE-MC
method, see \cite{loeper}, however, we were not able to obtain satisfactory
results via such a method.

\section{Pricing problem for call options\label{PricingCall}}

In this section we demonstrate how to use analytical and numerical methods
for solving the pricing problem for the call option. In Section \ref%
{CallAnal} we discuss its analytical solution for the standard Heston model
with arbitrary $\rho $; in other cases we present the analytical solution
for $\rho =0$. In Section \ref{CallNum} we describe how the transformed
Heston pricing problem can be solved numerically via the various methods
developed in Section \ref{NumSol}. In Section \ref{CallComp} we calibrate
our model to the market and use the corresponding parameters to calculate
call option prices explicitly both analytically and numerically. We conclude
that for call options numerical and analytical results agree well.

\subsection{Formulation\label{CallForm}}

Although the main topic of this paper is the efficient valuation of exotic
derivative in the LSV\ framework, it is clearly necessary to price vanilla
options first. For brevity, we concentrate on pricing calls. Puts can be
priced by put-call parity. As always, rather than pricing a call with
non-dimensional maturity $T$ and strike $K$, we price the so-called covered
call, whose payoff is of the form%
\begin{equation}
V\left( F\right) =\min \left\{ F,K\right\} ,  \label{cov_call}
\end{equation}%
and represent the price of a call as the difference between the spot and the
price of a covered call.

In Section \ref{LSVFormulation} we introduced four pricing equations (\ref%
{Heston_pde1}), (\ref{shifted_pde1}), (\ref{pde_im2}), (\ref{pde_re2}). We
need to augment each one of them with the corresponding terminal and
boundary conditions.

The corresponding initial condition for a call option has the form%
\begin{equation}
u^{s}\left( x_{1},x_{2}\right) =\left\{ 
\begin{array}{ll}
\begin{array}{ll}
e^{\QTOVERD. . {1}{2}x_{1}}, & x_{1}\in \left[ -\infty ,X_{K}^{H}\right] ,
\\ 
Ke^{-\QTOVERD. . {1}{2}x_{1}}, & x_{1}\in \left[ X_{K}^{H},\infty \right] ,%
\end{array}
& s=H, \\ 
\begin{array}{ll}
\frac{2\sqrt{1-\beta }}{\beta }\sinh \left( \QTOVERD. . {1}{2}\beta \left(
x_{1}-X_{0}^{DH}\right) \right) , & x_{1}\in \left[ X_{0}^{DH},X_{K}^{DH}%
\right] , \\ 
Ke^{-\QTOVERD. . {1}{2}\beta x_{1}}, & x_{1}\in \left[ X_{K}^{DH},\infty %
\right] ,%
\end{array}
& s=DH, \\ 
\begin{array}{ll}
\frac{\sqrt{\frac{\alpha }{2}\left( \mathsf{m}^{2}+\mathsf{n}^{2}\right) }}{%
\sqrt{\left\vert \omega ^{I}\right\vert }}\sin \left( \sqrt{\left\vert
\omega ^{I}\right\vert }\left( x_{1}-X_{0}^{I}\right) \right) , & x_{1}\in 
\left[ X_{0}^{I},X_{K}^{I}\right] , \\ 
\frac{\sqrt{\frac{\alpha }{2}}K}{\sqrt{\left\vert \omega ^{I}\right\vert }}%
\sin \left( \sqrt{\left\vert \omega ^{I}\right\vert }\left( X_{\infty
}^{I}-x_{1}\right) \right) , & x_{1}\in \left[ X_{K}^{I},X_{\infty }^{I}%
\right] ,%
\end{array}
& s=I, \\ 
\begin{array}{cc}
\frac{\sqrt{\frac{\alpha }{2}\mathsf{pq}}}{\sqrt{\omega ^{R}}}\sinh \left( 
\sqrt{\omega ^{R}}\left( x_{1}-X_{0}^{R}\right) \right) , & x_{1}\in \left[
X_{0}^{R},X_{K}^{R}\right] , \\ 
\frac{\sqrt{\frac{\alpha }{2}}K}{\sqrt{\omega ^{R}}}\sinh \left( \sqrt{%
\omega ^{R}}\left( X_{\infty }^{R}-x_{1}\right) \right) , & x_{1}\in \left[
X_{K}^{R},X_{\infty }^{R}\right] ,%
\end{array}
& s=R.%
\end{array}%
\right.  \label{call_payoff}
\end{equation}%
Here%
\begin{equation}
X_{K}^{s}=\left\{ 
\begin{array}{ll}
\ln \left( K\right) , & s=H, \\ 
\frac{1}{\beta }\ln \left( \beta \left( K-1\right) +1\right) , & s=DH, \\ 
\frac{1}{\sqrt{\left\vert \omega ^{I}\right\vert }}\left( \arctan \left( 
\frac{K-\mathsf{m}}{\mathsf{n}}\right) -\arctan \left( \frac{1-\mathsf{m}}{%
\mathsf{n}}\right) \right) , & s=I, \\ 
\frac{1}{2\sqrt{\omega ^{R}}}\ln \left( \frac{\left( 1-\mathsf{p}\right)
\left( K-\mathsf{q}\right) }{\left( 1-\mathsf{q}\right) \left( K-\mathsf{p}%
\right) }\right) , & s=R.%
\end{array}%
\right.  \label{call_XK}
\end{equation}%
It is clear that all the corresponding payoffs vanish at the boundaries $%
x=X_{0}^{s}$, $x=X_{\infty }^{s}$.

The boundary conditions in the $x_{1}$ direction are simple%
\begin{equation}
U^{s}\left( t,X_{0}^{s},x_{2}\right) =0,\ \ \ \ \ U^{s}\left( t,X_{\infty
}^{s},x_{2}\right) =0,  \label{call_bc}
\end{equation}%
where the equality is understood in the limiting sense when $\left\vert
X_{0,\infty }^{s}\right\vert =\infty $. The boundary conditions in the $%
x_{2} $ direction are naturally imposed.

\subsection{Analytical solution\label{CallAnal}}

In this section we consider possible (semi-)analytical solutions for the
pricing equations (\ref{Heston_pde1}), (\ref{shifted_pde1}), (\ref{pde_im2}%
), (\ref{pde_re2}), supplied with the initial condition (\ref{call_payoff}),
and boundary conditions (\ref{call_bc}).

\subsubsection{Heston model}

We start with the Heston model (\ref{Heston_pde1}), (\ref{call_payoff}), (%
\ref{call_bc}). It is well known that for this model the price of a covered
call can be represented in the form of a single Fourier integral via the
Lewis-Lipton formula, \cite{lewis-1}, \cite{lewis-2}, \cite{lipton-2}.
Additional information on computation of the corresponding Fourier integral
can be found in \cite{schmelzle}, \cite{janek}, and \cite{zeliade}.
Specifically, solution of the Heston problem can be written in the form%
\begin{equation}
U^{H}\left( \tau ,x_{1},x_{2}\right) =\frac{1}{2\pi }\int_{-\infty }^{\infty
}u^{H}\left( \tau ,k,x_{2}\right) \nu ^{H}\left( k\right) e^{ikX}dk,
\label{hest_int1}
\end{equation}%
where $\nu ^{H}\left( k\right) $ is the Fourier transform of $u^{H}\left(
x_{1}\right) $,%
\begin{equation}
\nu ^{H}\left( k\right) =\int_{-\infty }^{\infty }u^{H}\left( x_{1}\right)
e^{-ikx_{1}}dX=\frac{e^{-\left( ik-\QTOVERD. . {1}{2}\right) X_{K}^{H}}}{%
\lambda ^{H}\left( k\right) }=\frac{\sqrt{\sigma \left( K\right) }%
e^{-ikX_{K}^{H}}}{\lambda ^{H}\left( k\right) },  \label{hest_trans}
\end{equation}%
\begin{equation}
\lambda ^{H}\left( k\right) =k^{2}+\QTOVERD. . {1}{4},  \label{hest_lambda}
\end{equation}%
and $u^{H}\left( \tau ,k,x_{2}\right) $ satisfies the following equation%
\begin{equation}
\begin{array}{l}
\partial _{\tau }u^{H}\left( \tau ,k,x_{2}\right) +\QTOVERD. . {1}{2}\lambda
^{H}\left( k\right) x_{2}u^{H}\left( \tau ,k,x_{2}\right) \\ 
-\QTOVERD. . {1}{2}\varepsilon ^{2}x_{2}\partial _{x_{2}}^{2}u^{H}\left(
\tau ,k,x_{2}\right) -\left( \kappa -\left( \kappa -\rho \varepsilon \left(
ik+\QTOVERD. . {1}{2}\right) \right) x_{2}\right) \partial
_{x_{2}}u^{H}\left( \tau ,k,x_{2}\right) =0,%
\end{array}
\label{hest_PDE1}
\end{equation}%
the initial condition%
\begin{equation}
u^{H}\left( 0,k,x_{2}\right) =1,  \label{hest_ic}
\end{equation}%
and the regularity conditions in the $x_{2}$ direction, which are provided
by the equation itself. As usual, we can use the affine ansatz and write%
\begin{equation}
u^{H}\left( \tau ,k,x_{2}\right) =e^{\tilde{A}\left( \tau ,k\right) +\tilde{B%
}\left( \tau ,k\right) x_{2}},  \label{affine_ans}
\end{equation}%
\begin{equation}
\begin{array}{l}
\tilde{A}^{\prime }\left( \tau ,k\right) -\kappa \tilde{B}\left( \tau
,k\right) =0, \\ 
\tilde{B}^{\prime }\left( \tau ,k\right) -\QTOVERD. . {1}{2}\varepsilon ^{2}%
\tilde{B}^{2}\left( \tau ,k\right) +\left( \hat{\kappa}-\rho \varepsilon
ik\right) \tilde{B}\left( \tau ,k\right) +\QTOVERD. . {1}{2}\lambda
^{H}\left( k\right) =0,%
\end{array}
\label{Hest_ODEs}
\end{equation}%
\begin{equation}
\tilde{A}\left( 0,k\right) =0,\ \ \ \ \ \tilde{B}\left( 0,k\right) =0.
\label{Hest_bc}
\end{equation}%
The Riccati transform (\ref{Riccati}) yields the following equation%
\begin{equation}
\tilde{\gamma}^{\prime \prime }\left( \tau ,k\right) +\left( \hat{\kappa}%
-\rho \varepsilon ik\right) \tilde{\gamma}^{\prime }\left( \tau ,k\right)
-\QTOVERD. . {1}{4}\varepsilon ^{2}\lambda ^{H}\left( k\right) \tilde{\gamma}%
\left( \tau ,k\right) =0,  \label{linear_ODE1}
\end{equation}%
supplied with the initial conditions%
\begin{equation}
\tilde{\gamma}\left( 0,k\right) =1,\ \ \ \ \ \tilde{\gamma}^{\prime }\left(
0,k\right) =0.  \label{Riccati_bc}
\end{equation}%
Two linearly independent solutions are%
\begin{equation}
\tilde{\gamma}_{\pm }\left( \tau ,k\right) =e^{\pm \Xi _{\pm }\left(
k\right) \tau /2},  \label{gamma_sol1}
\end{equation}%
where%
\begin{equation}
\begin{array}{l}
\Xi _{\pm }\left( k\right) =\mp \left( \hat{\kappa}-\rho \varepsilon
ik\right) +\varpi \left( k\right) , \\ 
\varpi \left( k\right) =\sqrt{\left( \hat{\kappa}-\rho \varepsilon ik\right)
^{2}+\varepsilon ^{2}\lambda ^{H}\left( k\right) },%
\end{array}
\label{hest_params_gen}
\end{equation}%
so that $\tilde{\gamma}$ can be written in the form%
\begin{equation}
\tilde{\gamma}\left( \tau ,k\right) =\frac{e^{\Xi _{+}\left( k\right) \tau
/2}\Upsilon \left( \tau ,k\right) }{2\varpi \left( k\right) },
\label{gamma_sol2}
\end{equation}%
where%
\begin{equation}
\Upsilon \left( \tau ,k\right) =\Xi _{-}\left( k\right) +\Xi _{+}\left(
k\right) e^{-\varpi \left( k\right) \tau },
\end{equation}%
Accordingly,%
\begin{equation}
\begin{array}{l}
\tilde{A}\left( \tau ,k\right) =-\frac{\kappa }{\varepsilon ^{2}}\left( \Xi
_{+}\left( k\right) \tau +2\ln \left( \frac{\Upsilon \left( \tau ,k\right) }{%
2\varpi \left( k\right) }\right) \right) , \\ 
\tilde{B}\left( \tau ,k\right) =-\frac{\left( 1-e^{-\varpi \left( k\right)
\tau }\right) \lambda ^{H}\left( k\right) }{\Upsilon \left( \tau ,k\right) },%
\end{array}
\label{hest_A_B_gen}
\end{equation}%
Thus,%
\begin{equation}
U^{H}\left( \tau ,x_{1},x_{2}\right) =\frac{\sqrt{\sigma \left( K\right) }}{%
2\pi }\int_{-\infty }^{\infty }\frac{e^{\tilde{A}\left( \tau ,k\right) +%
\tilde{B}\left( \tau ,k\right) x_{2}+ik\left( x_{1}-X_{K}^{H}\right) }}{%
\lambda ^{H}\left( k\right) }dk.  \label{hest_int2}
\end{equation}%
In particular, when $\rho =0$ we have%
\begin{equation}
\Xi _{\pm }\left( k\right) =\mp \kappa +\varpi \left( k\right) ,\ \ \ \varpi
\left( k\right) =\sqrt{\kappa ^{2}+\varepsilon ^{2}\lambda ^{H}\left(
k\right) }.  \label{hest_A_B_0}
\end{equation}%
It is clear that all the relevant functions are even functions of $k$, so
that we can rewrite Eq. (\ref{hest_int2}) in the form%
\begin{equation}
\begin{array}{ll}
U^{H}\left( \tau ,x_{1},x_{2}\right) & =\frac{\sqrt{\sigma \left( K\right) }%
}{\pi }\int_{0}^{\infty }\frac{u^{H}\left( \tau ,k,x_{2}\right) }{\lambda
^{H}\left( k\right) }\cos \left( k\left( x_{1}-X_{K}^{H}\right) \right) dk
\\ 
& =\frac{\sqrt{\sigma \left( K\right) }}{\pi }\int_{0}^{\infty }\frac{%
u^{H}\left( \tau ,k,x_{2}\right) }{\lambda ^{H}\left( k\right) }\cos \left(
k\ln \left( \frac{F}{K}\right) \right) dk.%
\end{array}
\label{hest_int3}
\end{equation}

\subsubsection{Displaced Heston model\label{CallDH}}

Rather disappointingly, it is possible to find the price of a covered call
in a displaced Heston model only when $\rho =0$, \cite{lipton-2}. In
principle, it is possible to argue that one can assume that $\rho =0$ and
choose the scaling parameter $\beta $ in order to mimic the effects of
nonzero $\rho $.

For $\rho =0$ the solution of the displaced Heston problem can be written in
the form%
\begin{equation}
U^{DH}\left( \tau ,x_{1},x_{2}\right) =\frac{2}{\pi }\int_{0}^{\infty
}u^{DH}\left( \tau ,k,x_{2}\right) \nu ^{DH}\left( k\right) \sin \left(
k\left( x_{1}-X_{0}^{DH}\right) \right) dk,  \label{shifted_int1}
\end{equation}%
where $\nu ^{DH}\left( k\right) $ is the sine Fourier transform of $%
u^{DH}\left( x_{1}\right) $. Equation (\ref{A3}) of Appendix \ref{Fourier}
shows that%
\begin{equation}
\begin{array}{ll}
\nu ^{DH}\left( k\right) & =\int_{X_{0}^{DH}}^{\infty }u^{DH}\left(
x_{1}\right) \sin \left( k\left( x_{1}-X_{0}^{DH}\right) \right) dx_{1} \\ 
& =\frac{\sqrt{\sigma \left( K\right) }\sin \left( k\left(
X_{K}^{DH}-X_{0}^{DH}\right) \right) }{\lambda ^{DH}\left( k\right) },%
\end{array}
\label{shifted_transform}
\end{equation}%
and $u^{DH}\left( \tau ,k,x_{2}\right) $ has the form (\ref{affine_ans}),
with $\tilde{A}\left( \tau ,k\right) ,\tilde{B}\left( \tau ,k\right) $ given
by equations (\ref{hest_A_B_0}) with $\lambda ^{H}\left( k\right) $ replaced
by $\lambda ^{DH}\left( k\right) $ defined as follows%
\begin{equation}
\lambda ^{DH}\left( k\right) =k^{2}+\QTOVERD. . {1}{4}\beta ^{2}.
\label{shifted_lambda}
\end{equation}%
Accordingly,%
\begin{equation}
\begin{array}{lll}
U^{DH}\left( \tau ,x_{1},x_{2}\right) & = & \frac{2\sqrt{\sigma \left(
K\right) }}{\pi } \\ 
&  & \times \int_{0}^{\infty }\frac{u^{DH}\left( \tau ,k,x_{2}\right) }{%
\lambda ^{DH}\left( k\right) }\sin \left( k\left(
X_{K}^{DH}-X_{0}^{DH}\right) \right) \sin \left( k\left(
x_{1}-X_{0}^{DH}\right) \right) dk.%
\end{array}
\label{shifted_int2}
\end{equation}

\subsubsection{QLSV model\label{CallQSV}}

As before, we can only solve the problem (semi-) analytically when $\rho =0$%
, \cite{lipton-2}. In case (A), $s=I$,\ the solution of the corresponding
problem has the form%
\begin{equation}
U^{I}\left( \tau ,x_{1},x_{2}\right) =\sum\limits_{k=1}^{\infty
}u_{k}^{I}\left( \tau ,x_{2}\right) \nu _{k}^{I}\sin \left( \zeta
_{k}^{I}\left( x_{1}-X_{0}^{I}\right) \right) ,  \label{quad_sum1}
\end{equation}%
where 
\begin{equation}
\zeta _{k}^{I}=\frac{\pi k}{\Delta ^{I}}=\frac{\sqrt{\left\vert \omega
^{I}\right\vert }\pi k}{\frac{\pi }{2}+\arctan \left( \frac{\mathsf{m}}{%
\mathsf{n}}\right) }>\sqrt{\left\vert \omega ^{I}\right\vert }k,  \label{k1}
\end{equation}%
and $\nu _{k}^{I}$ are the corresponding Fourier coefficients%
\begin{equation}
\nu _{k}^{I}=\frac{2}{\Delta ^{I}}\int_{X_{0}^{I}}^{X_{\infty
}^{I}}u^{I}\left( x_{1}\right) \sin \left( \zeta _{k}^{I}\left(
x_{1}-X_{0}^{I}\right) \right) dx_{1}.  \label{quad_phi1}
\end{equation}%
The corresponding solution has the form (\ref{affine_ans}), (\ref{hest_A_B_0}%
) with $\lambda ^{H}\left( k\right) $ replaced by%
\begin{equation}
\lambda _{k}^{I}=\left( \zeta _{k}^{I}\right) ^{2}+\omega ^{I}.
\label{lambda1}
\end{equation}%
We note that $u_{k}$ are decaying functions of $\tau $. Equation (\ref{A6})
of Appendix \ref{Fourier} shows that%
\begin{equation}
\nu _{k}^{I}=\frac{2\sqrt{\sigma \left( K\right) }\sin \left( \zeta
_{k}^{I}\left( X_{K}^{I}-X_{0}^{I}\right) \right) }{\Delta ^{I}\lambda
_{k}^{I}}.  \label{quad_phi2}
\end{equation}%
Thus%
\begin{equation}
\begin{array}{lll}
U^{I}\left( \tau ,x_{1},x_{2}\right) & = & \frac{2\sqrt{\sigma \left(
K\right) }}{\Delta ^{I}} \\ 
&  & \times \sum\limits_{k=1}^{\infty }\frac{u_{k}^{I}\left( \tau
,x_{2}\right) }{\lambda _{k}^{I}}\sin \left( \zeta _{k}^{I}\left(
X_{K}^{I}-X_{0}^{I}\right) \right) \sin \left( \zeta _{k}^{I}\left(
x_{1}-X_{0}^{I}\right) \right) .%
\end{array}
\label{quad_sum2}
\end{equation}%
In case (B), $s=R$, the solution of the corresponding problem has the form%
\begin{equation}
U^{R}\left( \tau ,x_{1},x_{2}\right) =\sum\limits_{k=1}^{\infty
}u_{k}^{R}\left( \tau ,x_{2}\right) \nu _{k}^{R}\sin \left( \zeta _{k}\left(
x_{1}-X_{0}^{R}\right) \right) ,  \label{quad_sum3}
\end{equation}%
where 
\begin{equation}
\zeta _{k}^{R}=\frac{\pi l}{\Delta ^{R}}=\frac{2\sqrt{\omega ^{R}}\pi l}{\ln
\left( \frac{\mathsf{p}}{\mathsf{q}}\right) }>0,  \label{k2}
\end{equation}%
$\phi _{k}^{R}$ are the corresponding Fourier coefficients%
\begin{equation}
\nu _{k}^{R}=\frac{2}{\Delta ^{R}}\int_{X_{0}^{R}}^{X_{\infty
}^{R}}u^{R}\left( x_{1}\right) \sin \left( \zeta _{k}^{R}\left(
x_{1}-X_{0}^{R}\right) \right) dx_{1},  \label{quad_phi13}
\end{equation}%
and $u_{k}^{R}\left( \tau ,x_{2}\right) $ is given by(\ref{affine_ans}), (%
\ref{hest_A_B_0}) with $\lambda ^{H}\left( k\right) $ replaced by%
\begin{equation}
\lambda _{k}^{R}=\left( \zeta _{k}^{R}\right) ^{2}+\omega ^{R}.
\label{lambda2}
\end{equation}%
Equation (\ref{A4}) of Appendix \ref{Fourier} shows that%
\begin{equation}
\nu _{k}^{R}=\frac{2\sqrt{\sigma \left( K\right) }\sin \left( \zeta
_{k}^{R}\left( X_{K}^{R}-X_{0}^{R}\right) \right) }{\Delta ^{R}\lambda
_{k}^{R}},  \label{quad_phi4}
\end{equation}%
so that%
\begin{equation}
\begin{array}{lll}
U^{R}\left( \tau ,x_{1},x_{2}\right) & = & \frac{2\sqrt{\sigma \left(
K\right) }}{\Delta ^{R}} \\ 
&  & \times \sum\limits_{k=1}^{\infty }\frac{u_{k}^{R}\left( \tau
,x_{2}\right) }{\lambda _{k}^{R}}\sin \left( \zeta _{k}^{R}\left(
X_{K}^{R}-X_{0}^{R}\right) \right) \sin \left( \zeta _{k}^{R}\left(
x_{1}-X_{0}^{R}\right) \right) .%
\end{array}
\label{quad_sum4}
\end{equation}%
To summarize%
\begin{equation}
\begin{array}{lll}
U^{s}\left( \tau ,x_{1},x_{2}\right) & = & \frac{2\sqrt{\sigma \left(
K\right) }}{\Delta ^{s}} \\ 
&  & \times \sum\limits_{k=1}^{\infty }\frac{u_{k}^{s}\left( \tau
,x_{2}\right) }{\lambda _{k}^{s}}\sin \left( \zeta _{k}^{s}\left(
X_{K}^{s}-X_{0}^{s}\right) \right) \sin \left( \zeta _{k}^{s}\left(
x_{1}-X_{0}^{s}\right) \right) ,%
\end{array}
\label{quad_sum5}
\end{equation}%
$s=I,R.$

It is worth noting that integral (\ref{shifted_int2}) can be approximated by
a discrete infinite sum over an equidistant grid $\zeta _{k}=\pi k/\Delta $,
where $\Delta $ is an appropriately chosen discretization parameter, so that 
$U^{DH}\left( \tau ,x_{1},x_{2}\right) $ can be approximately written in the
form (\ref{quad_sum5}) as well. Similar, but much more complex formulas can
be found in \cite{andersen3}.

\subsection{Numerical solution\label{CallNum}}

For brevity, in this subsection we will restrict ourselves to the standard
Heston model governed by Eqs (\ref{Heston_pde1}), (\ref{call_payoff}), (\ref%
{call_bc}). Other cases can be analyzed along similar lines.

Application of ADI methods to the problem at hand is straightforward. All we
need to do is to specify the computational domain $-L_{1}<x_{1}<L_{1}$, $%
0\leq x_{2}<L_{2}$, in the $\left( x_{1},x_{2}\right) $-plane and define the
corresponding one-dimensional grids. We are prepared to trade speed for
accuracy in our calculations. Accordingly, we choose dense grids which are
uniform with respect to $x_{1}$ and $\sqrt{x_{2}}$, respectively.

Given the fact that for the Heston model the pricing problem is defined on
the entire axis $-\infty <x_{1}<\infty $, it is not natural (but not
impossible) to use the Galerkin method to solve it. To do so, one would need
to artificially cut the domain, assume that $-L_{1}<x_{1}<L_{1}$, and impose
zero boundary conditions at $x_{1}=\pm L_{1}$.\footnote{%
A\ choice of basis functions $e_{k}\left( x\right) =\cos \left( \zeta
_{k}x\right) $, where $\zeta _{k}=\cot \left( \zeta _{k}L\right) $, produces
better results. We leave it to the interested reader to pursue.} We do not
pursue this avenue of research here, and postpone the development of the
Galerkin method until the next section, where we use it to price
double-no-touch options with impressive efficacy.

We need to solve Eqs (\ref{expansion_0}), (\ref{expansion_1}), (\ref%
{expansion_2}), with $s=H$. Since the domain covers the entire axis, in this
case we have 
\begin{equation}
e_{k}\left( x_{1}\right) =e^{ikx_{1}}.  \label{e_k1}
\end{equation}%
It is easy to see that%
\begin{equation}
\lambda _{k}=\QTOVERD. . {1}{2}\left( k^{2}+\QTOVERD. . {1}{4}\right) ,\ \ \
\mu _{k,k^{\prime }}=\left( ik+\QTOVERD. . {1}{2}\right) \delta \left(
k-k^{\prime }\right) .  \label{lambda_mu}
\end{equation}%
For simplicity, we choose a single mode initial condition%
\begin{equation}
u\left( x_{1}\right) =e^{i\varkappa x_{1}},  \label{heston_ic_fourier}
\end{equation}%
so that%
\begin{equation}
\nu _{k}=\delta \left( k-\varkappa \right) .  \label{nu_k}
\end{equation}%
As we know, the actual boundary condition can decomposed into individual
modes.

By using Eq. (\ref{W0}), it is straightforward to see that%
\begin{equation}
\overrightarrow{W}^{\left( 0\right) }\left( \boldsymbol{\tau },x_{2}\right)
=e^{A\left( 0,1\right) +B\left( 0,1\right) x_{2}}\overrightarrow{e}%
_{k}=e^{A\left( 0,1\right) +B\left( 0,1\right) x_{2}+i\varkappa x_{1}},
\label{W0_1}
\end{equation}%
and, in general,%
\begin{equation}
\overrightarrow{W}^{\left( n\right) }\left( \boldsymbol{\tau },x_{2}\right)
=\left( i\varkappa +\QTOVERD. . {1}{2}\right) ^{n}\left(
\sum_{m=0}^{n}C_{n,m}x_{2}^{m}\right) e^{\boldsymbol{A}\left( 0,n+1\right)
+B\left( n,n+1\right) x_{2}+i\varkappa x_{1}},  \label{Wn_1}
\end{equation}%
where ALL $\lambda _{k_{i}}$ are the same, $\lambda _{k_{i}}=\lambda
_{\varkappa }$, $i=1,...,n+1$. Thus, what we need to check is that 
\begin{equation}
\begin{array}{l}
e^{\tilde{A}\left( \tau ,\varkappa \right) +\tilde{B}\left( \tau ,\varkappa
\right) x_{2}} \\ 
=e^{\boldsymbol{A}\left( 0,1\right) +B\left( 0,1\right) x_{2}} \\ 
+\left( \rho \varepsilon \left( i\varkappa +\QTOVERD. . {1}{2}\right)
\right) \int_{0}^{\tau }d\tau _{1}\left(
\sum_{m=0}^{1}C_{n,m}x_{2}^{m}\right) e^{\boldsymbol{A}\left( 0,2\right)
+B\left( 1,2\right) x_{2}} \\ 
+\left( \rho \varepsilon \left( i\varkappa +\QTOVERD. . {1}{2}\right)
\right) ^{2}\int_{0}^{\tau }\int_{\tau _{1}}^{\tau }d\tau _{1}d\tau
_{2}\left( \sum_{m=0}^{2}C_{n,m}x_{2}^{m}\right) e^{\boldsymbol{A}\left(
0,3\right) +B\left( 2,3\right) x_{2}} \\ 
+\left( \rho \varepsilon \left( i\varkappa +\QTOVERD. . {1}{2}\right)
\right) ^{3}\int_{0}^{\tau }\int_{\tau _{1}}^{\tau }\int_{\tau _{2}}^{\tau
}d\tau _{1}d\tau _{2}d\tau _{3}\left( \sum_{m=0}^{2}C_{n,m}x_{2}^{m}\right)
e^{\boldsymbol{A}\left( 0,4\right) +B\left( 3,4\right) x_{2}} \\ 
+...\ ,%
\end{array}
\label{Hest_tst0}
\end{equation}%
where $\tilde{A}\left( \tau ,\varkappa \right) $, $\tilde{B}\left( \tau
,\varkappa \right) $ are given by Equation (\ref{hest_A_B_gen}).
Equivalently, we can check that%
\begin{equation}
\begin{array}{lll}
\left. e^{\tilde{A}\left( \tau ,\varkappa \right) +\tilde{B}\left( \tau
,\varkappa \right) x_{2}}\right\vert _{\rho =0} & = & e^{\boldsymbol{A}%
\left( 0,1\right) +B\left( 0,1\right) x_{2}}, \\ 
\left. \frac{\partial e^{\tilde{A}\left( \tau ,\varkappa \right) +\tilde{B}%
\left( \tau ,\varkappa \right) x_{2}}}{\partial \rho }\right\vert _{\rho =0}
& = & \left( \varepsilon \left( i\varkappa +\QTOVERD. . {1}{2}\right)
\right) \int_{0}^{\tau }d\tau _{1}\left(
\sum_{m=0}^{1}C_{n,m}x_{2}^{m}\right) \\ 
&  & \times e^{\boldsymbol{A}\left( 0,2\right) +B\left( 1,2\right) x_{2}},
\\ 
\frac{1}{2!}\left. \frac{\partial ^{2}e^{\tilde{A}\left( \tau ,\varkappa
\right) +\tilde{B}\left( \tau ,\varkappa \right) x_{2}}}{\partial \rho ^{2}}%
\right\vert _{\rho =0} & = & \left( \varepsilon \left( i\varkappa +\QTOVERD.
. {1}{2}\right) \right) ^{2}\int_{0}^{\tau }\int_{\tau _{1}}^{\tau }d\tau
_{1}d\tau _{2}\left( \sum_{m=0}^{2}C_{n,m}x_{2}^{m}\right) \\ 
&  & \times e^{\boldsymbol{A}\left( 0,3\right) +B\left( 2,3\right) x_{2}},
\\ 
\frac{1}{3!}\left. \frac{\partial ^{3}e^{\tilde{A}\left( \tau ,\varkappa
\right) +\tilde{B}\left( \tau ,\varkappa \right) x_{2}}}{\partial \rho ^{3}}%
\right\vert _{\rho =0} & = & \left( \varepsilon \left( i\varkappa +\QTOVERD.
. {1}{2}\right) \right) ^{3}\int_{0}^{\tau }\int_{\tau _{1}}^{\tau
}\int_{\tau _{2}}^{\tau }d\tau _{1}d\tau _{2}d\tau _{3}\left(
\sum_{m=0}^{2}C_{n,m}x_{2}^{m}\right) \\ 
&  & \times e^{\boldsymbol{A}\left( 0,4\right) +B\left( 3,4\right) x_{2}},%
\end{array}
\label{Hest_tst1}
\end{equation}%
etc. Eqs (\ref{Hest_tst1}) can be checked numerically.

Application of the MC method to the pricing of call options is
straightforward and is performed along the lines outlined in Section \ref%
{MonteCarlo}.

\subsection{Comparison of analytical and numerical solutions for the call
problem\label{CallComp}}

In order to perform a comparison of analytical and numerical solutions, we
have to choose a concrete set of the relevant parameters. To this end, we
calibrate the Heston model to the set of market data used to produce Figure %
\ref{fig:AUDJPY}. Since we restrict ourselves to time-independent
parameters, we cannot match all market prices simultaneously. Rather then
performing calibration in the least-squares error sense, we choose one
representative maturity, say $T=1y$, and calibrate the model to the selected
market prices. The corresponding dimensional parameters are 
\begin{equation}
\kappa =2.580,\ \ \ \theta =0.043,\ \ \ \varepsilon =1.000,\ \ \ \rho
=-0.360,\ \ \ v=0.114,
\end{equation}%
and their non-dimensional counterparts are 
\begin{equation}
\bar{\kappa}=59.758,\ \ \ \bar{\varepsilon}=23.162,\ \ \ x_{2}=2.628.
\end{equation}%
We emphasize that $\vartheta =2\kappa \theta /\varepsilon ^{2}-1=2\bar{\kappa%
}/\bar{\varepsilon}^{2}-1=-0.7772$, so that the Feller condition is clearly
violated, as is usually the case in practice.

\begin{figure}[h]
\subfigure[$I_{1}\rightarrow \infty$] {\includegraphics[width=0.55\textwidth, angle=0]
{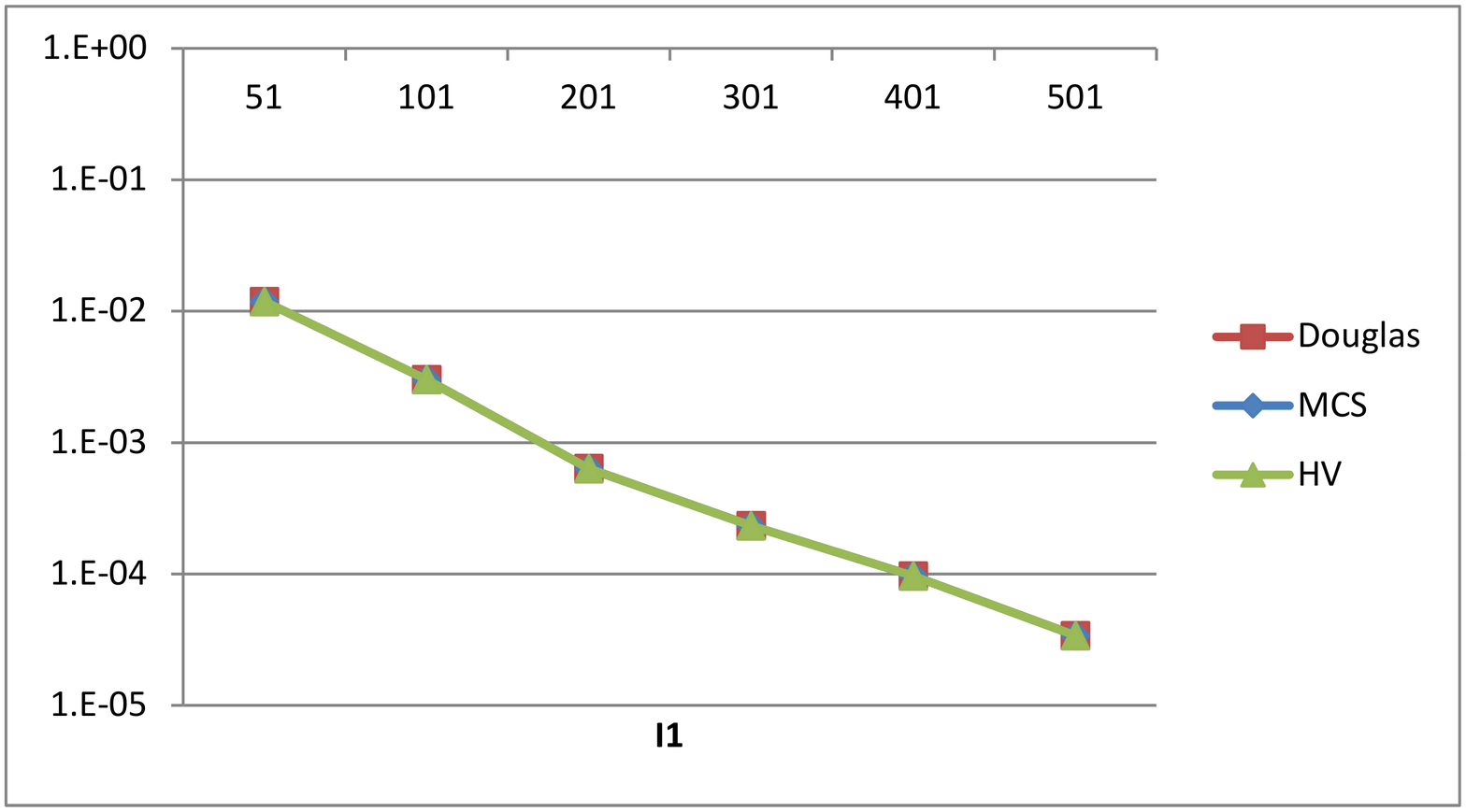}} 
\subfigure[$I_{2}\rightarrow \infty$] {\includegraphics[width=0.55\textwidth, angle=0]
{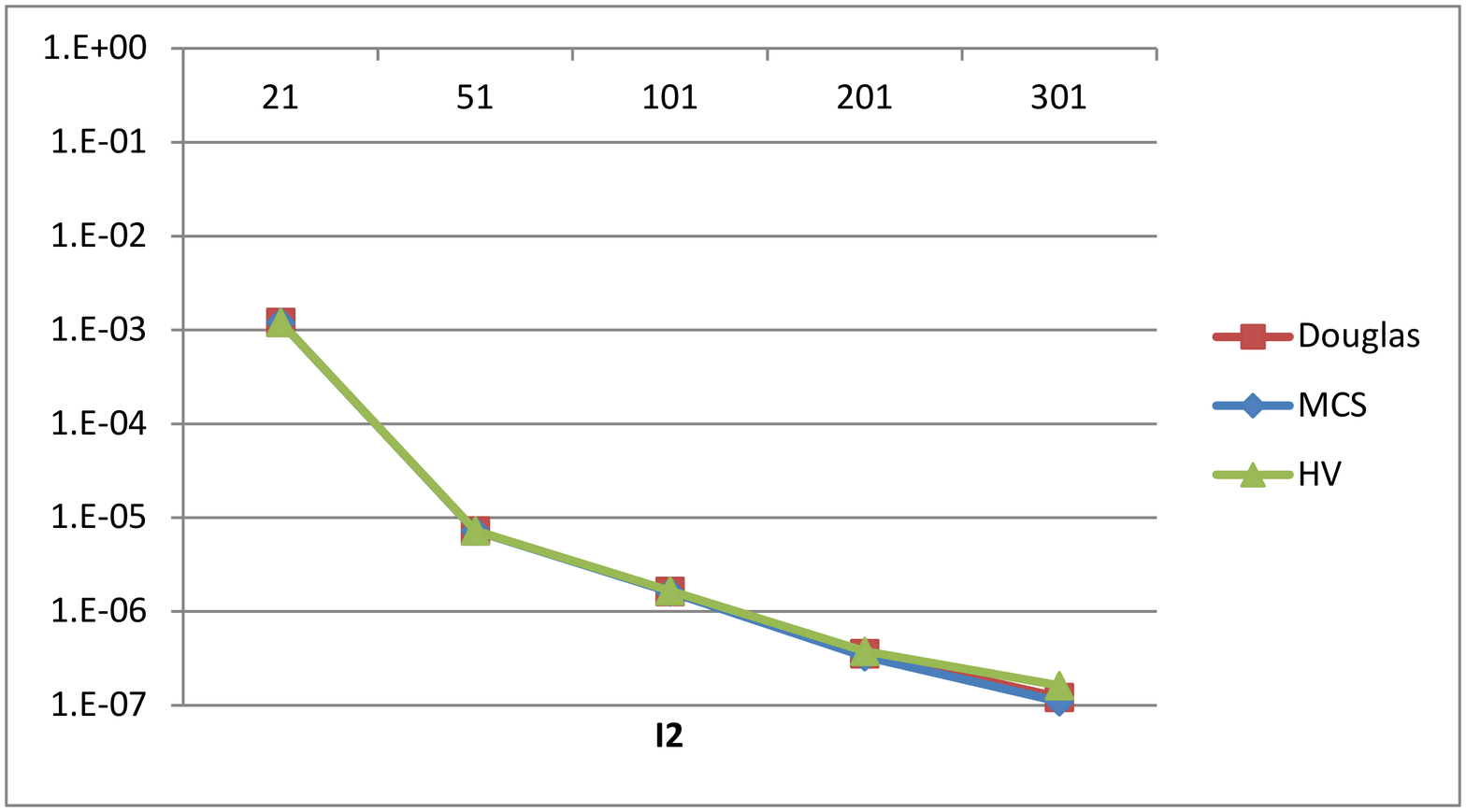}} 
\subfigure[$N\rightarrow \infty$] {\includegraphics[width=0.55\textwidth, angle=0]
{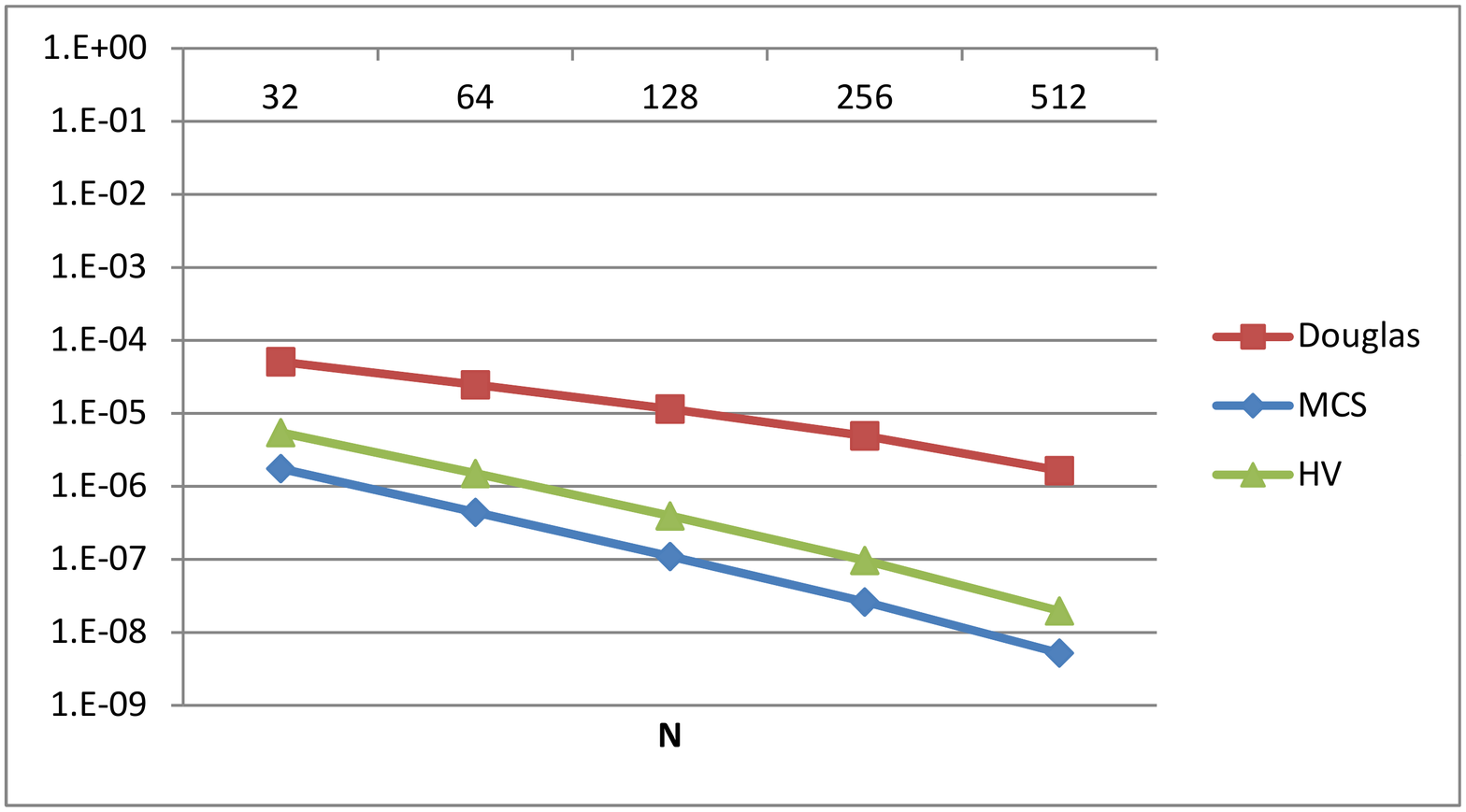}}
\caption{Convergence of the finite difference schemes for the ATM European
call pricing problem with $\protect\tau =1Y$. The default settings are as
follows: $N=1024$, $I_{1}=201$, $I_{2}=101$, $-5\leq x_{1}\leq 5$, $0\leq
x_{2}\leq 10$. In Figure (a) we vary $I_{1}$ from $51$ to $601$, while
keeping all other parameters fixed; we assume that the value corresponding
to $I_{1}=601$ is "exact". Similarly, in Figure (b) we vary $I_{2}$ from $21$
to $401$. Finally, in Figure (c) we vary $N$ from $32$ to $1024$. In the $%
x_{1}$-direction we use a uniform grid; in the $x_{2}$-direction we use a
grid which is uniform with respect to $\protect\sqrt{x_{2}}$.}
\label{fig:conv_eur}
\end{figure}

We use these parameters and compute the price of a call option via the ADI
methods discussed earlier. In Figure \ref{fig:conv_eur} we show the
convergence of these methods as a function of the number of steps in space
and time. It is clear that all the ADI method discussed in the paper
converge quadratically in space. The Do method converges linearly in time,
while the predictor-corrector methods \textit{a la} CS converge
quadratically in time.

\begin{equation*}
\text{Fig \ref{fig:conv_eur} near here}
\end{equation*}

\begin{figure}[ph]
\subfigure[] {\includegraphics[width=0.9\textwidth, angle=0]
{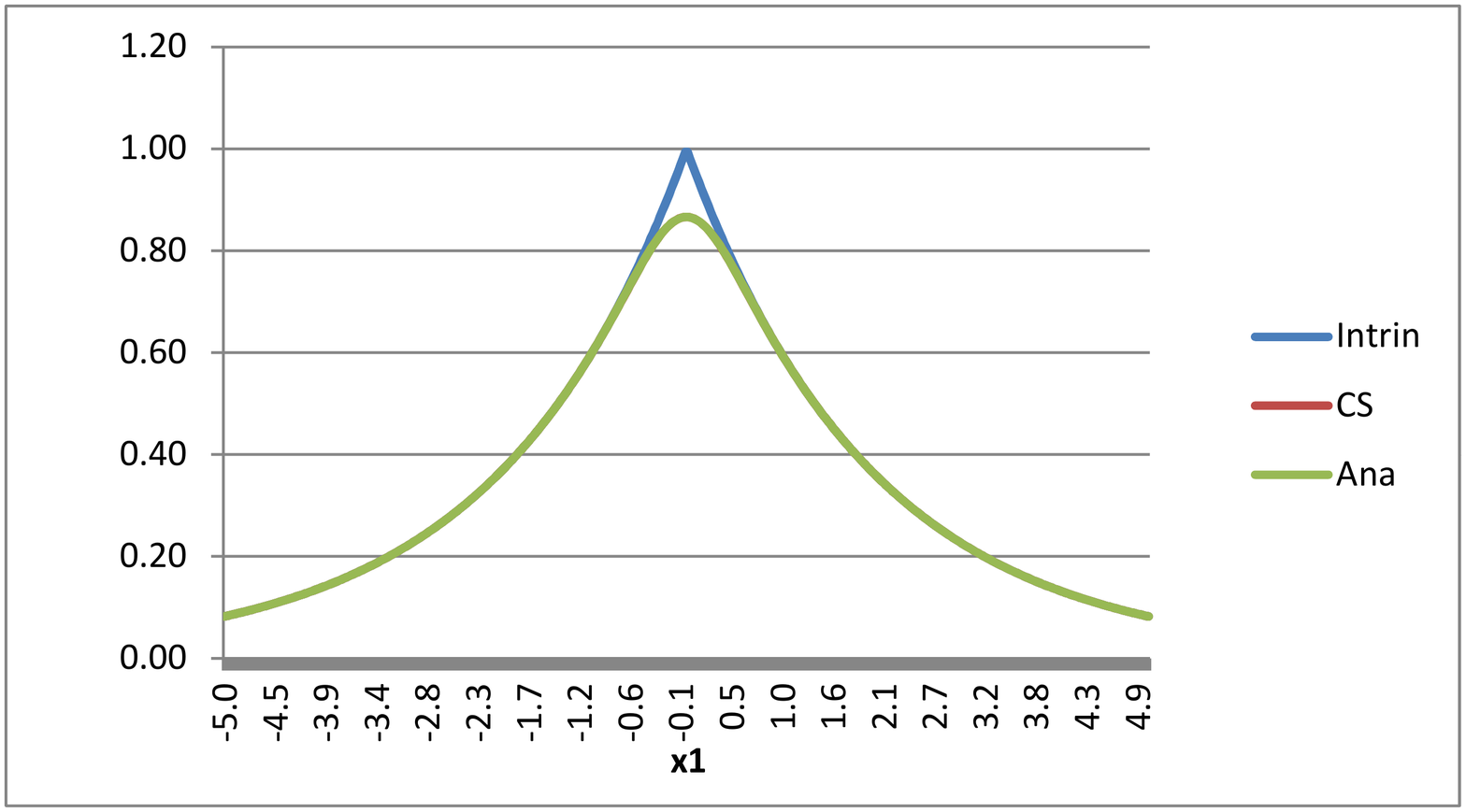}} 
\subfigure[] {\includegraphics[width=0.9\textwidth, angle=0]
{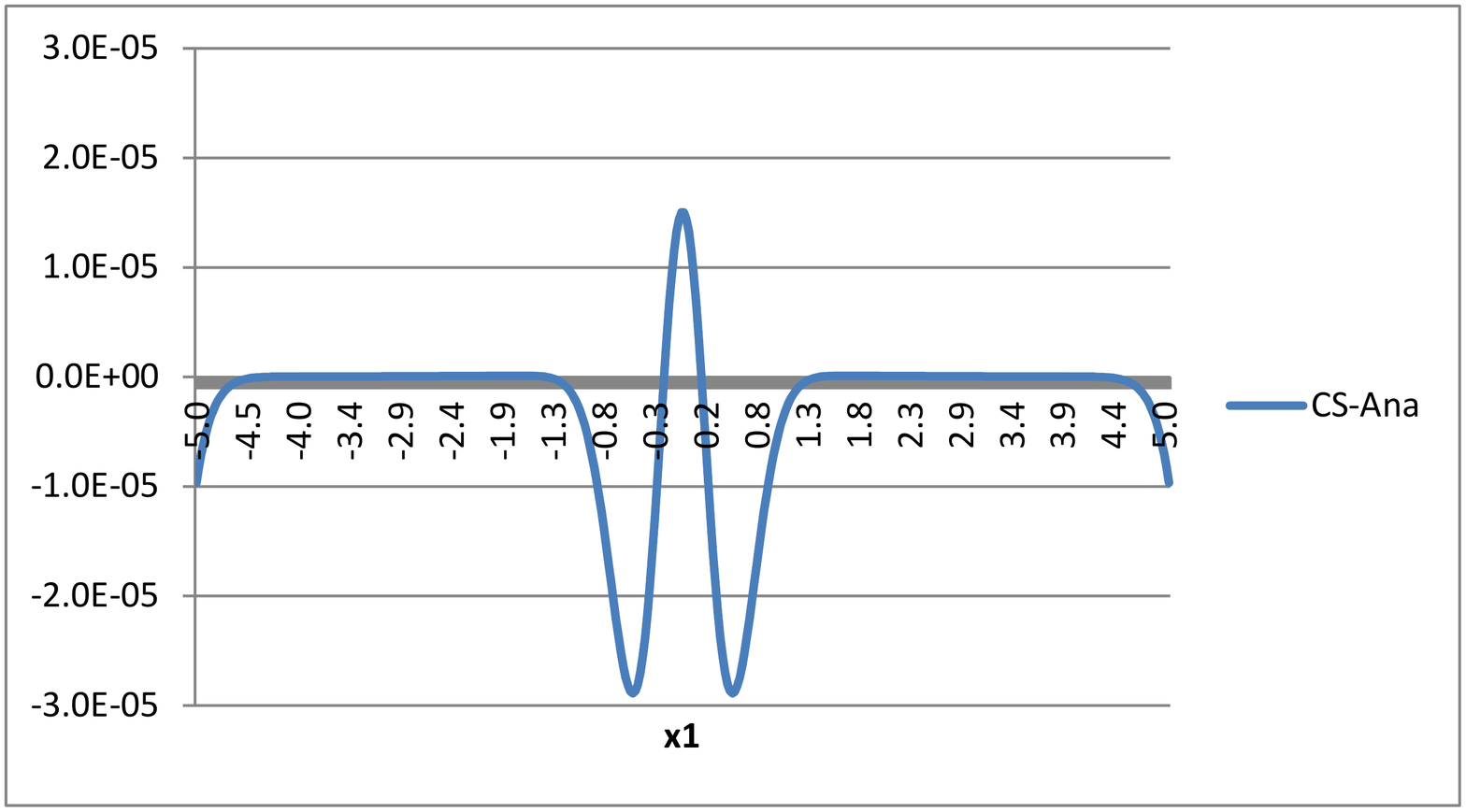}}
\caption{The profiles $U^{H}\left( \protect\tau ,x_{1},x_{2}\right) $ for
the call option obtained via the CS method and the Lewis-Lipton formula for $%
-5\leq x_{1}\leq 5$, $x_{2}=2.628$. In Figure (a) the intrinsic payoff of
the form $\exp \left( -\QTOVERD. . {1}{2}\left\vert x_{1}\right\vert \right) 
$ as well as the corresponding profiles are shown; for all practical
purposes these profiles overlap. In Figure (b) the difference between the
numerical and semi-analytical solutions is shown.}
\label{fig:eur_sol_vs_spot}
\end{figure}

In Figure \ref{fig:eur_sol_vs_spot} we show a snapshot of the price as a
function of $x_{1}$ with fixed $x_{2}=2.628$. It is clear that all the
numerical methods agree among themselves and converge to the semi-analytical
solution obtained via the Lewis-Lipton formula.

\begin{equation*}
\text{Fig \ref{fig:eur_sol_vs_spot} near here}
\end{equation*}

\section{Pricing problem for\ double no-touch options\label{DNT}}

DNTs are of particular interest for us. In this section, which is key to the
paper, we wish to compare various analytical and numerical methods for
solving the corresponding pricing problem. In Section \ref{DNT_Form} we
formulate the Liouville transformed pricing problem. In Section \ref%
{DNT_Anal} we solve this problem analytically for $\rho =0$. In Section \ref%
{DNT_num} we solve the Heston pricing problem numerically by using various
methods discussed in Section \ref{NumSol}. In Section \ref{DNT_Comp} we
compare solutions obtained by these methods and demonstrate that results
obtained by different numerical methods generally agree with each other very
well.

\subsection{Formulation\label{DNT_Form}}

So far, we have considered vanilla calls. Let us now study pricing of DNT
options paying a unit of currency at time $T$ provided that%
\begin{equation}
F_{L}<F_{t}<F_{U},\ \ \ 0\leq t\leq T,  \label{dnt_dom1}
\end{equation}%
and zero otherwise. There are other variations of the same basic product,
but, for the sake of brevity, we consider just this one. Clearly, very
little needs to be done to adapt our earlier findings to the problem at
hand. The interval of interest now becomes%
\begin{equation}
X_{L}^{s}<x_{1}<X_{U}^{s},  \label{dnt_dom2}
\end{equation}%
where $s=H,DH,I,R$. Depending on $\sigma \left( F\right) $ we have%
\begin{equation}
X_{\left\{ L,U\right\} }^{s}=\left\{ 
\begin{array}{ll}
\ln \left( F_{\left\{ L,U\right\} }\right) , & s=H, \\ 
\frac{1}{\beta }\ln \left( \beta \left( F_{\left\{ L,U\right\} }-1\right)
+1\right) , & s=DH, \\ 
\frac{1}{\sqrt{\left\vert \omega ^{I}\right\vert }}\left( \arctan \left( 
\frac{F_{\left\{ L,U\right\} }-\mathsf{m}}{\mathsf{n}}\right) -\arctan
\left( \frac{1-\mathsf{m}}{\mathsf{n}}\right) \right) , & s=I, \\ 
\frac{1}{2\sqrt{\omega ^{R}}}\ln \left( \frac{\left( 1-\mathsf{p}\right)
\left( F_{\left\{ L,U\right\} }-\mathsf{q}\right) }{\left( 1-\mathsf{q}%
\right) \left( F_{\left\{ L,U\right\} }-\mathsf{p}\right) }\right) . & s=R.%
\end{array}%
\right.  \label{dnt_dom3}
\end{equation}%
In all four cases the boundary conditions are clear%
\begin{equation}
U\left( \tau ,X_{L}^{s},x_{2}\right) =0,\ \ \ \ \ U\left( \tau
,X_{U}^{s},x_{2}\right) =0.  \label{dnt_bc}
\end{equation}%
The corresponding payoffs are%
\begin{equation}
u^{s}\left( x_{1}\right) =\left\{ 
\begin{array}{ll}
e^{-\QTOVERD. . {1}{2}x_{1}}, & s=H, \\ 
e^{-\QTOVERD. . {1}{2}\beta x_{1}}, & s=DH, \\ 
\frac{\sqrt{\frac{\alpha }{2}}}{\sqrt{\left\vert \omega ^{I}\right\vert }}%
\sin \left( \sqrt{\left\vert \omega ^{I}\right\vert }\left( X_{\infty
}^{I}-x_{1}\right) \right) , & s=I, \\ 
\frac{\sqrt{\frac{\alpha }{2}}}{\sqrt{\omega ^{R}}}\sinh \left( \sqrt{\omega
^{R}}\left( X_{\infty }^{R}-x_{1}\right) \right) . & s=R.%
\end{array}%
\right.  \label{dnt_payoff}
\end{equation}

\subsection{Analytical solution\label{DNT_Anal}}

When $\rho =0$ pricing of a DNT\ can be done (semi)-analytically, \cite%
{lipton-1}, \cite{lipton-3}. As before, we can represent the corresponding
solution in the form%
\begin{equation}
U^{s}\left( \tau ,x_{1},x_{2}\right) =\sum\limits_{k=1}^{\infty
}u_{k}^{s}\left( \tau ,x_{2}\right) \nu _{k}^{s}e_{k}\left( x_{1}\right) ,
\label{dnt_sol1}
\end{equation}%
where%
\begin{equation}
e_{k}\left( x_{1}\right) =\sin \left( \zeta _{k}^{s}\left(
x_{1}-X_{L}\right) \right) ,  \label{dnt_el}
\end{equation}%
\begin{equation}
\zeta _{k}^{s}=\frac{\pi k}{\left( X_{U}^{s}-X_{L}^{s}\right) },
\label{dnt_kl}
\end{equation}%
and $\nu _{k}^{s}$ are the Fourier coefficients of the initial condition $%
u^{s}\left( x_{1}\right) $:%
\begin{equation}
\nu _{k}^{s}=\frac{2\zeta _{k}^{s}\left( \frac{1}{\sqrt{\sigma \left(
F_{L}\right) }}+\frac{\left( -1\right) ^{k+1}}{\sqrt{\sigma \left(
F_{U}\right) }}\right) }{\left( X_{U}^{s}-X_{L}^{s}\right) \lambda _{k}^{s}}.
\label{dnt_phil}
\end{equation}%
Accordingly,%
\begin{equation}
U^{s}\left( \tau ,x_{1},x_{2}\right) =\frac{2}{\left(
X_{U}^{s}-X_{L}^{s}\right) }\sum\limits_{k=1}^{\infty }u_{k}^{s}\left( \tau
,x_{2}\right) \frac{\zeta _{k}^{s}\left( \frac{1}{\sqrt{\sigma \left(
F_{L}\right) }}+\frac{\left( -1\right) ^{k+1}}{\sqrt{\sigma \left(
F_{U}\right) }}\right) }{\lambda _{k}^{s}}e_{k}\left( x_{1}\right) .
\label{dnt_sol2}
\end{equation}

\subsection{Numerical solution\label{DNT_num}}

As before, in this subsection we will restrict ourselves to the standard
Heston model governed by Eqs (\ref{Heston_pde1}), (\ref{dnt_bc}), (\ref%
{dnt_payoff}). We solve the pricing problem via numerical methods developed
in Section \ref{NumSol} and compare the corresponding solutions.

Application of ADI methods to the case at hand is relatively
straightforward, especially because the corresponding boundary conditions
are imposed exogenously. We omit details.

The Galerkin method is ideally suited for solving the DNT option pricing
problem. Provided that the maturity of the option is not too short, it is
sufficient to consider very few modes. Discretization in the $x_{2}$%
-direction can be fairly sparse without affecting accuracy too strongly.

Performing the small $\rho $ expansion is simple as well, since it is
normally enough to consider only the first few terms.

As always, achieving high accuracy via the MC method is difficult. In
contrast to other methods, the presence of barriers makes it even more
elaborate and requires using very large number of paths and very small time
steps. To achieve acceptable accuracy, we use 200,000 paths and 3 time steps
per day. Needless to say, for the problem under consideration, the MC method
cannot compete with other methods of interest.

\subsection{Comparison of different numerical solutions for the DNT problem 
\label{DNT_Comp}}

In what follows, we value a double--barrier option with a 1 year maturity on
a unit interval. As an initial condition we take the function (\ref%
{dnt_payoff}), $s=H$.

\begin{figure}[h]
\subfigure[$I_{1}\rightarrow \infty$] {\includegraphics[width=0.55\textwidth, angle=0]
{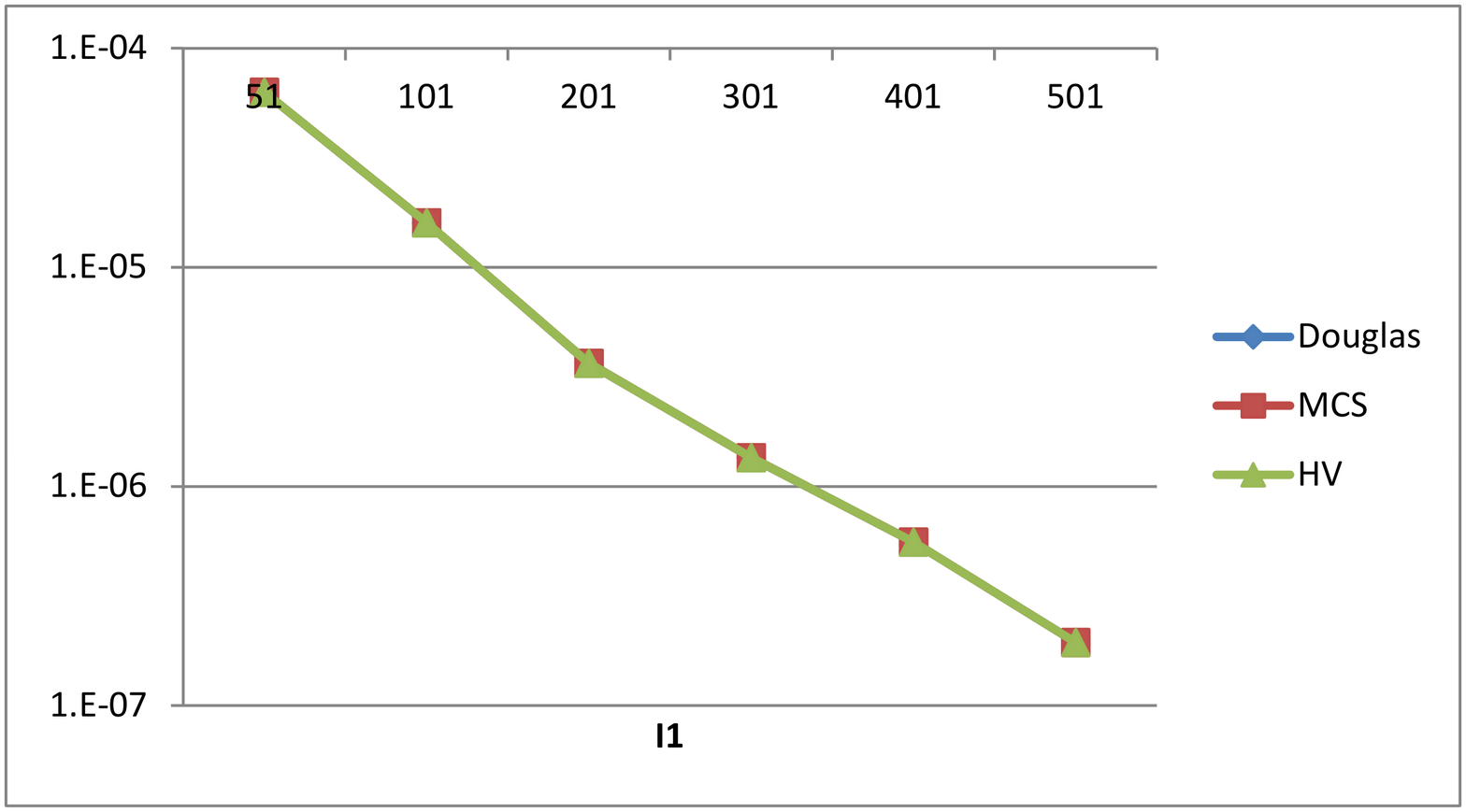}} 
\subfigure[$I_{2}\rightarrow \infty$] {\includegraphics[width=0.55\textwidth, angle=0]
{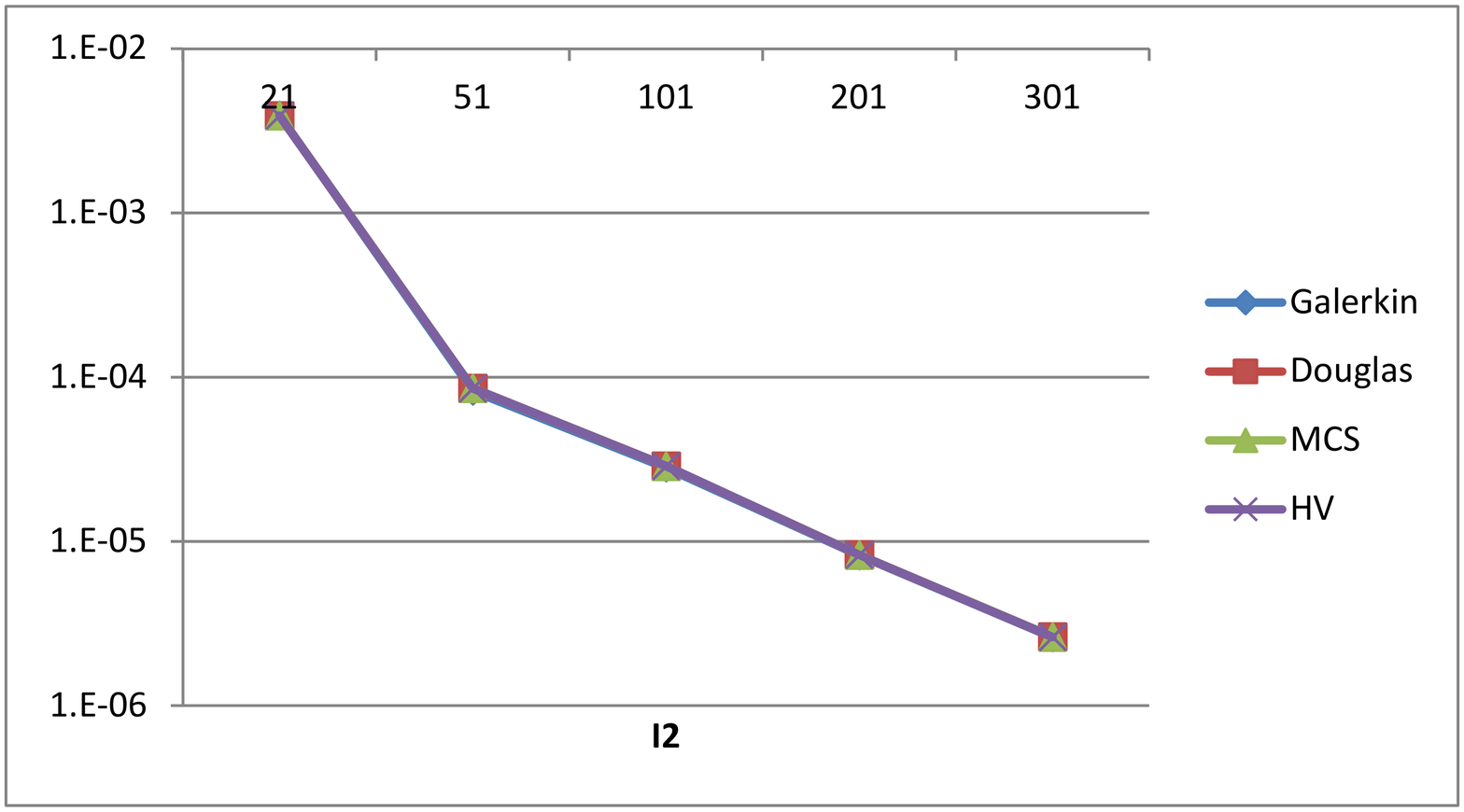}} 
\subfigure[$N\rightarrow \infty$] {\includegraphics[width=0.55\textwidth, angle=0]
{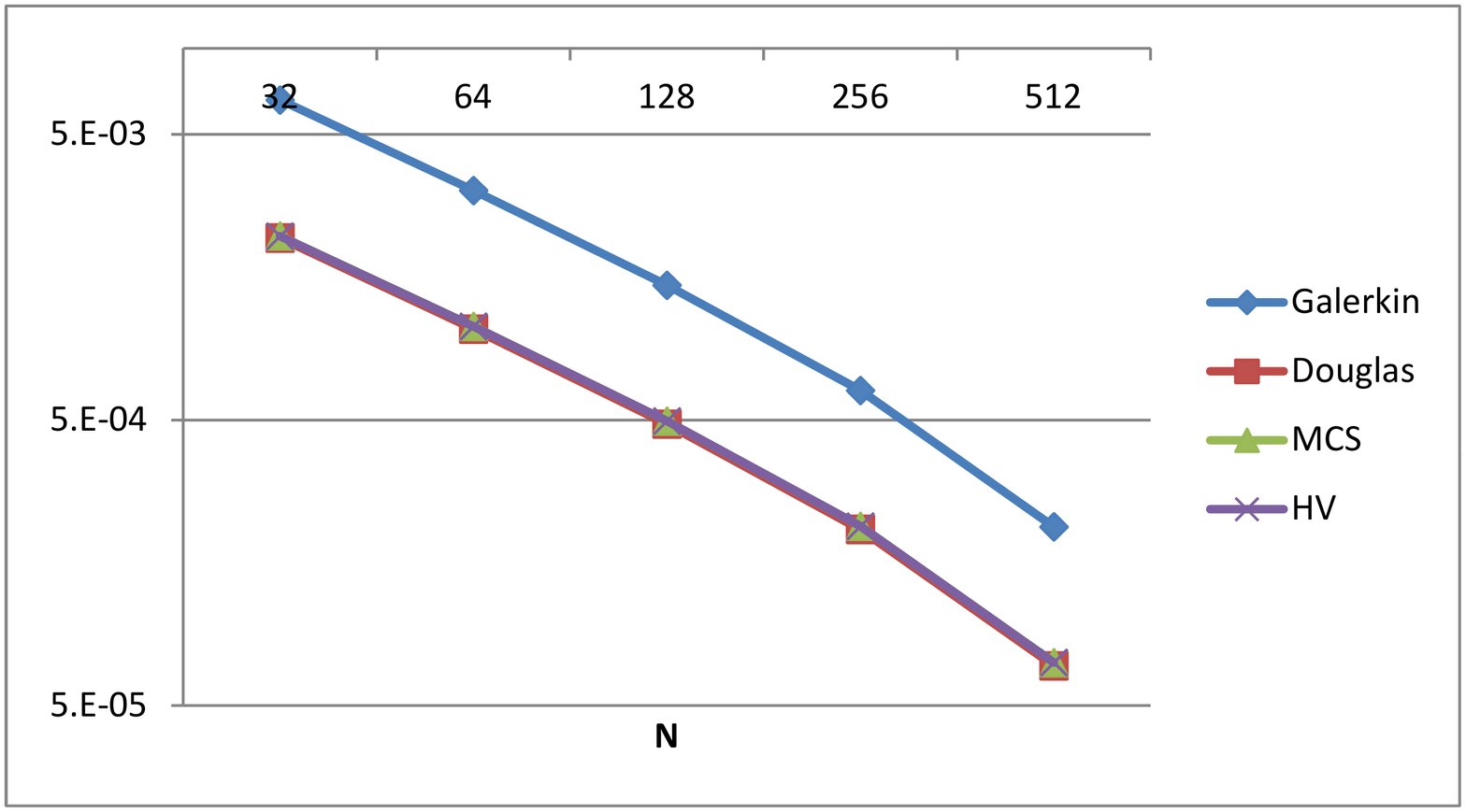}} 
\subfigure[$M\rightarrow \infty$] {\includegraphics[width=0.55\textwidth, angle=0]
{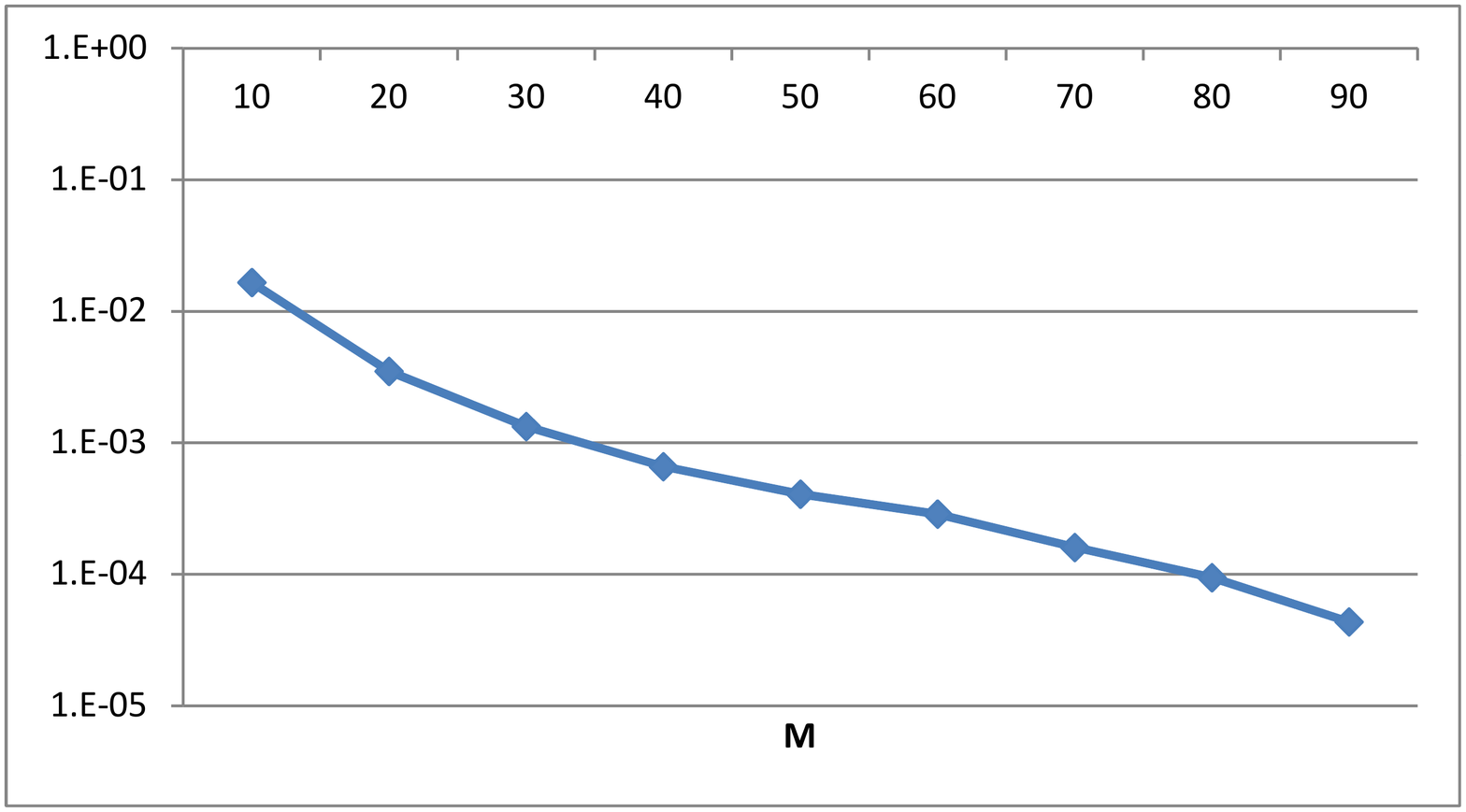}}
\caption{Convergence of the finite difference schemes for the DNT problem
with $\protect\tau =1Y$, $X_{L}=0$, $X_{U}=1$. Base settings are the same as
in Figure \protect\ref{fig:conv_eur}, in addition, $M=30$. In Figures (a),
(b), (c) we show results for $I_{1}\in \left[ 51,601\right] $, $I_{2}\in %
\left[ 21,401\right] $, and $N\in \left[ 32,1024\right] $, respectively. In
Figure (d) we show results for $M\in \left[ 10,100\right] $.}
\label{fig:conv_dnt}
\end{figure}

In Figure \ref{fig:conv_dnt} we review the convergence of the various ADI
methods.

\begin{equation*}
\text{Fig \ref{fig:conv_dnt} near here}
\end{equation*}%
This figure clearly shows that all ADI methods agree with each other. In
particular, the convergence is space is quadratic. However, it is clear that
the convergence in time is only linear for all the ADI methods. Thus, the
gain in accuracy related to the predictor-corrector step is not observed for
DNT options (at least in our calculations). We also show the quadratic
convergence of the Galerkin method with respect to the number of modes.

\begin{figure}[h]
\subfigure[] {\includegraphics[width=0.9\textwidth, angle=0]
{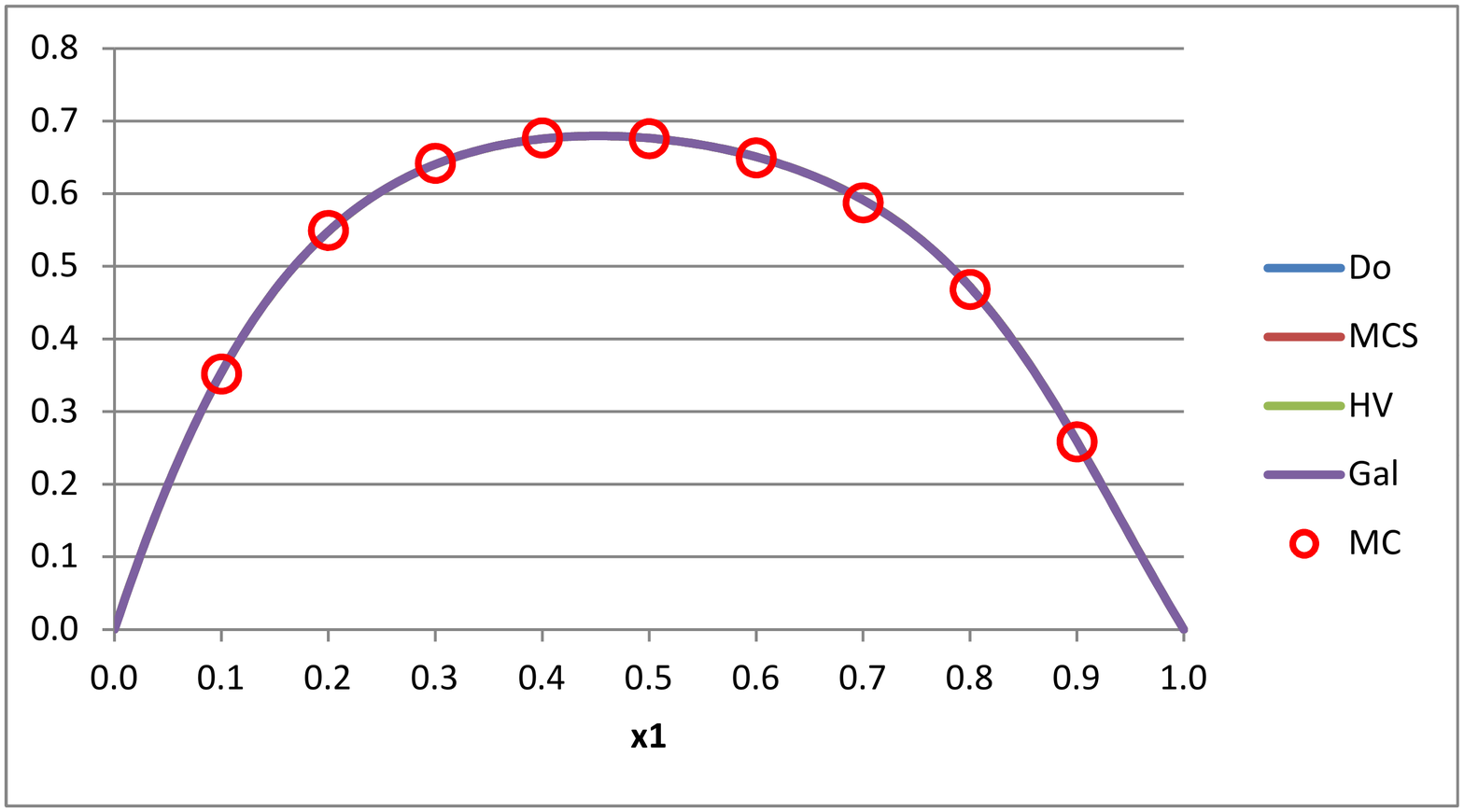}} 
\subfigure[] {\includegraphics[width=0.9\textwidth, angle=0]
{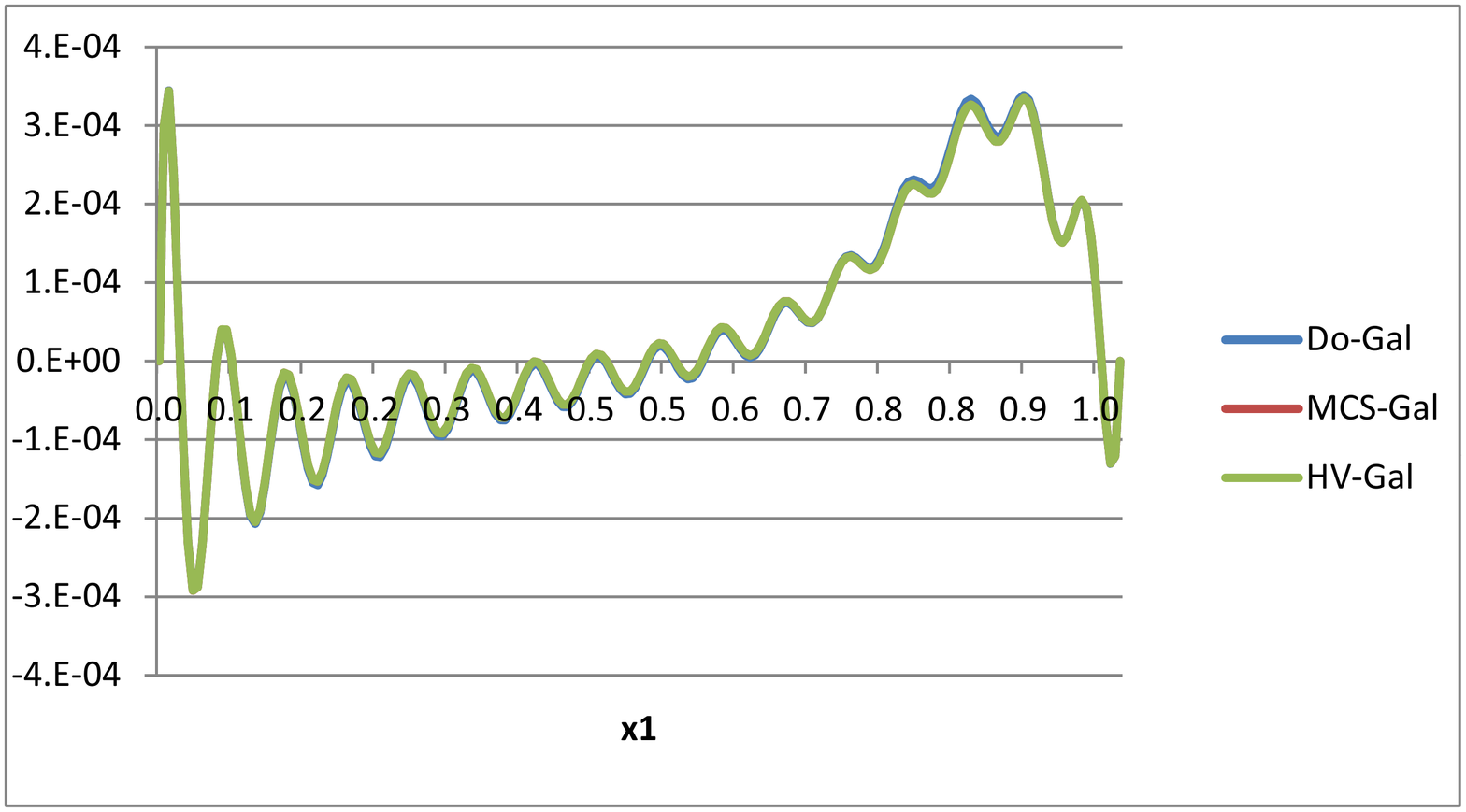}}
\caption{The profiles $U^{H}\left( \protect\tau ,x_{1},x_{2}\right) $ for
the DNT option obtained via the Galerkin and finite difference methods with $%
0\leq x_{1}\leq 1$, $x_{2}=2.628$. In Figure (a) the corresponding profiles
are shown; it is clear that they practically overlap. In addition, for the
sake of comparison, results of the MC simulation are shown as well. In
Figure (b) the differences between the ADI profiles and the Galerkin profile
are shown. The default parameters are used throughout.}
\label{fig:dnt_gal_vs_fd}
\end{figure}

In Figure \ref{fig:dnt_gal_vs_fd} we show the behavior of the DNT prices,
obtained via the numerical methods discussed earlier, for $X_{L}\leq
x_{1}\leq X_{U}$, and $x_{2}=2.628$. It is clear that all the methods
considered in the paper produce consistent prices. We see that, even with
thirty modes, the Galerkin method attains good convergence.

\begin{equation*}
\text{Fig \ref{fig:dnt_gal_vs_fd} near here}
\end{equation*}

\begin{figure}[h]
\subfigure[] {\includegraphics[width=0.9\textwidth, angle=0]
{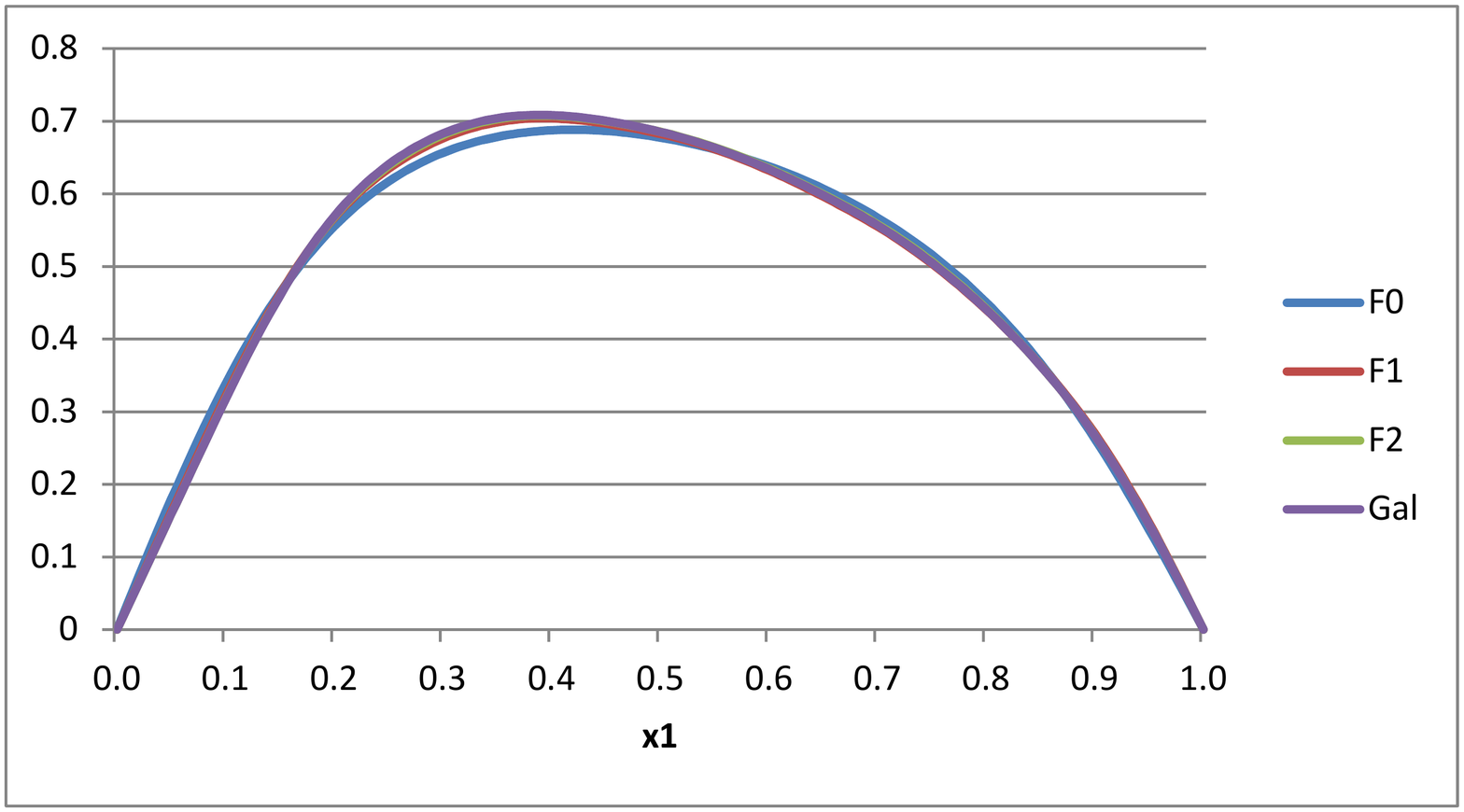}} 
\subfigure[] {\includegraphics[width=0.9\textwidth, angle=0]
{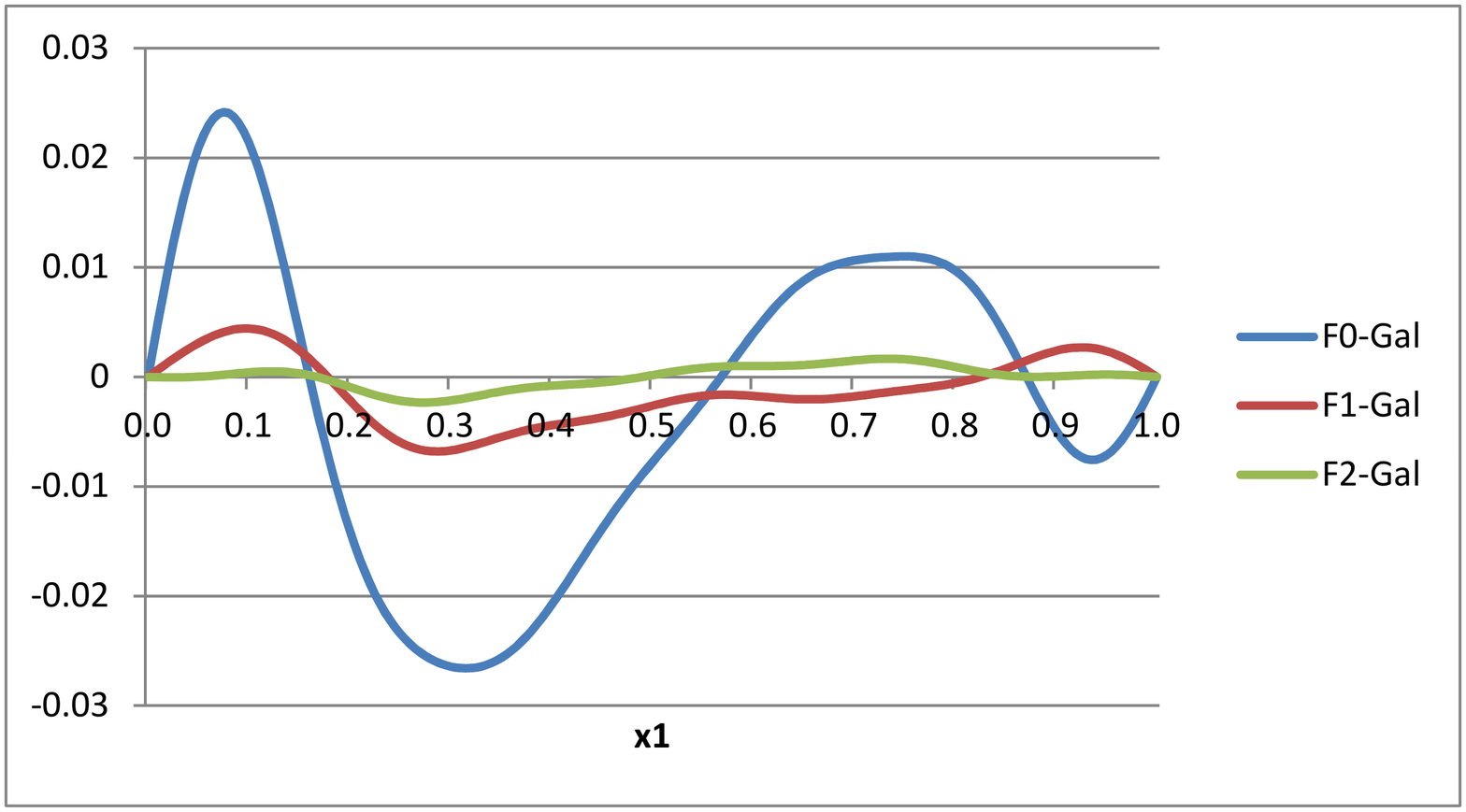}}
\caption{The profiles $U^{H}\left( \protect\tau ,x_{1},x_{2}\right) $ for
the DNT option obtained via the Galerkin and asymptotic expansion methods
for $0\leq x_{1}\leq 1$, $x_{2}=2.628$. In Figure (a) the corresponding
profiles are shown; while not overlapping, they are reasonably close. In
Figure (b) the differences between the asymptotic profiles and the Galerkin
profile are shown. The default parameters are used throughout. }
\label{fig:dnt_gal_vs_exp}
\end{figure}

Finally, we show the convergence of the method of analytical expansion
described in Figure \ref{fig:dnt_gal_vs_exp}. The graph depicts the price
for $X_{L}\leq x_{1}\leq X_{U}$, and $x_{2}=2.628$. We see that, even with
only three perturbations, we attain reasonable, but not perfect, convergence
relative to the solution obtained with the Galerkin method.

\begin{equation*}
\text{Fig \ref{fig:dnt_gal_vs_exp} near here}
\end{equation*}

\section{Two-dimensional Brownian motion\label{Two_corr_BM}}

Given the complex nature of the corresponding FD solutions, it is
instructive to look at a simpler problem. In this Section we consider
two-dimensional Brownian motions in a quadrant and a rectangle with
absorbing boundaries. The corresponding problems are of interest on their
own and can be viewed as the pricing problem for a dual single no-touch
option and a dual DNT option, respectively. In Section \ref{Quadrant} we
consider two-dimensional Brownian motion in a positive quadrant with
absorbing boundaries. This pricing problem can be solved both numerically
and analytically, so that we can benchmark the quality of the former by
using the latter. We conclude that for the problem under consideration
numerical methods work as expected. In Section \ref{Rectangle} we consider
two-dimensional Brownian motion in a rectangle with absorbing boundaries.\
While an analytical solution of the corresponding pricing problem is no
longer feasible, it can be solved numerically by applying all the methods of
Section \ref{NumSol}. Once again, agreement among different solutions is
good and the Galerkin method seems to be the most efficient.

\subsection{Two-dimensional Brownian motion in a positive quadrant with
absorbing boundaries\label{Quadrant}}

\subsubsection{Problem formulation\label{Quad_form}}

Consider two correlated Brownian motions in a positive quadrant. The
corresponding survival probability is governed by equation 
\begin{equation}
Q_{\tau }\left( \tau ,x_{1},x_{2}\right) -\QTOVERD. .
{1}{2}Q_{x_{1}x_{1}}\left( \tau ,x_{1},x_{2}\right) -\rho
Q_{x_{1}x_{2}}\left( \tau ,x_{1},x_{2}\right) -\QTOVERD. .
{1}{2}Q_{x_{2},x_{2}}\left( \tau ,x_{1},x_{2}\right) =0,  \label{two_brom}
\end{equation}%
\begin{equation}
Q\left( 0,x_{1},x_{2}\right) =1,  \label{two_brom_ic}
\end{equation}%
This is the simplest two-factor problem, which is useful for benchmarking
purposes. The boundary conditions have the form%
\begin{equation}
Q\left( \tau ,0,x_{2}\right) =0,\ \ \ \ \ Q\left( \tau ,x_{1},0\right) =0.
\label{two_brom_bc_quad}
\end{equation}%
The corresponding domain in the $\left( x_{1},x_{2}\right) $ plain is 
\begin{equation}
\mathfrak{D}_{q}=\left\{ \left( x_{1},x_{2}\right) ,\ \ \ 0\leq x_{1}<\infty
,\ \ \ 0\leq x_{2}<\infty \right\} .  \label{D_quadrant}
\end{equation}%
This problem is closely related to the DNT option pricing problem considered
earlier, but it does have some important distinctions.

\subsubsection{Analytical solution\label{Quad_anal}}

Problem (\ref{two_brom}), (\ref{two_brom_ic}), (\ref{two_brom_bc_quad}) can
be solved analytically. It can be shown that a change of variables%
\begin{equation}
\left( x_{1},x_{2}\right) \Rightarrow \left( y_{1},y_{2}\right) \Rightarrow
\left( r,\phi \right) ,  \label{change_var}
\end{equation}%
where%
\begin{equation}
y_{1}=x_{1},\ \ \ y_{2}=-\frac{1}{\bar{\rho}}\left( \rho x_{1}-x_{2}\right)
,\ \ \ y_{1}=r\sin \phi ,\ \ \ y_{2}=r\cos \phi ,  \label{new_var}
\end{equation}%
allows us to eliminate the cross derivative and transforms the pricing
problem in question into the following one%
\begin{equation}
Q_{\tau }\left( \tau ,r,\phi \right) -\QTOVERD. . {1}{2}\left( Q_{rr}\left(
\tau ,r,\phi \right) +\frac{1}{r}Q_{r}\left( \tau ,r,\phi \right) +\frac{1}{%
r^{2}}Q_{\phi \phi }\left( \tau ,r,\phi \right) \right) =0,  \label{rphi_DPE}
\end{equation}%
\begin{equation}
Q\left( 0,r,\phi \right) =1,  \label{rphi_ic}
\end{equation}%
\begin{equation}
Q\left( \tau ,r,0\right) =0,\ \ \ Q\left( \tau ,r,\varpi \right) =0,\ \ \
Q\left( \tau ,r,\phi \right) \underset{r\rightarrow 0}{\rightarrow }0,\ \ \
Q\left( \tau ,r,\phi \right) \underset{r\rightarrow \infty }{\rightarrow }1.
\label{rphi_bc}
\end{equation}%
Here $\varpi =\arccos \left( -\rho \right) $. Thus, we have managed to map
the positive quadrant onto a semi-strip%
\begin{equation}
\widetilde{\mathfrak{D}}_{q}=\left\{ \left( r,\phi \right) ,\ \ \ 0\leq
r<\infty ,\ \ \ 0\leq \phi \leq \varpi \right\} .  \label{rphi_dom}
\end{equation}%
Since coefficients of Eq. (\ref{rphi_DPE}) are $\phi $-independent, we can
use the Galerkin method to solve it, see \cite{he}, \cite{lipton-book}, \cite%
{zhou}. An elementary solution of Eq. (\ref{rphi_DPE}) satisfying boundary
conditions (\ref{rphi_bc}) in the $\phi $-direction can be written in the
form%
\begin{equation}
Q_{k}\left( \tau ,r,\phi \right) \sim g_{k}\left( \tau ,r\right) \sin \left(
\zeta _{k}\phi \right) ,  \label{rphi_elem_sol}
\end{equation}%
where, as often before, $\zeta _{k}=\pi l/\varpi $, and $g_{k}\left( \tau
,r\right) $ is a solution of the following problem%
\begin{equation}
g_{k,\tau }\left( \tau ,r\right) -\QTOVERD. . {1}{2}\left( g_{k,rr}\left(
\tau ,r\right) +\frac{1}{r}g_{k,r}\left( \tau ,r\right) -\frac{\zeta _{k}^{2}%
}{r^{2}}g_{k}\left( \tau ,r\right) \right) =0,  \label{pde_gl}
\end{equation}%
We write%
\begin{equation}
Q\left( \tau ,r,\phi \right) =\frac{4}{\pi }\dsum\limits_{k=1,k\
odd}^{\infty }\frac{g_{k}\left( \tau ,r\right) }{k}\sin \left( \zeta
_{k}\phi \right) ,  \label{rhi_decom2}
\end{equation}%
so that the corresponding boundary and initial condition for $g_{k}\left(
\tau ,r\right) $ are chosen to be of the form%
\begin{equation}
g_{k}\left( \tau ,r\right) \underset{r\rightarrow 0}{\rightarrow }0,\ \ \
g_{k}\left( \tau ,r\right) \underset{r\rightarrow \infty }{\rightarrow }1,
\label{bc_gl}
\end{equation}%
\begin{equation}
g_{k}\left( 0,r\right) =1.  \label{ic_gl}
\end{equation}%
It can be checked directly that $g_{k}$ is a self-similar function;%
\begin{equation}
g_{k}\left( \tau ,r\right) =\sqrt{\frac{\pi }{2}}\sqrt{\upsilon }%
e^{-\upsilon }\left( I_{\frac{1}{2}\left( \zeta _{k}-1\right) }\left(
\upsilon \right) +I_{\frac{1}{2}\left( \zeta _{k}+1\right) }\left( \upsilon
\right) \right) \equiv \sqrt{\frac{\pi }{2}}\mathsf{J}_{k}\left( \upsilon
\right) ,  \label{gl}
\end{equation}%
where $\upsilon =r^{2}/4\tau $, see Appendix \ref{eqgl}. Accordingly,%
\begin{equation}
Q\left( \tau ,r,\phi \right) =\sqrt{\frac{8}{\pi }}\dsum\limits_{k=1,k\
odd}^{\infty }\frac{\mathsf{J}_{k}\left( \upsilon \right) }{k}\sin \left(
\zeta _{k}\phi \right) .  \label{Qrphi}
\end{equation}%
Alternative derivation based on the integration of the Green's function can
be found in many papers, see, e.g., \cite{iyengar}, \cite{lipton-savescu},
and \cite{metzler} for further details. Finally, in order to compute $%
Q\left( \tau ,x_{1},x_{2}\right) $, all we need to do is to express $\left(
r,\phi \right) $ in terms of $\left( x_{1},x_{2}\right) $ via Eqs (\ref%
{new_var}).

\subsubsection{Numerical solution\label{Quad_num}}

We wish to solve the problem (\ref{two_brom}), (\ref{two_brom_ic}), (\ref%
{two_brom_bc_quad}) numerically. To this end we discretize Eq. (\ref%
{two_brom}), and the corresponding initial condition (\ref{two_brom_ic});
the boundary condition at the boundary is clear, at infinity we choose
natural boundary conditions for suitably large values of $x_{1},x_{2}$. We
solve the corresponding discrete problem via an ADI method.

\subsubsection{Comparison of analytical and numerical solutions for the
quadrant problem\label{Quad_comp}}

\begin{figure}[h]
\subfigure[] {\includegraphics[width=0.9\textwidth, angle=0]
{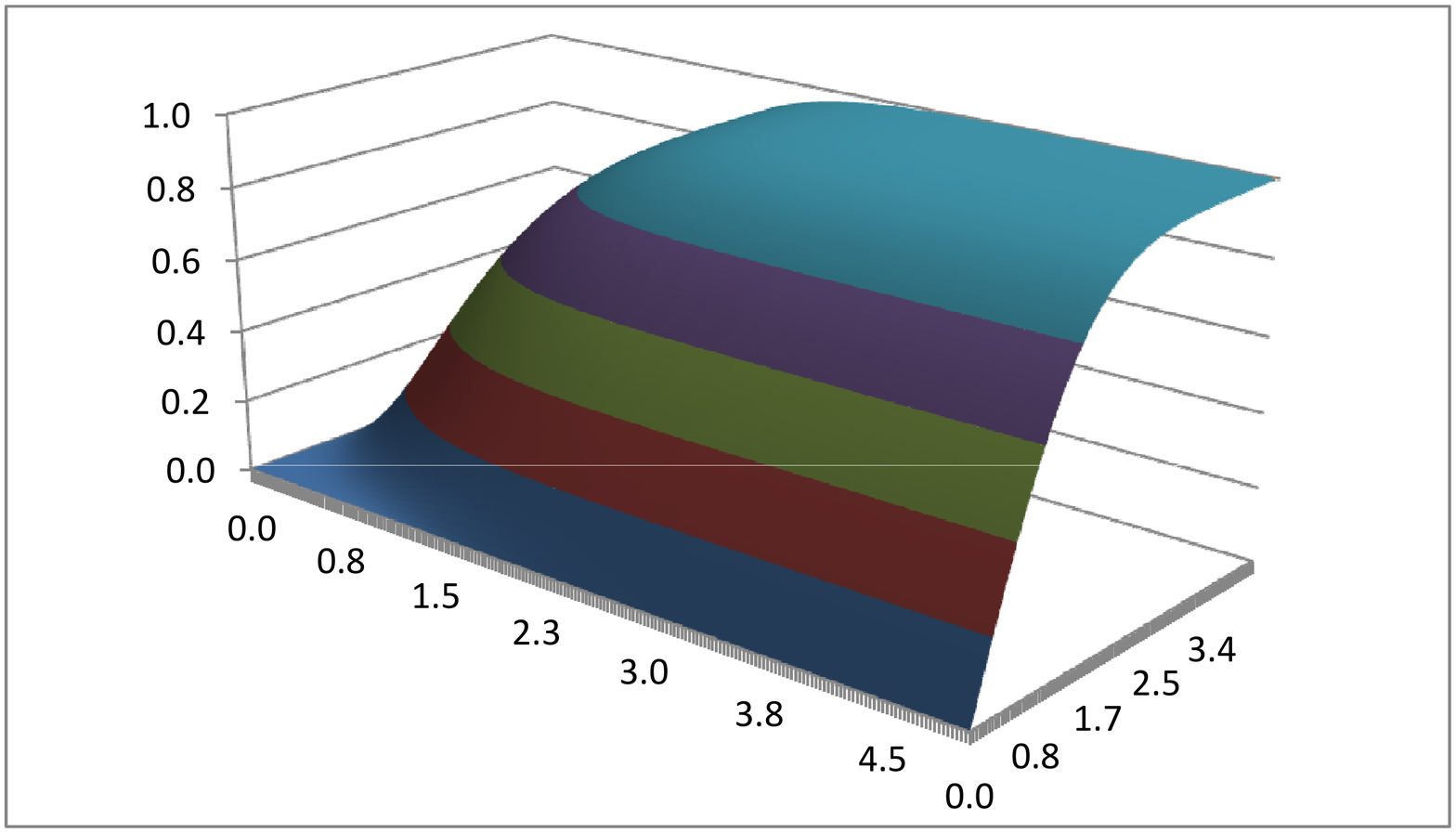}} 
\subfigure[] {\includegraphics[width=0.9\textwidth, angle=0]
{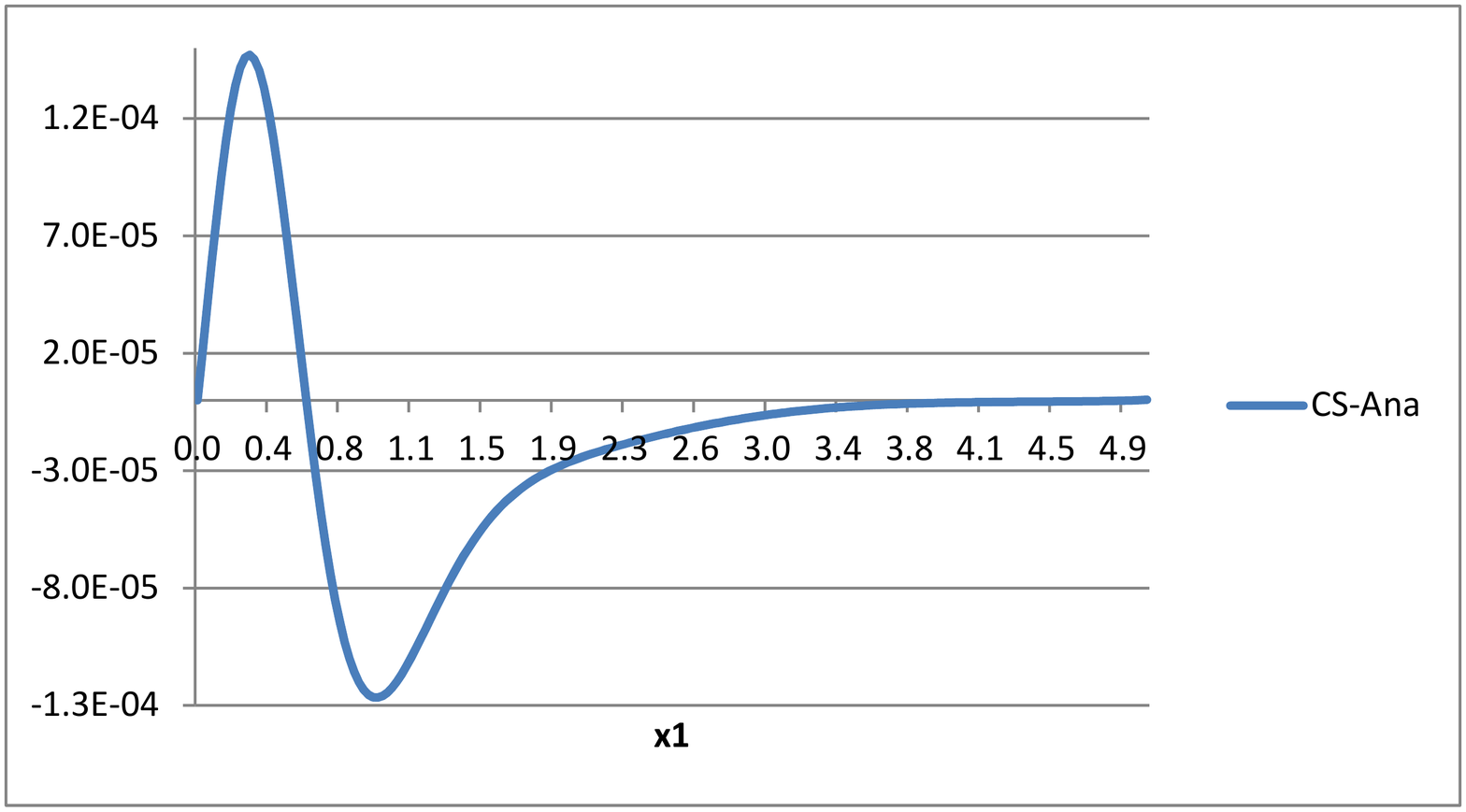}}
\caption{Comparison of the analytical and ADI solutions for the quadrant
problem with $\protect\tau =1$, $\protect\rho =-0.90$. The computational
domain is given by $0\leq x_{1}\leq 5$, $0\leq x_{2}\leq 4$. The following
parameters are used: $I_{1}=201$, $I_{2}=101$, $N=1000$, $M=10$. In Figure
(a) we show the survival probabilities obtained both analytically and
numerically, which clearly agree with each other. In Figure (b) we show the
difference between the two solutions as a function of $x_{1}$ for $x_{2}=1.0$%
.}
\label{fig:quadrant}
\end{figure}

Analytical and numerical solutions are compared in Figure \ref{fig:quadrant}%
. This figure shows that the ADI solution does converge to the analytical
one and that this convergence is good. Moreover, it makes clear that the
choice of the natural boundary conditions is appropriate. We emphasize that
choosing Dirichlet boundary conditions would cause major loss of accuracy.

\subsection{Two-dimensional Brownian motion in a rectangle with absorbing
boundaries\label{Rectangle}}

In this section we consider two correlated Brownian motions in a rectangle.
It can be viewed as a pricing problem for a quadruple no-touch option. Its
solution along the lines described below was proposed by Lipton and Little, 
\cite{lipton-4}, and discussed in more detail in \cite{lipton-book}, Section
12.9.

\subsubsection{Problem formulation\label{Rect_form}}

The survival probability for two correlated Brownian motions in a rectangle
is governed by Eq. (\ref{two_brom}) augmented with the initial condition (%
\ref{two_brom_ic}), and the boundary conditions of the form%
\begin{equation}
Q\left( \tau ,0,x_{2}\right) =0,\ \ \ Q\left( \tau ,L_{1},x_{2}\right) =0,\
\ \ Q\left( \tau ,x_{1},0\right) =0,\ \ \ Q\left( \tau ,x_{1},L_{2}\right)
=0,  \label{two_brom_bc_rec}
\end{equation}%
The corresponding domain in the $\left( x_{1},x_{2}\right) $ plain is 
\begin{equation}
\mathfrak{D}_{r}=\left\{ \left( x_{1},x_{2}\right) ,\ \ \ 0\leq
x_{1}<L_{1},\ \ \ 0<x_{2}<L_{2}\right\} .  \label{d_rec}
\end{equation}

\subsubsection{Numerical solution\label{Rect_num}}

Numerical solution of the problem (\ref{two_brom}), (\ref{two_brom_ic}), (%
\ref{two_brom_bc_rec}) is relatively simple. It can be solved by any of the
methods developed in Section \ref{NumSol}; to be concrete, we use the
standard CS method. Since all the relevant boundary conditions are of the
Dirichlet type, the application of the CS method is straightforward,
especially in the light of our previous discussion, and is left to the
reader as an exercise.

The small $\rho $ expansion is more interesting, so we discuss it in some
detail. As before, we can $Q\left( \tau ,x_{1},x_{2}\right) $ as a vector
function%
\begin{equation}
Q\left( \tau ,x_{1},x_{2}\right) \boldsymbol{=}\dsum%
\limits_{k_{1}=1,k_{2}=1}^{\infty }Q_{k_{1}k_{2}}\left( \tau \right)
e_{k_{1}k_{2}}\left( x_{1},x_{2}\right) ,  \label{Qx1x2_vec}
\end{equation}%
where $e_{k_{1}k_{2}}$ are orthogonal (but not normal) basis vectors of the
form%
\begin{equation}
e_{k_{1},k_{2}}\left( x_{1},x_{2}\right) =\sin \left( \frac{\pi k_{1}x_{1}}{%
L_{1}}\right) \sin \left( \frac{\pi k_{2}x_{2}}{L_{2}}\right) \equiv \sin
\left( \zeta _{k_{1}}x_{1}\right) \sin \left( \zeta _{k_{2}}x_{2}\right) .
\label{ek1k2}
\end{equation}%
It is clear that $Q_{k_{1}k_{2}}\left( \tau \right) $ is a \emph{matrix}
rather than a \emph{vector}, so one way to deal with this fact is to use a
tensor-based formalism, as was done in Section \ref{Explicit} above.
However, for the sake of variety, we describe how to use a matrix-based
techniques instead. To this end, we assume that $1\leq k_{i}\leq N$, map
each pair $\left( k_{1},k_{2}\right) $ into a single number $K$ (and back)
as follows%
\begin{equation}
\begin{array}{l}
K=\left( k_{1}-1\right) +\left( k_{2}-1\right) N, \\ 
k_{2}=\left[ \frac{K}{N}\right] +1,\ \ \ \ \ k_{1}=K-\left( k_{2}-1\right)
N+1.%
\end{array}
\label{k_to_K}
\end{equation}%
and write%
\begin{equation}
e_{K}\left( x_{1},x_{2}\right) =\sin \left( \zeta _{k_{1}}x_{1}\right) \sin
\left( \zeta _{k_{2}}x_{2}\right) .  \label{e_K}
\end{equation}%
This allows us to think of $Q\left( \tau ,x_{1},x_{2}\right) $ as a \emph{%
vector function} of $\tau $, 
\begin{equation}
Q\left( \tau ,x_{1},x_{2}\right) \Rightarrow \left\{ Q_{K}\left( \tau
\right) \right\} \equiv \overrightarrow{Q}\left( \tau \right) .
\label{Qx1x2_vec1}
\end{equation}

We can write the pricing equation as follows%
\begin{equation}
\frac{d\overrightarrow{Q}}{d\tau }-\mathbb{A}\overrightarrow{Q}-\rho \mathbb{%
B}\overrightarrow{Q}=0,\ \ \ \ \ \overrightarrow{Q}\left( 0\right) =%
\overrightarrow{\nu }.  \label{ODE_vec}
\end{equation}%
Here%
\begin{equation}
\begin{array}{lll}
\mathbb{A}e_{K} & = & \QTOVERD. . {1}{2}\left(
e_{K,x_{1}x_{1}}+e_{K,x_{2}x_{2}}\right) \\ 
& = & -\QTOVERD. . {1}{2}\left( \zeta _{k_{1}}^{2}+\zeta _{k_{2}}^{2}\right)
e_{K}\equiv -\lambda _{K}e_{K},%
\end{array}
\label{A_B_mat0}
\end{equation}%
\begin{equation}
\begin{array}{lll}
\mathbb{B}e_{K} & = & e_{K,x_{1}x_{2}} \\ 
& = & \dsum\limits_{L=0,L\neq K}^{N^{2}-1}\frac{4}{L_{1}L_{2}}\frac{%
k_{1}k_{2}l_{1}l_{2}\left( 1-\left( -1\right) ^{k_{1}-l_{1}}\right) \left(
1-\left( -1\right) ^{k_{2}-l_{2}}\right) }{\left( k_{1}^{2}-l_{1}^{2}\right)
\left( k_{2}^{2}-l_{2}^{2}\right) }e_{_{L}}\equiv
\dsum\limits_{L=0}^{N^{2}-1}\mu _{K,L}e_{L},%
\end{array}
\label{A_B_mat1}
\end{equation}%
\begin{equation}
\begin{array}{lll}
1 & = & \dsum\limits_{K=0}^{N^{2}-1}\frac{4}{L_{1}L_{2}}\frac{\left(
1+\left( -1\right) ^{k_{1}+1}\right) \left( 1+\left( -1\right)
^{k_{2}+1}\right) }{\zeta _{1}\zeta _{2}}e_{K} \\ 
& = & \dsum\limits_{K=0}^{N^{2}-1}\frac{4}{\pi ^{2}}\frac{\left( 1+\left(
-1\right) ^{k_{1}+1}\right) \left( 1+\left( -1\right) ^{k_{2}+1}\right) }{%
k_{1}k_{2}}e_{K}\equiv \dsum\limits_{K=0}^{N^{2}-1}\nu _{K}e_{K}.%
\end{array}
\label{A_B_mat2}
\end{equation}%
It is clear that%
\begin{equation}
\nu _{K}=\left\{ 
\begin{array}{ll}
\frac{16}{\pi ^{2}k_{1}k_{2}}, & k_{1},k_{2}\text{\ }\mathrm{odd}, \\ 
0, & \mathrm{otherwise}.%
\end{array}%
\right.  \label{nu_vec1}
\end{equation}%
Here, by definition, $\mu _{K,L}=0$ if $\left( k_{1}-l_{1}\right) \left(
k_{2}-l_{2}\right) =0$.

We assume that $\rho $ is small and use it as an expansion parameter. Then%
\begin{equation}
\overrightarrow{Q}=\overrightarrow{Q}^{\left( 0\right) }+\rho 
\overrightarrow{Q}^{\left( 1\right) }+\rho ^{2}\overrightarrow{Q}^{\left(
2\right) }+\rho ^{3}\overrightarrow{Q}^{\left( 3\right)
}...=\sum_{n=0}^{\infty }\rho ^{n}\overrightarrow{Q}^{\left( n\right) },
\label{Q_hro_exp}
\end{equation}%
where%
\begin{equation}
\frac{d\overrightarrow{Q}^{\left( 0\right) }}{d\tau }-\mathbb{A}%
\overrightarrow{Q}^{\left( 0\right) }=0,\ \ \ \ \ \overrightarrow{Q}^{\left(
0\right) }\left( 0\right) =\overrightarrow{\nu },  \label{ODE_Q0}
\end{equation}%
\begin{equation}
\frac{d\overrightarrow{Q}^{\left( 1\right) }}{d\tau }-\mathbb{A}%
\overrightarrow{Q}^{\left( 1\right) }=\mathbb{B}\overrightarrow{Q}^{\left(
0\right) },\ \ \ \ \ \overrightarrow{Q}^{\left( 1\right) }\left( 0\right) =0,
\label{ODE_Q1}
\end{equation}%
\begin{equation}
\frac{d\overrightarrow{Q}^{\left( 2\right) }}{d\tau }-\mathbb{A}%
\overrightarrow{Q}^{\left( 2\right) }=\mathbb{B}\overrightarrow{Q}^{\left(
1\right) },\ \ \ \ \ \overrightarrow{Q}^{\left( 2\right) }\left( 0\right) =0,
\label{ODE_Q2}
\end{equation}%
etc. In general,%
\begin{equation}
\frac{d\overrightarrow{Q}^{\left( n\right) }}{d\tau }-\mathbb{A}%
\overrightarrow{Q}^{\left( n\right) }=\mathbb{B}\overrightarrow{Q}^{\left(
n-1\right) },\ \ \ \ \ \overrightarrow{Q}^{\left( n\right) }\left( 0\right)
=0.  \label{ODE_Qn}
\end{equation}%
It is clear that%
\begin{equation}
\overrightarrow{Q}^{\left( 0\right)
}=\dsum\limits_{K_{1}=0}^{N^{2}-1}e^{-\lambda _{K_{1}}\tau }\nu
_{K_{1}}e_{K_{1}}\equiv \dsum\limits_{K_{1}=0}^{N^{2}-1}\upsilon
_{K_{1}}^{\left( 0\right) }\left( \tau \right) e_{K_{1}}.  \label{Q0_exp}
\end{equation}%
We can write $\overrightarrow{Q}^{\left( 1\right) }$ in the form%
\begin{equation}
\overrightarrow{Q}^{\left( 1\right)
}=\dsum\limits_{K_{2}=0}^{N^{2}-1}\upsilon _{K_{2}}^{\left( 1\right) }\left(
\tau \right) e_{K_{2}}.  \label{Q1_exp}
\end{equation}%
Substitution of this expression into the pricing equation yields%
\begin{equation}
\partial _{\tau }\upsilon _{K_{2}}^{\left( 1\right) }\left( \tau \right)
+\lambda _{K_{2}}\upsilon _{K_{2}}^{\left( 1\right) }\left( \tau \right)
=\dsum\limits_{K_{1}=0}^{N^{2}-1}e^{-\lambda _{K_{1}}\tau }\nu _{K_{1}}\mu
_{K_{1},K_{2}},\ \ \ \ \ \upsilon _{K_{2}}\left( 0\right) =0,
\label{ODE_Q11}
\end{equation}%
so that%
\begin{equation}
\upsilon _{K_{2}}^{\left( 1\right) }\left( \tau \right)
=\dsum\limits_{K_{1}=0}^{N^{2}-1}\Theta _{\lambda _{K_{1}},\lambda
_{K_{2}}}^{\left( 1\right) }\left( \tau \right) \nu _{K_{1}}\mu
_{K_{1},K_{2}}.
\end{equation}%
where $\Theta _{\lambda _{1},\lambda _{2}}^{\left( 1\right) }\left( \tau
\right) $ is the solution of the problem%
\begin{equation}
\partial _{\tau }\Theta _{\lambda _{1},\lambda _{2}}^{\left( 1\right)
}\left( \tau \right) +\lambda _{2}\Theta _{\lambda _{1},\lambda
_{2}}^{\left( 1\right) }\left( \tau \right) =e^{-\lambda _{1}\tau },\ \ \ \
\ \Theta _{\lambda _{1},\lambda _{2}}^{\left( 1\right) }\left( 0\right) =0,
\end{equation}%
which we represent in the form%
\begin{equation}
\Theta _{\lambda _{1},\lambda _{2}}^{\left( 1\right) }\left( \tau \right)
=e^{-\lambda _{1}\tau }\phi _{\lambda _{1}-\lambda _{2}}\left( \tau \right) ,
\end{equation}%
\begin{equation}
\phi _{\mu }\left( \tau \right) =\left\{ 
\begin{array}{cc}
\frac{e^{\mu \tau }-1}{\mu }, & \mu \neq 0, \\ 
\tau , & \mu =0.%
\end{array}%
\right.
\end{equation}%
Finally,%
\begin{equation}
\overrightarrow{Q}^{\left( 1\right)
}=\dsum\limits_{K_{1}=0,K_{2}=0}^{N^{2}-1}\Theta _{\lambda _{K_{1}},\lambda
_{K_{2}}}^{\left( 1\right) }\left( \tau \right) \nu _{K_{1}}\mu
_{K_{1},K_{2}}e_{K_{2}}.
\end{equation}%
By the same token,%
\begin{equation}
\overrightarrow{Q}^{\left( 2\right)
}=\dsum\limits_{K_{3}=0}^{N^{2}-1}\upsilon _{K_{3}}^{\left( 2\right) }\left(
\tau \right) e_{K_{3}},
\end{equation}%
where%
\begin{equation}
\partial _{\tau }\upsilon _{K_{3}}^{\left( 2\right) }\left( \tau \right)
+\lambda _{K_{3}}\upsilon _{K_{3}}^{\left( 2\right) }\left( \tau \right)
=\dsum\limits_{K_{1}=0,K_{2}=0}^{N^{2}-1}\Theta _{\lambda _{K_{1}},\lambda
_{K_{2}}}^{\left( 1\right) }\nu _{K_{1}}\mu _{K_{1},K_{2}}\mu
_{K_{2},K_{3}},\ \ \ \ \ \upsilon _{K_{3}}^{\left( 2\right) }\left( 0\right)
=0,
\end{equation}%
so that%
\begin{equation}
\upsilon _{K_{3}}^{\left( 2\right) }\left( \tau \right)
=\dsum\limits_{K_{1}=0,K_{2}=0}^{N^{2}-1}\Theta _{\lambda _{K_{1}},\lambda
_{K_{2}},\lambda _{K_{3}}}^{\left( 2\right) }\left( \tau \right) \nu
_{K_{1}}\mu _{K_{1},K_{2}}\mu _{K_{2},K_{3}}.
\end{equation}%
where $\Theta _{\lambda _{1},\lambda _{2},\lambda _{3}}^{\left( 2\right) }$
is the solution of the problem%
\begin{equation}
\partial _{\tau }\Theta _{\lambda _{1},\lambda _{2},\lambda _{3}}^{\left(
2\right) }\left( \tau \right) +\lambda _{3}\Theta _{\lambda _{1},\lambda
_{2},\lambda _{3}}^{\left( 2\right) }\left( \tau \right) =\Theta _{\lambda
_{1},\lambda _{2}}^{\left( 1\right) }\left( \tau \right) ,\ \ \ \ \ \Theta
_{\lambda _{1},\lambda _{2},\lambda _{3}}^{\left( 2\right) }\left( 0\right)
=0,
\end{equation}%
or, equivalently,%
\begin{equation}
\partial _{\tau }\Theta _{\lambda _{1},\lambda _{2},\lambda _{3}}^{\left(
2\right) }\left( \tau \right) +\lambda _{3}\Theta _{\lambda _{1},\lambda
_{2},\lambda _{3}}^{\left( 2\right) }\left( \tau \right) =e^{-\lambda
_{1}\tau }\phi _{\lambda _{1}-\lambda _{2}}\left( \tau \right) ,\ \ \ \ \
\Theta _{\lambda _{1},\lambda _{2},\lambda _{3}}^{\left( 2\right) }\left(
0\right) =0.
\end{equation}%
We write 
\begin{equation}
\Theta _{\lambda _{1},\lambda _{2},\lambda _{3}}^{\left( 2\right) }\left(
\tau \right) =e^{-\lambda _{1}\tau }\psi _{\lambda _{1}-\lambda _{2},\lambda
_{1}-\lambda _{3}}\left( \tau \right) .
\end{equation}%
A simple calculation yields%
\begin{equation}
\psi _{\mu _{1},\mu _{2}}\left( \tau \right) =\left\{ 
\begin{array}{cc}
\frac{\phi _{\mu _{1}}\left( \tau \right) -\phi _{\mu _{2}}\left( \tau
\right) }{\left( \mu _{1}-\mu _{2}\right) }, & \mu _{1}\neq \mu _{2}, \\ 
\frac{\tau e^{\mu _{1}\tau }-\phi _{\mu _{1}}\left( \tau \right) }{\mu _{1}},
& \mu _{1}=\mu _{2}\neq 0, \\ 
\frac{\tau ^{2}}{2}, & \mu _{1}=\mu _{2}=0.%
\end{array}%
\right.
\end{equation}%
In general,%
\begin{equation}
\Theta _{\lambda _{1},...,\lambda _{M+1}}^{\left( M\right) }\left( \tau
\right) =e^{-\lambda _{1}\tau }\psi _{\lambda _{1}-\lambda _{2},...,\lambda
_{1}-\lambda _{M+1}}\left( \tau \right) \equiv e^{-\lambda _{1}\tau }\psi
_{\mu _{1},...,\psi _{M}}\left( \tau \right) ,
\end{equation}%
\begin{equation}
\psi _{\mu _{1},...,\psi _{M}}\left( \tau \right) =\dsum\limits_{i=1}^{M}%
\frac{\phi _{\mu _{i}}\left( \tau \right) }{\dprod\limits_{i^{\prime }\neq
i}\left( \mu _{i}-\mu _{i^{\prime }}\right) },
\end{equation}%
where the limiting behavior of the above expression is calculated via
l'Hospital's rule.

\subsubsection{Comparison of different numerical solutions for the rectangle
problem\label{Rect_comp}}

\begin{figure}[h]
\subfigure[] {\includegraphics[width=0.9\textwidth, angle=0]
{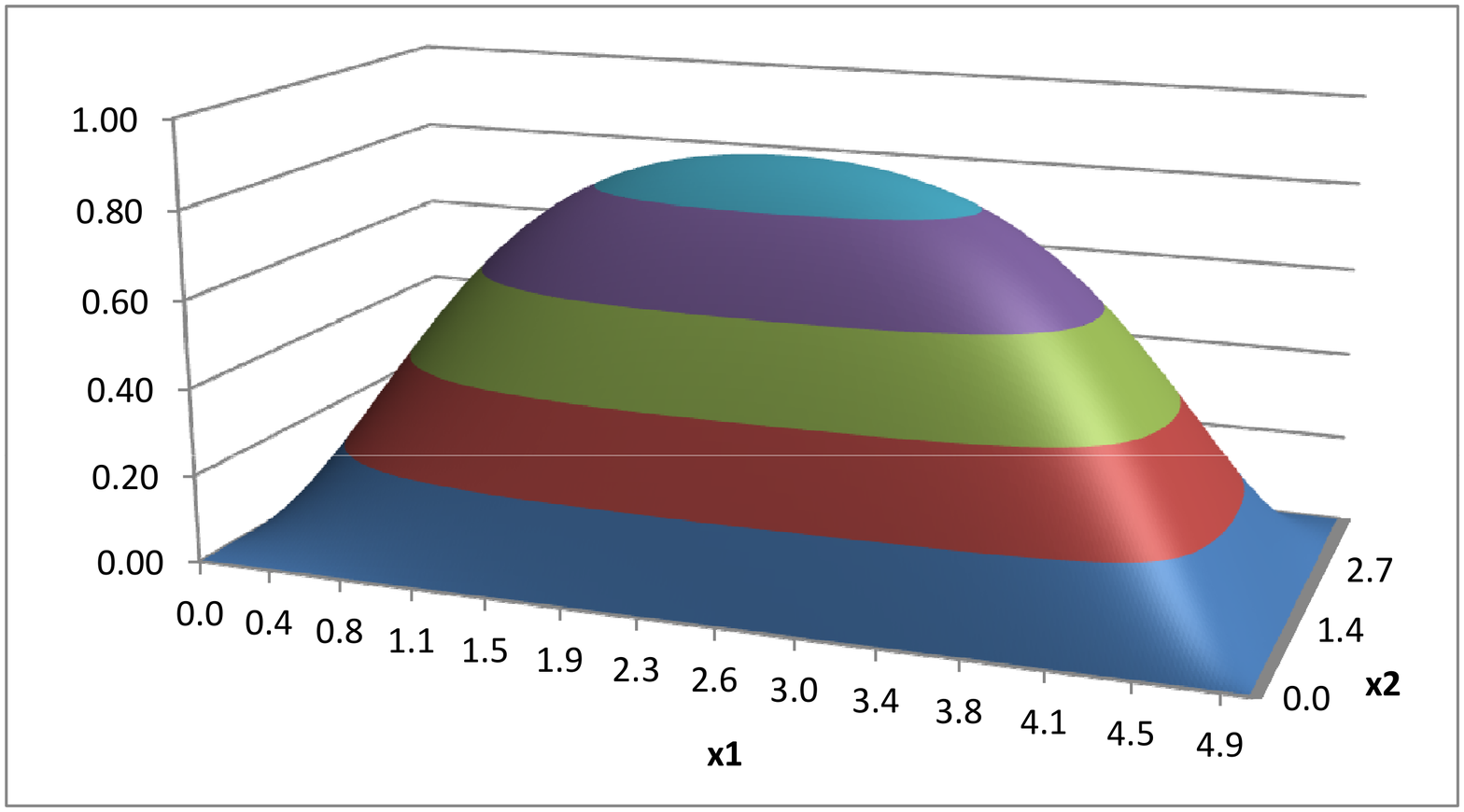}} 
\subfigure[] {\includegraphics[width=0.9\textwidth, angle=0]
{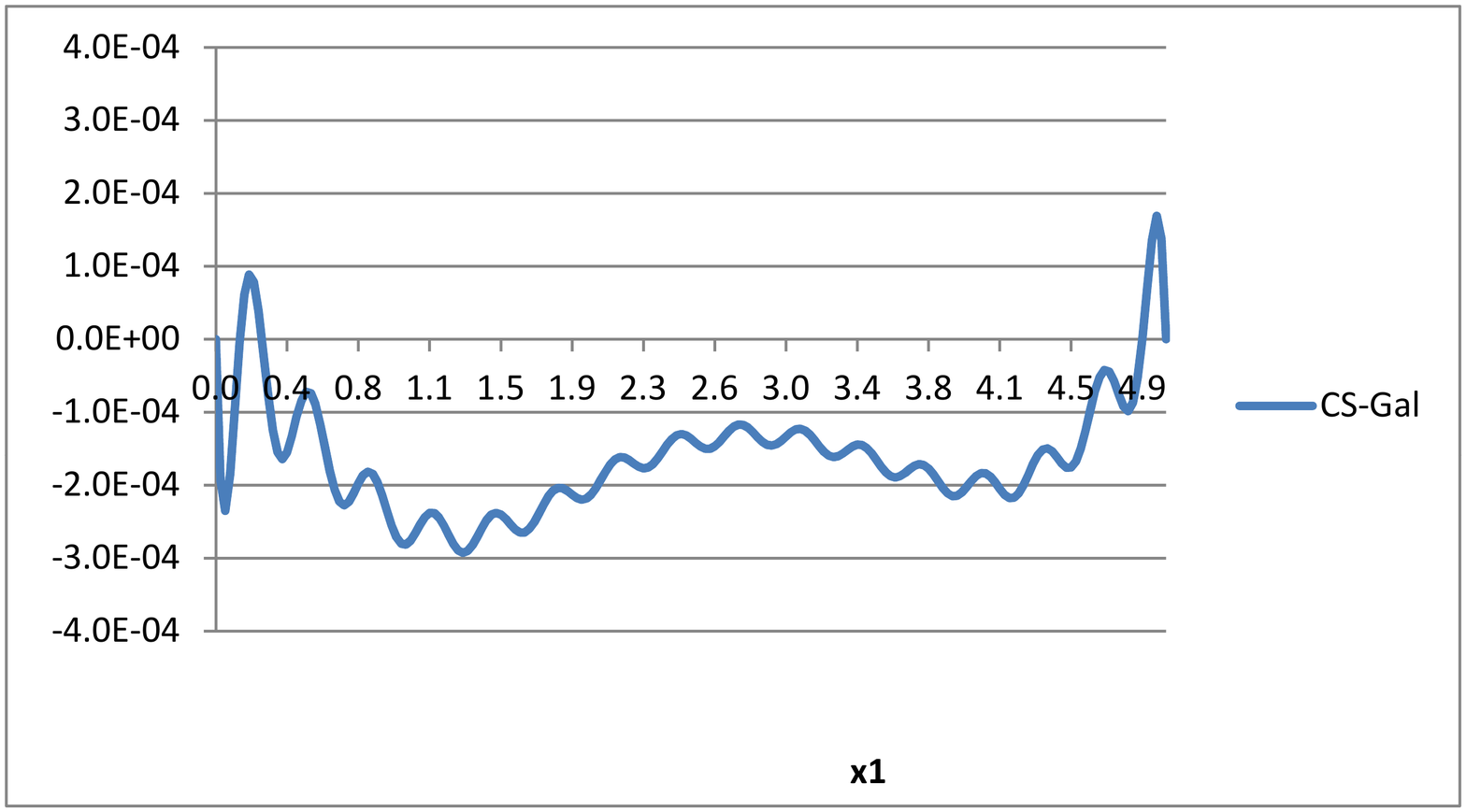}}
\caption{Comparison of the Galerkin and ADI solutions for the rectangle
problem with $\protect\tau =1$, $\protect\rho =-0.900$. The rectangle is
defined as follows $0\leq x_{1}\leq 5$, $0\leq x_{2}\leq 4$. We use the same
parameters as in Figure \protect\ref{fig:quadrant}.}
\label{fig:rectangle}
\end{figure}

Analytical and numerical solutions are compared in Figure \ref{fig:rectangle}%
. This figure shows that the ADI solution does converge to the Galerkin one
and that this convergence is good. Thus, for the rectangle problem the
Galerkin and ADI methods produce consistent results.

\begin{equation*}
\text{Fig \ref{fig:rectangle} near here}
\end{equation*}

\section{Conclusions and recommendations\label{Conclusion}}

In this paper we considered the pricing problem for vanilla and exotic
options in the LSV (more specifically QLSV) framework. We described several
known numerical methods for solving the corresponding problem with a special
emphasis on the choice of the proper boundary conditions. We observed that
for call options the CS method and its modifications have better convergence
in time than the simple Do method. However, for DNT options this advantage
disappears. In addition, we proposed a novel Galerkin-Ritz inspired method
and convincingly demonstrated that, when applicable, it is very efficient
and fast.\ This is due to the fact that the Galerkin method allows one to
reduce the amount of computations required for a typical ADI method by
treating the $x_{1}$-direction in a more natural fashion. We also emphasized
close links between the Galerkin method and the method of expansion in
powers of $\rho $. We showed that for $\rho =0$ the solutions produced via
the Galerkin method are exact. Whenever possible, we used analytical
solutions for benchmarking purposes and showed that numerical solutions
converge to the analytical ones in the limit.

We wish to thank Leif Andersen, Nicolas Hutchings, Stewart Inglis, Marsha
Lipton, Artur Sepp, and David Shelton for useful discussions.

\newpage

\appendix

\section{Brief comments on the Monte Carlo method\label{Broadie}}

A version of the Monte Carlo method exploiting formula (\ref{x_tT2}) was
proposed by Broadie and Kaya, \cite{broadie}. \ While it is well-known that $%
\chi _{\tau }\left( \left. v_{T}\right\vert v_{t}\right) $ is the so-called
non-central chi-square distribution given by Eq. (\ref{xi_vtvT}), the
conditional probability $\chi _{\tau }\left( \left. I_{t}^{T}\right\vert
v_{t},v_{T}\right) $ is more difficult to compute. By using general formulas
of \cite{lipton-book}, Section 13.11, where the general transitional
probability density for the Heston process $\left( x_{t},v_{t}\right) $ was
computed for the first time, and the augmentation techniques of Section
13.2, one can easily show that the characteristic function of the
conditional distribution of $I_{t}^{T}$ is given by%
\begin{equation}
Q\left( l,\tau ,v_{t},v_{T}\right) =\frac{P\left( R\left( l\right) ,\tau
,v_{t},v_{T}\right) }{P\left( \kappa ,\tau ,v_{t},v_{T}\right) },
\label{char_func_ItT}
\end{equation}%
where%
\begin{equation}
P\left( \kappa ,\tau ,v_{t},v_{T}\right) =\psi \left( \kappa ,\tau \right)
\exp \left( -\bar{\psi}\left( \kappa ,\tau \right) \left( v_{t}+v_{T}\right)
\right) I_{\vartheta }\left( 2\psi \left( \kappa ,\tau \right) \sqrt{%
v_{t}v_{T}}\right) ,  \label{aux_func}
\end{equation}%
\begin{equation}
\bar{\psi}\left( \kappa ,\tau \right) =\psi \left( \kappa ,\tau \right)
\cosh \left( \frac{\kappa \tau }{2}\right) ,  \label{psi_alpha_tau3}
\end{equation}%
\begin{equation}
R\left( l\right) =\sqrt{\kappa ^{2}-2i\varepsilon ^{2}l}.  \label{R(l)}
\end{equation}%
Accordingly,%
\begin{equation}
\chi _{\tau }\left( \left. I_{t}^{T}\right\vert v_{t},v_{T}\right) =\frac{1}{%
2\pi }\int_{-\infty }^{\infty }Q\left( l,\tau ,v_{t},v_{T}\right)
e^{-ilI_{t}^{T}}dl.  \label{I_cond_dist}
\end{equation}%
It is interesting to note that, in contrast to $\chi _{\tau }\left( \left.
v_{T}\right\vert v_{t}\right) $, $\chi _{\tau }\left( \left.
I_{t}^{T}\right\vert v_{t},v_{T}\right) $ is symmetric with respect to the
transposition $v_{t}\leftrightarrow v_{T}$. A similar formula is given in by
Broadie and Kaya, \cite{broadie}; however, their derivation, which is based
on the reduction of the square-root process to the Bessel process, is rather
indirect and unnecessarily complex. Thus, in order to find $\chi _{\tau
}\left( \left. I_{t}^{T}\right\vert v_{t},v_{T}\right) $, we need to
calculate the inverse Fourier transform of the corresponding characteristic
function. Needless to say that this is a difficult (but not insurmountable)
task, and should be avoided if possible.

\section{Derivation of equations (\protect\ref{shifted_transform}), (\protect
\ref{quad_phi2}), (\protect\ref{quad_phi13})\label{Fourier}}

In order to compute $\nu ^{DH}\left( k\right) $ we use the formulas%
\begin{equation}
\begin{array}{lll}
\int_{0}^{x}e^{cx^{\prime }}\sin \left( dx^{\prime }\right) dx^{\prime } & =
& \frac{e^{cx}\left[ c\sin \left( dx\right) -d\cos \left( dx\right) \right]
+d}{c^{2}+d^{2}}, \\ 
\int_{0}^{x}\sinh \left( cx^{\prime }\right) \sin \left( dx^{\prime }\right)
dx^{\prime } & = & \frac{c\cosh \left( cx\right) \sin \left( dx\right)
-d\sinh \left( cx\right) \cos \left( dx\right) }{c^{2}+d^{2}},%
\end{array}%
\end{equation}%
and get%
\begin{equation}
\begin{array}{lll}
\nu ^{DH}\left( k\right) & = & \int_{X_{0}^{DH}}^{\infty }u^{DH}\left(
x_{1}\right) \sin \left( k\left( x_{1}-X_{0}^{DH}\right) \right) dx_{1} \\ 
& = & \frac{2\sqrt{1-\beta }}{\beta }\int_{X_{0}^{DH}}^{X_{K}^{DH}}\sinh
\left( \QTOVERD. . {1}{2}\beta \left( x_{1}-X_{0}^{DH}\right) \right) \sin
\left( k\left( x_{1}-X_{0}^{DH}\right) \right) dx_{1} \\ 
&  & +K\int_{X_{K}^{DH}}^{\infty }e^{-\QTOVERD. . {1}{2}\beta x_{1}}\sin
\left( k\left( x_{1}-X_{0}^{DH}\right) \right) dx_{1} \\ 
& = & \frac{2\sqrt{1-\beta }}{\beta }\int_{0}^{Y_{K0}^{DH}}\sinh \left(
\QTOVERD. . {1}{2}\beta x\right) \sin \left( kx\right) dx+\frac{K}{\sqrt{%
1-\beta }}\int_{Y_{K0}^{DH}}^{\infty }e^{-\QTOVERD. . {1}{2}\beta x}\sin
\left( kx\right) dx \\ 
& = & \frac{\sqrt{1-\beta }}{\beta \lambda ^{DH}\left( k\right) }\left(
e^{\QTOVERD. . {1}{2}\beta Y_{K0}^{DH}}\left( \QTOVERD. . {1}{2}\beta \sin
\left( kY_{K0}^{DH}\right) -k\cos \left( kY_{K0}^{DH}\right) \right) \right.
\\ 
&  & \left. -e^{-\QTOVERD. . {1}{2}\beta Y_{K0}^{DH}}\left( -\QTOVERD. .
{1}{2}\beta \sin \left( kY_{K0}^{DH}\right) -k\cos \left(
kY_{K0}^{DH}\right) \right) \right) \\ 
&  & -\frac{K}{\sqrt{1-\beta }\lambda ^{DH}\left( k\right) }e^{-\QTOVERD. .
{1}{2}\beta Y_{K0}^{DH}}\left( -\QTOVERD. . {1}{2}\beta \sin \left(
kY_{K0}^{DH}\right) -k\cos \left( kY_{K0}^{DH}\right) \right) \\ 
& = & \left( \QTOVERD. . {1}{2}\sqrt{1-\beta }e^{\QTOVERD. . {1}{2}\beta
Y_{K0}^{DH}}+\QTOVERD. . {1}{2}\left( \sqrt{1-\beta }+\frac{\beta K}{\sqrt{%
1-\beta }}\right) e^{-\QTOVERD. . {1}{2}\beta Y_{K0}^{DH}}\right) \frac{\sin
\left( kY_{K0}^{DH}\right) }{\lambda ^{DH}\left( k\right) } \\ 
&  & +\left( -\frac{\sqrt{1-\beta }}{\beta }e^{\QTOVERD. . {1}{2}\beta
Y_{K0}^{DH}}+\left( \frac{\sqrt{1-\beta }}{\beta }+\frac{K}{\sqrt{1-\beta }}%
\right) e^{-\QTOVERD. . {1}{2}\beta Y_{K0}^{DH}}\right) \frac{k\cos \left(
kY_{K0}^{DH}\right) }{\lambda ^{DH}\left( k\right) } \\ 
& = & \frac{\sqrt{\sigma \left( K\right) }\sin \left( kY_{K0}^{DH}\right) }{%
\lambda ^{DH}\left( k\right) },%
\end{array}
\label{A3}
\end{equation}%
where $Y_{K0}^{DH}=X_{K}^{DH}-X_{0}^{DH}$. Similarly,%
\begin{equation}
\begin{array}{lll}
\nu _{k}^{R} & = & \frac{2}{\Delta ^{R}}\int_{X_{0}^{R}}^{X_{\infty
}^{R}}u^{R}\left( x_{1}\right) \sin \left( \zeta _{k}\left(
x_{1}-X_{0}^{R}\right) \right) dx_{1} \\ 
& = & \frac{2\sqrt{\frac{\alpha }{2}}}{\Delta ^{R}\sqrt{\omega ^{R}}}\left( 
\sqrt{\mathsf{pq}}\int_{X_{0}^{R}}^{X_{K}^{R}}\sinh \left( \sqrt{\omega ^{R}}%
\left( x_{1}-X_{0}^{R}\right) \right) \sin \left( \zeta _{k}\left(
x_{1}-X_{0}^{R}\right) \right) dx_{1}\right. \\ 
&  & \left. +K\int_{X_{K}^{R}}^{X_{\infty }^{R}}\sinh \left( \sqrt{\omega
^{R}}\left( X_{\infty }^{R}-x_{1}\right) \right) \sin \left( \zeta
_{k}\left( x_{1}-X_{0}^{R}\right) \right) dx_{1}\right) \\ 
& = & \frac{2\sqrt{\frac{\alpha }{2}}}{\Delta ^{R}\sqrt{\omega ^{R}}}\left( 
\sqrt{\mathsf{pq}}\int_{0}^{Y_{K0}^{R}}\sinh \left( \sqrt{\omega ^{R}}%
x\right) \sin \left( \zeta _{k}x\right) dx\right. \\ 
&  & \left. +\left( -1\right) ^{k+1}K\int_{0}^{Y_{\infty K}^{R}}\sinh \left( 
\sqrt{\omega ^{R}}x\right) \sin \left( \zeta _{k}x\right) dx\right) \\ 
& = & \frac{2\sqrt{\frac{\alpha }{2}}}{\Delta ^{R}\sqrt{\omega ^{R}}\lambda
_{k}^{R}}\left( \sqrt{\mathsf{pq}}\left( \sqrt{\omega ^{R}}\cosh \left( 
\sqrt{\omega ^{R}}Y_{K0}^{R}\right) \sin \left( \zeta _{k}Y_{K0}^{R}\right)
\right. \right. \\ 
&  & \left. -\sinh \left( \sqrt{\omega ^{R}}Y_{K0}^{R}\right) \zeta _{k}\cos
\left( \zeta _{k}Y_{K0}^{R}\right) \right) \\ 
&  & +\left( -1\right) ^{k+1}K\left( \sqrt{\omega ^{R}}\cosh \left( \sqrt{%
\omega ^{R}}Y_{\infty K}^{R}\right) \sin \left( \zeta _{k}Y_{\infty
K}^{R}\right) \right. \\ 
&  & \left. \left. -\sinh \left( \sqrt{\omega ^{R}}Y_{\infty K}^{R}\right)
\zeta _{k}\cos \left( \zeta _{k}Y_{\infty K}^{R}\right) \right) \right) \\ 
& = & \frac{2\sqrt{\frac{\alpha }{2}}}{\Delta ^{R}\lambda _{k}^{R}}\left( 
\sqrt{\mathsf{pq}}\left( \sqrt{\frac{\left( K-\mathsf{q}\right) \mathsf{p}}{%
\left( K-\mathsf{p}\right) \mathsf{q}}}+\sqrt{\frac{\left( K-\mathsf{p}%
\right) \mathsf{q}}{\left( K-\mathsf{q}\right) \mathsf{p}}}\right) \right.
\\ 
&  & \left. +K\left( \sqrt{\frac{\left( K-\mathsf{p}\right) }{\left( K-%
\mathsf{q}\right) }}+\sqrt{\frac{\left( K-\mathsf{q}\right) }{\left( K-%
\mathsf{p}\right) }}\right) \right) \sin \left( \zeta _{k}Y_{K0}^{R}\right)
\\ 
&  & +\frac{2\sqrt{\frac{\alpha }{2}}\zeta _{k}}{\Delta ^{R}\sqrt{\omega ^{R}%
}\lambda _{k}^{R}}\left( -\sqrt{\mathsf{pq}}\left( \sqrt{\frac{\left( K-%
\mathsf{q}\right) \mathsf{p}}{\left( K-\mathsf{p}\right) \mathsf{q}}}-\sqrt{%
\frac{\left( K-\mathsf{p}\right) \mathsf{q}}{\left( K-\mathsf{q}\right) 
\mathsf{p}}}\right) \right. \\ 
&  & \left. +K\left( \sqrt{\frac{\left( K-\mathsf{p}\right) }{\left( K-%
\mathsf{q}\right) }}-\sqrt{\frac{\left( K-\mathsf{q}\right) }{\left( K-%
\mathsf{p}\right) }}\right) \right) \cos \left( \zeta _{k}Y_{K0}^{R}\right)
\\ 
& = & \frac{2\sqrt{\frac{\alpha }{2}}}{\Delta ^{R}\lambda _{k}^{R}}\sqrt{%
\left( K-\mathsf{p}\right) \left( K-\mathsf{q}\right) }\sin \left( \zeta
_{k}Y_{K0}^{R}\right) \\ 
& = & \frac{2\sqrt{\sigma \left( K\right) }\sin \left( \zeta
_{k}Y_{K0}^{R}\right) }{\Delta ^{R}\lambda _{k}^{R}}.%
\end{array}
\label{A4}
\end{equation}%
where $Y_{K0}^{R}=X_{K}^{R}-X_{0}^{R}$, $Y_{\infty K}^{R}=X_{\infty
}^{R}-X_{K}^{R}$. Finally, in order to compute $\nu _{k}^{I}$ we use the
formula%
\begin{equation}
\int_{0}^{x}\sin \left( cx^{\prime }\right) \sin \left( dx^{\prime }\right)
dx^{\prime }=\frac{c\cos \left( cx\right) \sin \left( dx\right) -d\sin
\left( cx\right) \cos \left( dx\right) }{\left( d^{2}-c^{2}\right) },
\end{equation}%
and get%
\begin{equation}
\begin{array}{lll}
\nu _{k}^{I} & = & \frac{2}{\Delta ^{I}}\int_{X_{0}^{I}}^{X_{\infty
}^{I}}u^{I}\left( x_{1}\right) \sin \left( \zeta _{k}\left(
x_{1}-X_{0}^{I}\right) \right) dx_{1} \\ 
& = & \frac{2\sqrt{\frac{\alpha }{2}}}{\Delta ^{I}\sqrt{\left\vert \omega
^{I}\right\vert }}\left( \sqrt{\mathsf{m}^{2}+\mathsf{n}^{2}}%
\int_{X_{0}^{I}}^{X_{K}^{I}}\sin \left( \sqrt{\left\vert \omega
^{I}\right\vert }\left( x_{1}-X_{0}^{I}\right) \right) \sin \left( \zeta
_{k}\left( x_{1}-X_{0}^{I}\right) \right) dx_{1}\right. \\ 
&  & \left. +K\int_{X_{K}^{I}}^{X_{\infty }^{I}}\sin \left( \sqrt{\left\vert
\omega ^{I}\right\vert }\left( X_{\infty }^{I}-x_{1}\right) \right) \sin
\left( \zeta _{k}\left( x_{1}-X_{0}^{I}\right) \right) dx_{1}\right) \\ 
& = & \frac{2\sqrt{\frac{\alpha }{2}}}{\Delta ^{I}\sqrt{\left\vert \omega
^{I}\right\vert }}\left( \sqrt{\mathsf{m}^{2}+\mathsf{n}^{2}}%
\int_{0}^{Y_{K0}^{I}}\sin \left( \sqrt{\left\vert \omega ^{I}\right\vert }%
x\right) \sin \left( \zeta _{k}x\right) dx\right. \\ 
&  & \left. +\left( -1\right) ^{k+1}K\int_{0}^{Y_{\infty K}^{I}}\sin \left( 
\sqrt{\left\vert \omega ^{I}\right\vert }x\right) \sin \left( \zeta
_{k}x\right) dx\right) \\ 
& = & \frac{2\sqrt{\frac{\alpha }{2}}}{\Delta ^{I}\sqrt{\left\vert \omega
^{I}\right\vert }\lambda _{k}^{I}}\left( \sqrt{\mathsf{m}^{2}+\mathsf{n}^{2}}%
\left( \sqrt{\left\vert \omega ^{I}\right\vert }\cos \left( \sqrt{\left\vert
\omega ^{I}\right\vert }Y_{K0}^{I}\right) \sin \left( \zeta
_{k}Y_{K0}^{I}\right) \right. \right. \\ 
&  & \left. -\zeta _{k}\sin \left( \sqrt{\left\vert \omega ^{I}\right\vert }%
Y_{K0}^{I}\right) \cos \left( \zeta _{k}Y_{K0}^{I}\right) \right) \\ 
&  & +\left( -1\right) ^{k+1}K\left( \sqrt{\left\vert \omega ^{I}\right\vert 
}\cos \left( \sqrt{\left\vert \omega ^{I}\right\vert }Y_{\infty
K}^{I}\right) \sin \left( \zeta _{k}Y_{\infty K}^{I}\right) \right. \\ 
&  & \left. \left. -\zeta _{k}\sin \left( \sqrt{\left\vert \omega
^{I}\right\vert }Y_{\infty K}^{I}\right) \cos \left( \zeta _{k}Y_{\infty
K}^{I}\right) \right) \right) \\ 
& = & \frac{2\sqrt{\frac{\alpha }{2}}}{\Delta ^{I}\lambda _{k}^{I}}\left( 
\sqrt{\mathsf{m}^{2}+\mathsf{n}^{2}}\cos \left( \sqrt{\left\vert \omega
^{I}\right\vert }Y_{K0}^{I}\right) +K\cos \left( \sqrt{\left\vert \omega
^{I}\right\vert }Y_{\infty K}^{I}\right) \right) \sin \left( \zeta
_{k}Y_{K0}^{I}\right) \\ 
&  & -\frac{2\sqrt{\frac{\alpha }{2}}\zeta _{k}}{\Delta ^{I}\sqrt{\left\vert
\omega ^{I}\right\vert }\lambda _{k}^{I}}\left( \sqrt{\mathsf{m}^{2}+\mathsf{%
n}^{2}}\sin \left( \sqrt{\left\vert \omega ^{I}\right\vert }%
Y_{K0}^{I}\right) -K\sin \left( \sqrt{\left\vert \omega ^{I}\right\vert }%
Y_{\infty K}^{I}\right) \right) \cos \left( \zeta _{k}Y_{K0}^{I}\right) \\ 
& = & \frac{2\sqrt{\frac{\alpha }{2}}\sqrt{\left( K-\mathsf{m}\right) ^{2}+%
\mathsf{n}^{2}}}{\Delta ^{I}\lambda _{k}^{I}}\sin \left( \zeta
_{k}Y_{K0}^{I}\right) \\ 
& = & \frac{2\sqrt{\sigma \left( K\right) }}{\Delta ^{I}\lambda _{k}^{I}}%
\sin \left( \zeta _{k}Y_{K0}^{I}\right) ,%
\end{array}
\label{A6}
\end{equation}%
where $Y_{K0}^{I}=X_{K}^{I}-X_{0}^{I}$, $Y_{\infty K}^{I}=X_{\infty
}^{I}-X_{K}^{I}$.

\section{Derivation of equation (\protect\ref{gl})\label{eqgl}}

According to Eq. (\ref{gl}). we have%
\begin{equation}
g_{k,\tau }\left( \tau ,r\right) =-\frac{\upsilon \mathsf{J}_{k}^{\prime
}\left( \upsilon \right) }{\tau },\ \ \ g_{k,r}\left( \tau ,r\right) =\frac{r%
\mathsf{J}_{k}^{\prime }\left( \upsilon \right) }{2\tau },\ \ \
g_{k,rr}\left( \tau ,r\right) =\frac{\upsilon \mathsf{J}_{k}^{\prime \prime
}\left( \upsilon \right) }{\tau }+\frac{\mathsf{J}_{k}^{\prime }\left(
\upsilon \right) }{2\tau },
\end{equation}%
so that%
\begin{equation}
\begin{array}{l}
g_{k,\tau }\left( \tau ,r\right) -\QTOVERD. . {1}{2}\left( g_{k,rr}\left(
\tau ,r\right) +\frac{1}{r}g_{k,r}\left( \tau ,r\right) -\frac{\zeta _{k}^{2}%
}{r^{2}}g_{k}\left( \tau ,r\right) \right) \\ 
=-\frac{1}{\tau }\left( \upsilon \mathsf{J}_{k}^{\prime \prime }\left(
\upsilon \right) +\left( 2\upsilon +1\right) \mathsf{J}_{k}^{\prime }\left(
\upsilon \right) -\QTOVERD. . {1}{4}\frac{\zeta _{k}^{2}}{\upsilon }\mathsf{J%
}_{k}\left( \upsilon \right) \right) .%
\end{array}%
\end{equation}%
Thus, we need to prove that 
\begin{equation}
\upsilon \mathsf{J}_{k}^{\prime \prime }\left( \upsilon \right) +\left(
2\upsilon +1\right) \mathsf{J}_{k}^{\prime }\left( \upsilon \right)
-\QTOVERD. . {1}{4}\frac{\zeta _{k}^{2}}{\upsilon }\mathsf{J}_{k}\left(
\upsilon \right) =0.
\end{equation}%
Since%
\begin{equation}
\mathsf{J}_{k}\left( \upsilon \right) =\sqrt{\upsilon }e^{-\upsilon }\left(
I_{\frac{1}{2}\left( \zeta _{k}-1\right) }\left( \upsilon \right) +I_{\frac{1%
}{2}\left( \zeta _{k}+1\right) }\left( \upsilon \right) \right) =\sqrt{%
\upsilon }e^{-\upsilon }\mathsf{K}_{k}\left( \upsilon \right) ,
\end{equation}%
we can easily derive an equivalent equation for $\mathsf{K}_{k}\left(
\upsilon \right) $:%
\begin{equation}
\upsilon ^{2}\mathsf{K}_{k}^{\prime \prime }\left( \upsilon \right)
+2\upsilon K_{k}^{\prime }\left( \upsilon \right) -\left( \upsilon
^{2}+\upsilon +\QTOVERD. . {1}{4}\left( \zeta _{k}^{2}-1\right) \right) 
\mathsf{K}_{k}\left( \upsilon \right) =0.
\end{equation}%
By definition of the modified Bessel function we have%
\begin{equation}
\begin{array}{l}
\upsilon ^{2}I_{\frac{1}{2}\left( \zeta _{k}\pm 1\right) }^{\prime \prime
}\left( \upsilon \right) +2\upsilon I_{\frac{1}{2}\left( \zeta _{k}\pm
1\right) }^{\prime }\left( \upsilon \right) -\left( \upsilon ^{2}+\upsilon
+\QTOVERD. . {1}{4}\left( \zeta _{k}^{2}-1\right) \right) I_{\frac{1}{2}%
\left( \zeta _{k}\pm 1\right) }\left( \upsilon \right) \\ 
=\upsilon I_{\frac{1}{2}\left( \zeta _{k}\pm 1\right) }^{\prime }\left(
\upsilon \right) -\left( \upsilon -\QTOVERD. . {1}{2}\left( 1\pm \zeta
_{k}\right) \right) I_{\frac{1}{2}\left( \zeta _{k}\pm 1\right) }\left(
\upsilon \right) .%
\end{array}%
\end{equation}%
Summation of these expressions yields%
\begin{equation}
\begin{array}{l}
\upsilon ^{2}\mathsf{K}_{k}^{\prime \prime }\left( \upsilon \right)
+2\upsilon \mathsf{K}_{k}^{\prime }\left( \upsilon \right) -\left( \upsilon
^{2}+\upsilon +\QTOVERD. . {1}{4}\left( \zeta _{k}^{2}-1\right) \right) 
\mathsf{K}_{k}\left( \upsilon \right) \\ 
=\upsilon \mathsf{K}_{k}^{\prime }\left( \upsilon \right) -\left( \upsilon
-\QTOVERD. . {1}{2}\right) \mathsf{K}_{k}\left( \upsilon \right) +\QTOVERD.
. {1}{2}\zeta _{k}\left( I_{\frac{1}{2}\left( \zeta _{k}+1\right) }\left(
\upsilon \right) -I_{\frac{1}{2}\left( \zeta _{k}-1\right) }\left( \upsilon
\right) \right) \\ 
=\upsilon I_{\frac{1}{2}\left( \zeta _{k}+3\right) }\left( \upsilon \right)
+\left( \zeta _{k}+1\right) I_{\frac{1}{2}\left( \zeta _{k}+1\right) }\left(
\upsilon \right) -\upsilon I_{\frac{1}{2}\left( \zeta _{k}-1\right) }\left(
\upsilon \right) \\ 
=0,%
\end{array}%
\end{equation}%
as claimed. Here we use the fact that%
\begin{equation}
I_{\nu -1}\left( \upsilon \right) -I_{\nu +1}\left( \upsilon \right) =\frac{%
2\nu }{\upsilon }I_{\nu }\left( \upsilon \right) .
\end{equation}%
By using asymptotic expressions for the modified Bessel functions, it is
easy to check that $\mathsf{J}\left( \upsilon \right) $ satisfies the
corresponding boundary and initial conditions.

We notice in passing that Equation (\ref{gl}) is similar to the familiar
expression for the survival probability of the standard Brownian motion on
the positive semi-axis with absorbing boundary, which can be written as
follows:%
\begin{equation}
g\left( \tau ,x_{1}\right) =\Phi \left( \upsilon \right) -\Phi \left(
-\upsilon \right) ,
\end{equation}%
where $\upsilon =x_{1}/\sqrt{\tau }$.

\clearpage

\clearpage

\clearpage

\clearpage

\clearpage

\clearpage

\clearpage

\clearpage

\clearpage

\end{document}